\documentclass[twocolumn]{aastex701}

\shorttitle{Icy Volatile Enhancements in Evolving Protoplanetary Disks}
\shortauthors{Yunerman, Price, and \"Oberg}

\received{November 25, 2025}
\revised{March 24, 2025}
\revised{March 31, 2025}
\accepted{April 14, 2026}

\submitjournal{APJ}

\usepackage[T1]{fontenc}
\usepackage[utf8]{inputenc}

\begin{document}

\title{Icy Volatile Enhancements in Evolving Protoplanetary Disks}

\author[orcid=0009-0008-4937-3314,sname='Yunerman']{Elizabeth S. Yunerman}
\affiliation{Center for Astrophysics $|$ Harvard \& Smithsonian, 60 Garden Street, Cambridge, MA 02138, USA}
\email[show]{eyunerman@cfa.harvard.edu}  

\author[orcid=0000-0002-3286-3543, sname='Price']{Ellen M. Price} 
\affiliation{Department of Astronomy, University of Chicago, 5640 South Ellis Ave., Chicago, IL 60637, USA}
\email{emprice@uchicago.edu}

\author[orcid=0000-0001-8798-1347, sname='\"Oberg']{Karin I. \"Oberg}
\affiliation{Center for Astrophysics $|$ Harvard \& Smithsonian, 60 Garden Street, Cambridge, MA 02138, USA}
\email{koberg@cfa.harvard.edu}


\begin{abstract}

Protoplanetary disk ice lines shape a multitude of planet formation processes, setting the environmental composition through evolution.
Ice line locations depend on molecular sublimation and deposition properties, but in dynamic disks where temperature and density structures change, so do the expected compositions of planets and planetesimals. 
In turbulent viscous disks with particle drift, thermal evolution, and desorption/adsorption, \cite{price_ice-coated_2021} demonstrated that the CO/H$_2$O ice ratio beyond the CO ice line can become enhanced by $\sim10\times$.  
We expand on their work by incorporating additional carbon, nitrogen, and oxygen species, more particle sizes, and a broader disk parameter exploration.
We find that before $\sim0.5$~Myr, volatile ices are enhanced relative to H$_2$O as the outer disk is desiccated by drift, while at later disk times outward advection and volatile deposition further increase relative volatile icy enhancements beyond the evolving critical disk radius. 
The outcome of these combined relative icy enhancement to H$_2$O mechanisms is solid C/O $\sim$ N/O $\sim1$ beyond the hypervolatile ice lines, much higher than expected in static disks.
Hypervolatiles (N$_2$, CO, and CH$_4$) robustly increase to $\sim100\times$ across the explored parameter space, while mid-volatiles (CO$_2$ and NH$_3$) are sensitive to model choices, with enhancements ranging from $\sim2-50\times$. 
Together these results demonstrate that coupling disk dynamics with simple sublimation and deposition chemistry is fundamental to predicting grain, planetesimal, and planetary compositions, particularly the role of advection in redistributing volatiles across disk radii.
\end{abstract}



\section{Introduction}\label{sec:intro}

Compositions of planets and planetesimals are built up by a combination of dust and ice, and in the case of planets, gas, acquired from their planet-forming, or protoplanetary, disk \citep{oberg_astrochemistry_2021}. 
Two outcomes of this process that are of special interest to this paper are the formation of gas giants, whose elemental ratios should encode information about the gas and solid compositions at their formation location \citep[e.g.][]{oberg_effects_2011}, and the formation of comets, which should present near-pristine records of the icy volatile composition in the outer regions of our own disk, the Solar Nebula \citep[e.g.][]{mumma_chemical_2011}.
Giant planet atmospheres form primarily through the accretion of gas, which is likely enriched by dust and ice pollution \citep{ikoma_formation_2025,youdin_formation_2025,venturini_planet_2016}, and these two volatile origins, together, set the relative elemental abundance ratios, commonly C/N/O/H \citep[e.g.,][]{madhusudhan_exoplanetary_2019,jiang_chemical_2023,wang_early_2025}. 
Comets, left-over icy planetesimals, offer complementary constraints on planet formation if their volatile inventories can be linked to specific disk regions or disk processes \citep{blum_formation_2022,altwegg_cometary_2019,seligman_volatile_2022}. 
In both cases, linking present-day abundance observations to formation locations depends on the development of accurate models of disk volatile distributions \citep{turrini_tracing_2021}, complicated by dynamical disk processes including the fluid dynamics of small gas and solid particles, disk and atmospheric chemistry, and planetary migration, often modeled independently \citep[][and references therein]{molliere_interpreting_2022}.

Volatile distributions in static disk models are governed by species ice lines, the disk radii where major volatiles transition from the gas to the ice phase \citep{oberg_effects_2011}. 
Ice line locations are determined by the assumed disk temperature and species-specific kinetics \citep[e.g.][]{hollenbach_water_2008}.
Modeling ice lines and distributions has provided a popular framework for interpreting gas giant C/O and N/O ratios \citep{turrini_tracing_2021}, potentially situating where a planet accreted its gas relative to ice lines of water and carbon- and nitrogen-bearing species.
Interpreting  C/O, C/H and O/H ratios in gas giants may be quite complex, since these ratios may also reflect the degree of solid enrichment during planet formation and evolution \citep{bitsch_how_2022}. 
Secondly, the predicted C/O of the formed planet changes when considering particle size, growth and drift \citep{piso_co_2015,booth_chemical_2017,schneider_how_2021}, or for example whether the planet accreted at early or late disk times \citep{mah_close-ice_2023}.
While attractive in its simplicity, attempting to link elemental ratios to formation location remains difficult, and with the increasing number of observations of disks \citep{andrews_disk_2018,oberg_molecules_2021,banzatti_jwst_2023} and (exo)planets \citep{bitsch_how_2022,penzlin_bowie-align_2024} that are in tension with expectations from the static disk model \citep{pacetti_planet_2025}, we are motivated to revisit the role of disk dynamics on the disk volatile distribution.

Observed comet abundances present an analogous story, where a static disk framework appears sufficient to explain many, but not all measured volatile compositions. Most solar system comets are abundant in H$_2$O ice with variable, but significant NH$_3$, CO$_2$, CO, and CH$_3$OH ice contents, which can be qualitatively explained by  comet formation beyond the H$_2$O ice line \citep{mumma_chemical_2011, ahearn_cometary_2012, altwegg_cometary_2019}. There are, however, an increasing number of comets that do not fit this template.
An observed population of H$_2$O depleted comets \citep{bockelee-morvan_composition_2017}, often times enriched in CO and CO$_2$ \citep{mckay_co2_2016}, likely require a more dynamical disk or formation scenario to explain their compositions \citep{rubin_origin_2020,blum_formation_2022,morbidelli_formation_2024}.
A small number of comets with extreme CO abundances, such as C/2016 R2 (PanSTARRS) \citep{biver_extraordinary_2018,mckay_peculiar_2019} and C/2023 H2 (Lemmon) \citep{lippi_chemical_2024}, are best explained by a dynamical depletion of water in the outermost disk, perhaps coupled with ice chemistry  \citep{eistrup_cometary_2019,price_ice-coated_2021,seligman_volatile_2022}.
Finally, the three interstellar comets identified so far---1I/`Oumuamua \citep{meech_brief_2017}, 2I/Borisov \citep{jewitt_initial_2019}, and 3I/ATLAS \citep{bolin_interstellar_2025}---reveal yet another population of evolved planetesimals that may have compositional links to the cold, dynamical, outer disk \citep{jewitt_interstellar_2023}.

To address these observations, there are now a number of different modeling approaches that explore how different disk processes reshape the ice and gas volatile distributions in disks \citep[see][and references therein]{oberg_protoplanetary_2023,drazkowska_planet_2023, miotello_setting_2023}.
First, pebble drift, followed by the return of volatiles to the gas phase after crossing their ice lines, readily produces local vapor enhancements interior to ice lines, though the size of these enhancements depends strongly on the effectiveness of diffusion \citep{cuzzi_material_2004, oberg_excess_2016,schneider_how_2021,bitsch_how_2022}.
Time-dependent viscous diffusion can also transport vapor and solids coupled to the bulk gas around the H$_2$O ice line, enabling outward recondensation and local solid enrichment just beyond the ice line, perhaps sufficient to promote planetesimal formation \citep{ciesla_evolution_2006,booth_chemical_2017,schoonenberg_planetesimal_2017,drazkowska_planetesimal_2017,ros_effect_2019}.
The relative importance of pebble- and gas-mediated volatile distributions depends on the solid particle size distribution, and how it evolves through collisional growth and fragmentation \citep{schoonenberg_planetesimal_2017,hyodo_formation_2019,estrada_global_2022}.
In the vertical dimension, particle settling and turbulent stirring together with volatile diffusion can substantially deplete volatiles in the gas, and enhance them in the solid phase, below the relevant volatile ice surfaces \citep{meijerink_radiative_2009,krijt_transport_2018,krijt_co_2020}.
While these modeling efforts only represent a fraction of the ongoing work, they already demonstrate that coupling disk dynamics and ice sublimation and deposition can change the volatile compositions by orders-of-magnitude across the disk, motivating further exploration of when and where to expect dynamic volatile enhancements and depletions in protoplanetary disks.

We pursue one such exploration, building on the work of \cite{price_ice-coated_2021}, who developed a disk model that includes turbulent viscous diffusion, advection, drift, desorption, and adsorption to explain the existence of CO-dominated comets.
In particular, we increase the number of C-, N-, and O-bearing disk species, expand the dust particle treatment, and numerically solve for icy (and gas-phase) volatile enhancements and depletions beyond the H$_2$O ice line for a range of parametric disk conditions.
The disk model is described in Section~\ref{sec:physical_disk_model}, and the physics governing ice lines are detailed in Section~\ref{sec:molecular_physics}.
Fiducial model results are presented in Section~\ref{sec:results}, followed by a discussion of variations from the fiducial scenario in Section~\ref{sec:variations_from_fiducial}.
Final ice abundances in the context of icy comets, and ice and gas C/O and N/O ratios in the context of gas giant atmosphere formation, are discussed in Sections \ref{sec:comets} and \ref{sec:gas_giants}, respectively.
A summary of this work and main conclusions are provided in Section~\ref{sec:summary_conc}.

\section{Disk Model} \label{sec:physical_disk_model}

We build on the work of \cite{price_ice-coated_2021} to compute radially and temporally evolving gas and solid surface densities, $\Sigma(r,t)$, to trace species-dependent abundances. Subscripts ``$g$'' and ``$s$'' will denote ``gas'' and ``solid'' throughout the entirety of this work. We focus on the evolution of a low-mass star of $M_{\star}=0.5M_{\odot}$ with a surrounding protoplanetary disk that is 5\% of the stellar mass, $M_{\rm disk} = 0.05M_{\star}$, close to the typical star and disk masses of observed star--disk systems \citep{gaidos_minimum_2017,winters_solar_2019,andrews_observations_2020}. Our fiducial disk midplane model is actively evolving via accretion through turbulent viscosity, with a classic collisional-cascade of solid particles ranging from 0.1 {\textmu}m to 10 cm. In Section~\ref{sec:variations_from_fiducial}, we compare and discuss the fiducial results with those that vary from this initial condition, testing our model over an array of assumed parameter values.


\subsection{Temperature Structure} \label{ssec:disk_temp}

\begin{figure}[!htp]
 \centering
 \includegraphics[width=8.5cm]{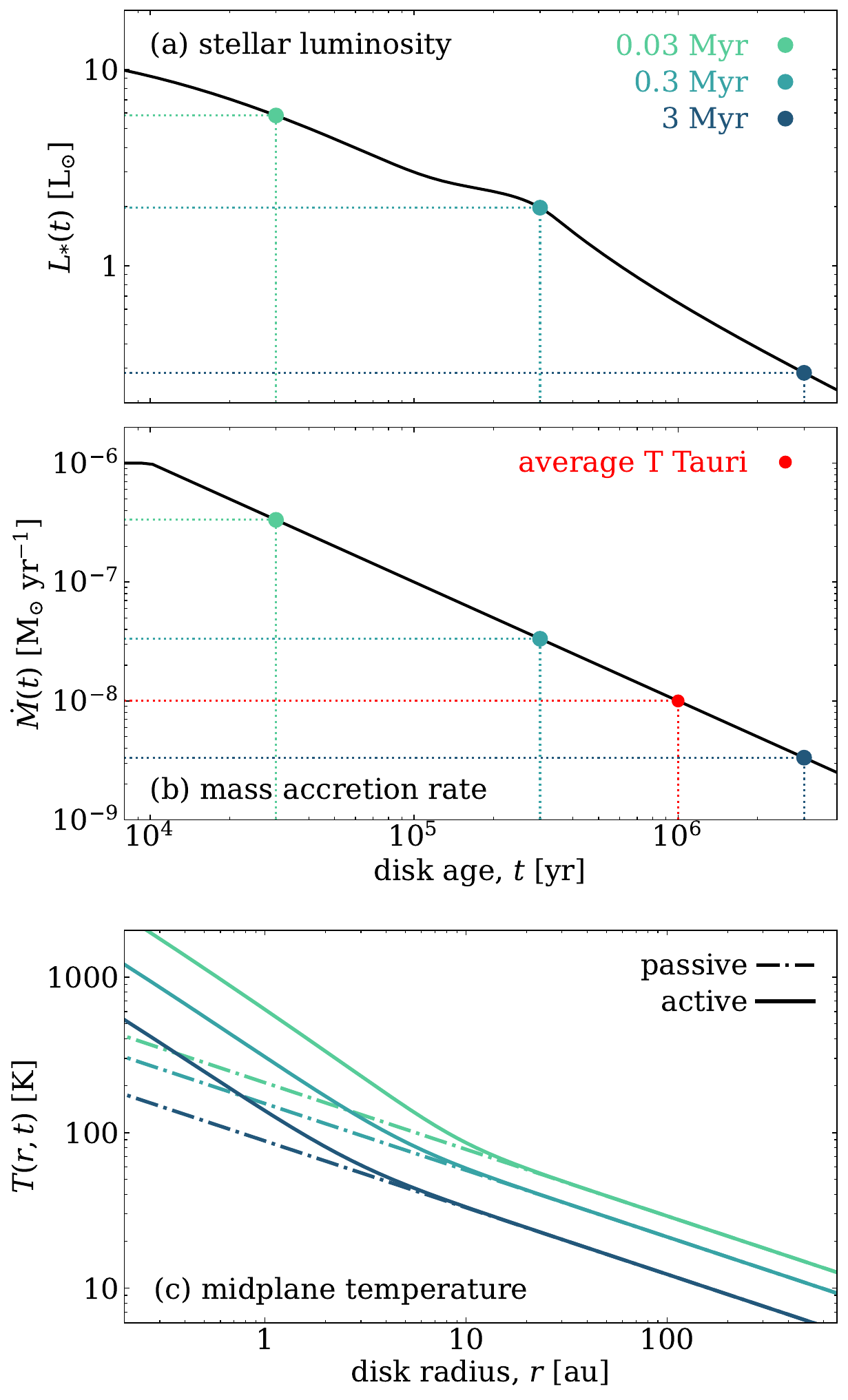}
 \caption{The disk midplane temperature may evolve significantly over the lifetime of a disk, in particular when incorporating an evolving stellar luminosity and mass accretion rate. Our model uses the MIST stellar evolution track for stellar luminosity with time, $L_{\star}(t)$, of a $0.5M_{\odot}$ M dwarf shown in panel (a), and uses the relation $\dot{M}_{\star}(t) \propto t^{-1}$ normalized to an average T Tauri star (red point) for an evolving mass accretion rate shown in panel (b). Panel (c) compares passive and active disk temperature profiles while stepping through times of  0.03, 0.3, and 3 Myr, using corresponding colors denoting time through all panels.}
 \label{fig:temps}
\end{figure}

The disk temperature profile, $T(r,t)$, propagates through nearly every equation involved in modeling planet formation processes, including setting desorption rates and influencing the fluid dynamics. We test several different temperature profiles, namely the analytic forms of a passive and active disk, while considering a non-evolving (static) and evolving stellar luminosity, $L_{\star}(t)$, and mass accretion rate, $\dot{M}_{\star}(t)$. Assuming the two-layer disk approximation of dust continuum radiative transfer, as in \cite{chiang_spectral_1997}, a passive disk that is primarily heated by stellar irradiation has a midplane temperature profile following the power law
\begin{equation}
    T_{\mathrm{irr}}(r,t) = T_0(t) \times \left(\frac{r}{1~\text{au}}\right)^{-3/7}
\end{equation}
where $r$ is the disk orbital radius and 
\begin{equation}
    T_0(t) = \bigg(\frac{2}{7}\bigg)^{1/4} \bigg(\frac{L_{\star}(t)}{4\pi\sigma_{\mathrm{SB}}}\bigg)^{2/7} \bigg(\frac{k_B}{\mu m_{\mathrm{H}} GM_{\star}}\bigg)^{1/7}
\end{equation}
is the temperature at 1~au, $\sigma_{\mathrm{SB}}$ is the Stefan-Boltzmann constant, $k_B$ is the Boltzmann constant, $G$ is Newton's gravitational constant, $m_{\mathrm{H}}$ is the atomic mass of hydrogen, and $\mu=2.35$ is the mean molecular weight for a hydrogen-helium solar composition. 
The stellar luminosity will decrease over the first few million years during its pre-main sequence phase \citep[e.g.,][]{siess_internet_2000,fischer_herschel_2017,fischer_accretion_2023}. 
We use the MESA Isochrones \& Stellar Tracks (MIST) generated stellar evolution for a $ 0.5M_{\sun}$ star, to incorporate a time-dependent stellar luminosity, assuming that the disk age is that of its host star, $t_{\rm disk} = t_{\star}$, where the pre-main sequence phase is by default the point in the MESA evolution when the central temperature of the star reaches $10^5$ K, and is below the temperature for sustained hydrogen burning \citep{dotter_mesa_2016,choi_mesa_2016,paxton_modules_2011,paxton_modules_2013,paxton_modules_2015}. We note that the additional luminosity from disk accretion is not included in MESA, such that we may be slightly underestimating the disk temperature for $t < 1$ Myr, although this temperature change is likely to be minor. The resulting interpolated luminosity evolution is displayed in Figure~\ref{fig:temps}a.

In active disks, there is an additional heat source due to the energy dissipated through viscous accretion of disk material onto the host star. The midplane temperature due to accretion heating is 
\begin{equation} \label{eq.13}
    T_{\text{acc}}(r,t) = \bigg(\frac{9}{32\pi}\frac{\tau_{\text{vert}}}{\sigma_{\mathrm{SB}}}\dot{M}_{\star}(t)\Omega(r)^2\bigg)^{1/4}
\end{equation}
\citep[e.g.,][]{lee_viscous_1991,garaud_effect_2007,rosenthal_how_2020} where $\Omega(r) = \sqrt{GM_{\star}/r^3}$ is the Keplerian orbital frequency. The vertical optical depth, $\tau_{\text{vert}}=1/2\; \Sigma_{\rm g}\kappa$, depends on the evolving gas disk surface density, $\Sigma_{\rm g}(r,t)$, discussed in detail in Section~\ref{ssec:gas_evol}, along with the dust opacity, $\kappa$.
We choose a constant dust opacity of 0.5 cm$^2$ g$^{-1}$ for simplicity, although we note that the opacity is expected to vary with temperature, particle size, structure, and composition  \citep{semenov_rosseland_2003,cuzzi_utilitarian_2014,piso_minimum_2015, woitke_consistent_2016}.

The mass accretion rate through the disk, $\dot{M}$, may also evolve over a disk lifetime depending on accretion variability and stellar mass, such that $\dot{M} \propto t^{-1}$ \citep{hartmann_accretion_2016,testi_protoplanetary_2022,grant_dotm_2023,fischer_accretion_2023}. On average, T Tauri stars at around $t_{\rm TT} \sim 1$ Myr have $\dot{M}_{\rm TT} \approx 10^{-8} \: \text{M}_{\odot}\:\text{yr}^{-1}$ \citep{hartmann_accretion_1998}, and by normalizing to this value, we assume that
\begin{equation}
    \dot{M}_{\star}(t) = \dot{M}_{\rm TT} \times \bigg( \frac{t_{\rm TT}}{t} \bigg)
\end{equation}
where for $t < 1$ kyr, $\dot{M}$ is capped at $10^{-6} \: {\rm M_{\odot} yr^{-1}}$ ensuring that the accretion rate is not too high at early times for a roughly solar mass star \citep{lee_young_2020}. The time-evolving accretion rate is displayed in Figure~\ref{fig:temps}b.

The complete analytic temperature profile for an active disk is then 
\begin{equation}\label{eq:temperature}
    T(r,t) = \left[T_{\rm irr}(r,t)^4 + T_{\rm acc}(r,t)^4\right]^{1/4}
\end{equation}
where both irradiation heating and accretion heating components are made time-dependent by considering the evolving stellar luminosity and mass accretion rate. We do not take into account the higher temperatures expected at elevated disk heights since we only consider the evolution of volatiles at the midplane. Figure~\ref{fig:temps}c compares the passive and active model temperature profiles for several timesteps. The passive model assumes $T_{\rm acc}=0$. Active disks will have an inner disk dominated by accretion heating and an outer disk primarily dominated by irradiation heating. For a disk around a low-mass star, this switch moves from $\sim$5~au to $\sim$1~au during the first few million years of evolution, while the overall magnitude of the temperature decreases by $\sim3\times$ over the 3 Myr evolution. 

\subsection{Gas Disk Density Structure} \label{ssec:gas_evol}
We assume that the gaseous accretion disk is viscously evolving, such that the motion of the bulk gas in the absence of sources and sinks is given by  
\begin{equation}\label{eq:diff_eq_gas}
    \frac{\partial \Sigma_{\rm g}}{\partial t} - \frac{3}{r}\frac{\partial}{\partial r}\bigg[\sqrt{r}\frac{\partial}{\partial r}\bigg(\nu_{\rm t} \sqrt{r} \Sigma_{\rm g} \bigg)\bigg] = 0
\end{equation}
The gas surface density, $\Sigma_{\rm g}$, is the vertically integrated volumetric density, $\rho_{\rm g}$, assuming that the disk is vertically isothermal \citep{lynden-bell_evolution_1974}. 
The dissipative force for an accretion disk is driven by turbulent kinematic viscosity 
\begin{equation}
    \nu_{\rm t}(r,t) = \alpha c_{\rm s} H_{\rm g} =  \alpha c_{\rm s}^2 / \Omega
\end{equation}
where $c_{\rm s}=\sqrt{k_B T /\mu m_{\rm H}}$ is the isothermal sound speed of the gas, and $H_{\rm g}=c_{\rm s}/\Omega$ is the gas pressure scale height. The constant $\alpha$ is a parametrization of the accretion disk turbulence 
\citep{shakura_black_1973}, where typical values range from $10^{-5}$ to $10^{-2}$, although disk turbulence is likely to vary radially and vertically \citep{flaherty_turbulence_2018,andrews_observations_2020,rosotti_empirical_2023,jiang_grain-size_2024}.  
For the fiducial model we choose $\alpha=10^{-3}$ and test a less efficient case of $\alpha=10^{-4}$ in Section \ref{ssec:modelparamdepend}.
The solution presented in \cite{lynden-bell_evolution_1974} follows the functional form 
\begin{equation}\label{eq:self_similar}
    \Sigma_{\rm g,ss}(r) = \Sigma_{\rm c} \bigg(\frac{r}{r_{\rm c}}\bigg)^{-\gamma} \exp\bigg[-\bigg(\frac{r}{r_{\rm c}}\bigg)^{2-\gamma}\bigg]
\end{equation}
where $\Sigma_{\rm c}=(2-\gamma)M_{\rm disk}/(2\pi r_{\rm c}^2)$ is the surface density normalization constant based on the the total disk gas mass, $M_{\rm disk}$, at critical radius, $r_{\rm c}$, and is set by the power, $\gamma$ \citep{krijt_transport_2018}. For our fiducial model with $M_{\rm disk}=0.025M_{\odot}$, and choosing an initial $r_{\rm c}=20$~au and $\gamma=1$, results in $\Sigma_{\rm c}\approx88.38$g cm$^{-2}$.

For an accretion disk where turbulent kinematic viscosity is the dominant diffusive force, $r_{\rm c}$ will evolve outwards from the initial critical radius, $r_{\rm c,0}=r_{\rm c}(t=0)$, to larger radii, following the advection of momentum as a result of disk material accreting onto the host star. While much of the material will be accreted on disk lifetimes, a portion of the disk must also be advected outwards, and can be traced by $r_{\rm c}(t)$. The gas accretion disk flattens and spreads out on viscous timescales 
\begin{equation}\label{eq:t_viscous}
    t_{\rm viscous}(r,t) \approx {r^2}/{\nu_{\rm t}}
\end{equation}
which itself is radially and temporally evolving due to turbulent viscous diffusion, such that the gas surface density decreases interior to $r_{\rm c}$ and increases beyond it. If the angular momentum from mass accretion is transported through other mechanisms, such as through disk winds or interrupted through disk substructure at later disk times, $r_{\rm c}(t)$ will evolve differently. In Section \ref{sec:variations_from_fiducial}, we compare our fiducial result with a model that is less efficient at transporting momentum over 3 Myr, as well as models which start with different $r_{\rm c,0}$.

\begin{figure}[!htb]
 \centering
 \includegraphics[width=\linewidth]{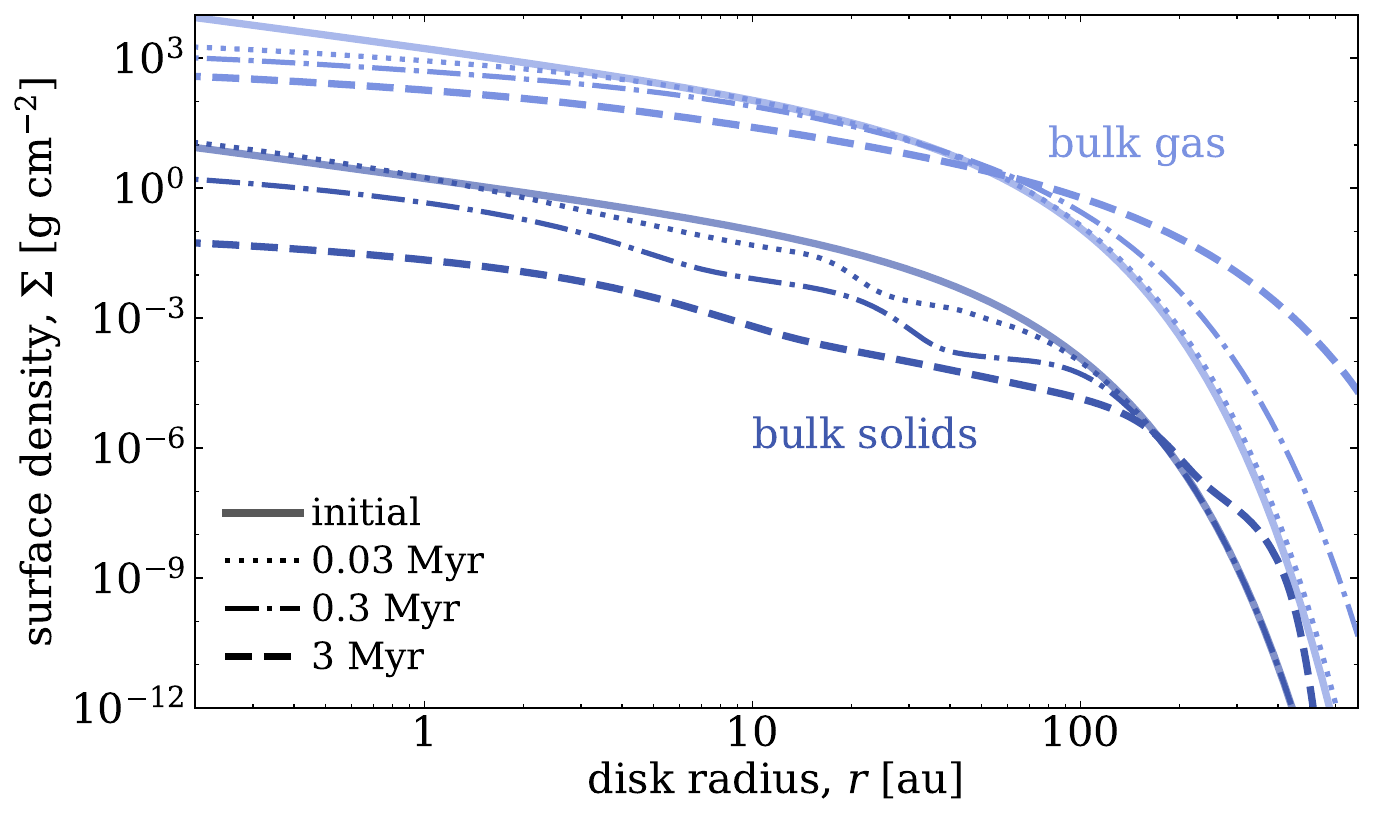}
 \caption{Bulk gas surface density (light blue) decreases and viscously spreads outward over 3 Myr, while bulk solid surface density (dark blue) significantly decreases, with a small portion in the outer disk that increases. The line styles correspond to different times in the evolution with solid, dotted, dotted-dashed, and dashed taken at initial (1 yr), 0.03, 0.3, and 3 Myr, respectively. }
 \label{fig:bulksurfdens}
\end{figure}

Following \cite{price_ice-coated_2021}, we use a smooth interpolation between $\Sigma_{\rm g,ss}$ and a flat line
\begin{equation}
    \Sigma_{\rm g}(r,t=0) = \big[\Sigma_{\rm g,ss}(r)^{-p} + \Sigma_{\rm g,ss}(r_t)^{-p} \big]^{-1/p}
\end{equation}
at a transition radius, $r_{\rm t}$, near the minimum disk radius,  $r_{\rm min}$, and power, $p$, for the time-dependent gas surface density solution.
We adopt the values $p=5$ and $r_t=0.1$~au, and assume a minimum disk radius of 0.01~au and a maximum disk radius of 800~au. Figure~\ref{fig:bulksurfdens} displays the bulk gas surface density evolution (light blue) for an actively evolving disk at four different times. 

\subsection{Solid Particle Evolution} \label{ssec:accretion}

We model the time-dependent radial evolution of solid particles in protoplanetary disks following \cite{weidenschilling_aerodynamics_1977}, \cite{birnstiel_gas-_2010}, and \cite{price_ice-coated_2021}, and describe the adopted particle size distributions in Section~\ref{ssec:size_dist}. 
For more recent reviews on solid particle evolution in protoplanetary disks see \cite{drazkowska_planet_2023} and \cite{birnstiel_dust_2024}.
The solid surface density evolution in the absence of sources and sinks is 
\begin{equation}\label{eq:diff_eq_solid}
    \frac{\partial\Sigma_{\rm s}}{\partial t} + \frac{1}{r}\frac{\partial}{\partial r}\bigg[rF_{\rm tot}\bigg] = 0
\end{equation}
where the total flux of solid particles is set by a combination of diffusion and advection. 
Added source and sink terms due to adsorption and desorption are detailed in Section~\ref{sec:molecular_physics}.
The diffusive component
\begin{equation}
    F_{\rm diff} = -\frac{\nu_{\rm t}\Sigma_{\rm g}}{\tau_{\rm s}^2 + 1}\frac{\partial}{\partial r}\bigg(\frac{\Sigma_{\rm s}}{\Sigma_{\rm g}}\bigg)
\end{equation}
is a result of the turbulent motion of the gas, which carries with it the particles that are well-coupled, and is hence set by $\nu_{\rm t}$. 
The Stokes number, or dimensionless stopping time $\tau_{\rm s}$, represents how well-coupled the solid particles are to the surrounding gas and is determined by comparing the particle's radius, $s$, to roughly the mean free path of the particle in the gas, $9/4 \;\lambda_{\rm mfp}$. We assume that all particles are in the Epstein regime with 
\begin{equation}
    \tau_{\rm s} = \frac{\pi}{2}\frac{s \rho_{\rm int}}{\Sigma_{\rm g}}
\end{equation}
where we choose a constant, silicate-like internal solid particle density of  $\rho_{\rm int} = 2.74$ g cm$^{-3}$.
The Epstein regime describes the motion of particles well in the outer disk \citep{birnstiel_dust_2024}. We do not include the Stokes drag regime as we focus this study on disk regions beyond the H$_2$O ice line. We also do not include the ram pressure regime, which becomes important for larger, kilometer-sized particles, much larger than those considered in this study \citep{perets_wind-shearing_2011, demirci_destruction_2020, yunerman_pathway_2024}.

The advective component\footnote{For more details on the velocities of the advective flux component, see \cite{birnstiel_gas-_2010} and \cite{price_ice-coated_2021}.} is 
\begin{equation}
    F_{\rm adv} = \Sigma_{\rm s} u_{r,s} \:
\end{equation}
and depends on the radial velocity of the particle,
\begin{equation}
    u_{\rm r,s} = \frac{u_{\rm g}}{\tau_{\rm s}^2 + 1} - \frac{2u_{\rm grad}}{\tau_{\rm s} + \tau_{\rm s}^{-1}} \:
\end{equation}
The first term describes the drag imposed by the sub-Keplerian gas with velocity, 
\begin{equation}
    u_{\rm g} = - \frac{3}{\Sigma_{\rm g}r^{1/2}} \frac{\partial}{\partial r}\big[\Sigma_{\rm g}\nu_{\rm t} r^{1/2}\big]
\end{equation}
and the second term is the particle-gas relative velocity, 
\begin{equation}
    u_{\rm grad} = -\frac{\epsilon_{\rm d}}{2\rho_{\rm g} \Omega} \frac{\partial p_{g}}{\partial r}
\end{equation}
set by the radial gas pressure gradient,  ${\partial p_{\rm g}}/{\partial r}$, and inward radial drift efficiency, $\epsilon_{\rm d}=0.1$.
At the midplane, the volumetric density of the gas is $\rho_{\rm g} = \Sigma_{\rm g}/(\sqrt{2\pi}H_{\rm g})$, and $p_{\rm g} = \rho_{\rm g}c_{\rm s}^2$ is the gas pressure.
The fiducial model assumes a dust-to-gas mass ratio of $10^{-3}$.
Figure~\ref{fig:bulksurfdens} displays the bulk solid surface density evolution (dark blue lines) summed over all particle sizes and species in our fiducial model for several different timesteps. The bulk solid surface density is governed primarily by depletion through particle drift (causing the dip in bulk $\Sigma_{\rm s}$ near 10~au before 1 Myr) and by the outward advection of small, well-coupled particles that follow the motion of the viscously diffusing gas.

\subsection{Particle Size Distribution} \label{ssec:size_dist}

We trace several discrete particle sizes, motivated by steady state particle collisional growth and fragmentation models, rather than implementing particle size evolution. 
The numerical integration in space and time of the solid particles is done over discrete size bins that make up fractions of the total solid disk mass. 
Following \cite{dohnanyi_collisional_1969}, we assume that the particle mass distribution is of the form 
\begin{equation}
    \frac{dn}{dm} \propto m_{\rm s}^{-\alpha_{\rm s}}
\end{equation}
where $m_{\rm s}=(4\pi/3) \: \rho_{\rm int} s^3$ is the solid particle mass.
For a self-similar cascade, \cite{tanaka_steady-state_1996} find $\alpha_{\rm s} = 11/6$, which we adopt as our fiducial solid size distribution. 
By skewing the mass distribution by $\pm5\%$ from the fiducial $\alpha_{s}$ towards either large or small particle sizes, we can approximate the balance of various physical size evolution processes (i.e., collisional growth, fragmentation, erosion, bouncing, etc.) into $\alpha_{\rm s}$ without modeling the size evolution in detail. 
Distributions skewed towards large sizes may represent efficient particle growth and/or inefficient collisional fragmentation, while distributions skewed towards smaller sizes may represent efficient fragmentation. 
We note that our fiducial size distribution power law is quite similar to what is seen in more complex models at mid to large disk radii \citep{birnstiel_dust_2024}.

\begin{figure}
    \centering
    \includegraphics[width=\linewidth]{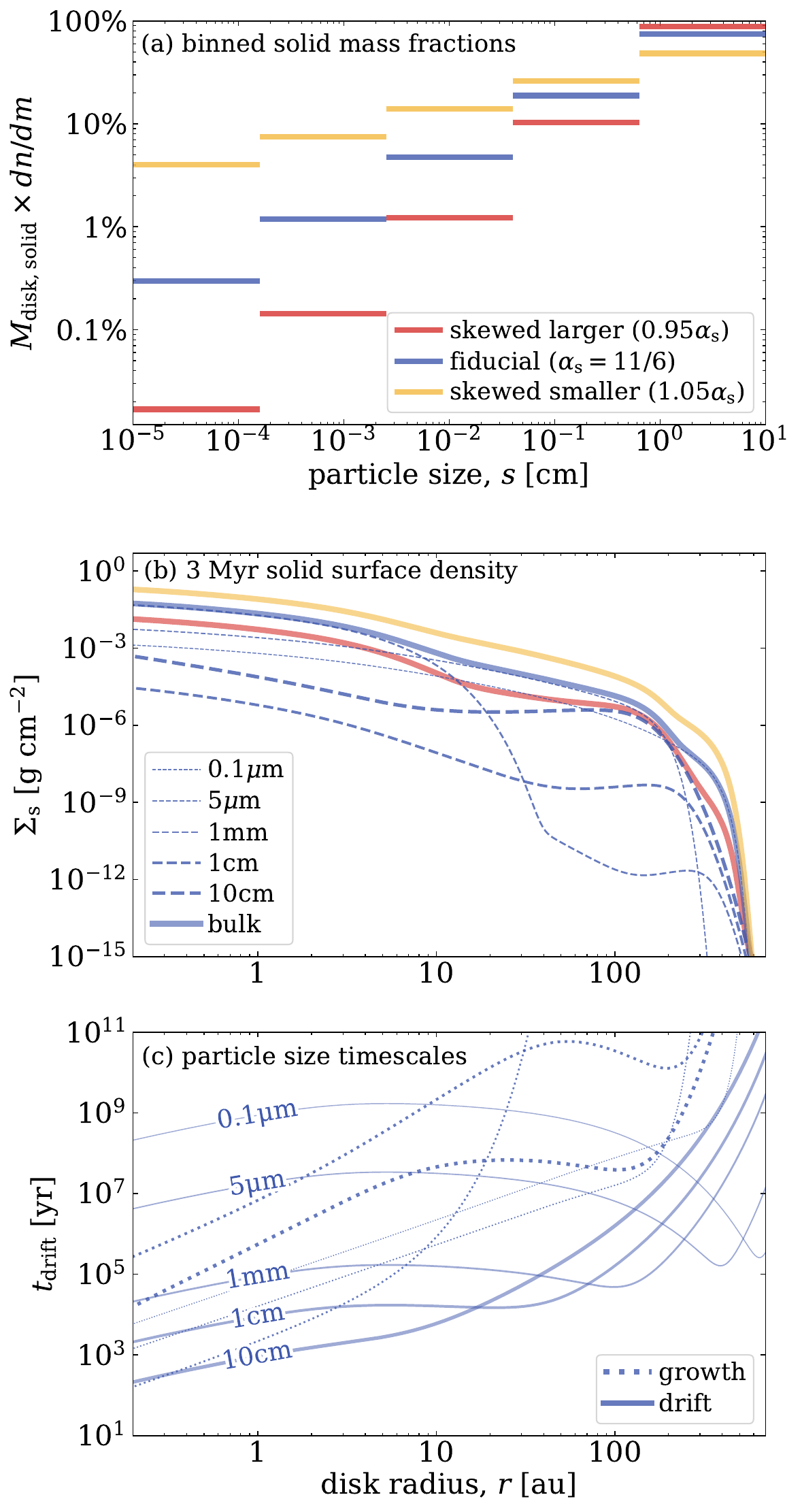}
    \caption{Panel (a) displays the normalized mass distributions for particles in a collisional cascade, showing how the bulk solid disk mass is partitioned across five particle size bins from $s_{\min} = 0.1$ \textmu m to $s_{\max} = 10$ cm in our model. Varying the fiducial slope $\alpha_{\rm s} = 11/6$ (blue) by $\pm$5\% skews the bulk of the solid mass either towards larger (red) or smaller (yellow) particles. Panel (b) then shows the evolved solid surface densities, $\Sigma_{\rm s}$, at 3 Myr for representative particle sizes in each bin (dashed) and for the total bulk (solid). Bulk solid surface densities are color-coded to match the corresponding assumed size distributions in panel (a). Panel (c) presents the fiducial radial drift (solid) and growth (dotted) timescales (Eq. \ref{eq:t_drift} and \ref{eq:t_growth}) across the disk for each modeled particle size using the 3 Myr surface densities, illustrating how larger particles decouple from the gas and drift inward rapidly as compared to the small well-coupled particles.}
    \label{fig:solid_particle_modeling}
\end{figure}

In Figure~\ref{fig:solid_particle_modeling}a, we show how the total solid disk mass, $M_{\rm disk}$, is distributed into a number of particle size bins between $s_{\rm min}$ and $s_{\rm max}$ following a collisional cascade, to obtain the solid mass fractions for each bin's discrete representative particle size.
The smallest particle size, which we take to be the average size of an ISM dust grain \citep[$s_{\rm min} = 0.1$ {\textmu}m,][]{bohlin_survey_1978,cuppen_laboratory_2024}, is the most abundant by number and provides the most surface area for molecular adsorption (detailed in Section~\ref{ssec:ads_and_des}). 
The maximum particle size is expected to change depending on the time and place in the disk through processes such as collisional growth, fragmentation, bouncing, erosion, etc., and further complicated by detailed relative velocity calculations that may depend on particle composition and structure \citep{birnstiel_simple_2012,powell_new_2019,yunerman_pathway_2024,birnstiel_dust_2024}. 
We therefore assume a maximum size $s_{\rm max} = 10$ cm for simplicity. The maximum size dominates the overall mass of the disk solid surface density distribution. 
For a particle evolution model consisting of two size bins, the mass fractions for $s_{\rm max}$ and $s_{\rm min}$ will be around 90-99\% and 1-10\%, respectively \citep{birnstiel_simple_2012}. Increasing the number of bins to 5 reduces the mass fractions for the smaller sizes as more mass is found in intermediate-sized particles. Increasing the number of bins further has only a minor impact on the solid particle dynamics, so we adopt a 5-bin particle size distribution with logarithmically-spaced sizes. 
The skewed size distribution mass fractions are also displayed in comparison with the fiducial collisional cascade, $\alpha_{\rm s}=11/6$. 
The ``skewed larger'' case, scaling $\alpha_{\rm s}$ down by $5\%$ (i.e., $\alpha_{\rm s}\times0.95$), redistributes the total solid disk mass towards the largest particles size.

Figure~\ref{fig:solid_particle_modeling}b compares the bulk solid surface densities between the fiducial and skewed mass fractions and decomposes the fiducial bulk solid surface density evolution into its different particle size components. 
Each solid particle size's surface density is initially a fraction of the bulk solid mass. 
Evolved bulk solid surface densities between the fiducial and skewed distribution models are different because there is a different dynamical balance between the small, well-coupled particles (0.1 and 5 {\textmu}m), and the large, marginally-coupled particles (1 mm, 1 cm, and 10 cm).
For some intuition of how quickly particles are depleted throughout the disk, Figure~\ref{fig:solid_particle_modeling}c displays the approximate size-dependent radial drift timescale
\begin{equation}\label{eq:t_drift}
    t_{\rm drift} =  \left| \frac{r}{\dot{r}} \right| \approx \frac{1}{2\eta \Omega}\bigg(\frac{1+\tau_{\rm s}^2}{\tau_{\rm s}}\bigg),
\end{equation}
where we assume $\dot{r}\approx u_{\rm r,s} \approx {-2\tau_{\rm s}u_{\rm grad}}/({1+\tau_{\rm s}^2})$.
Here, $\eta = -1/2\; (\partial \log P/\partial \log r) (H_{\rm g}/r)^2\approx 1/2\; (c_{\rm s}/v_{\rm K})^2$ is the gas-pressure support parameter \citep{ormel_catching_2018}, reflective of the gas disk pressure gradient within $u_{\rm grad}$, and $v_{\rm K} = \Omega r$ is Keplerian velocity.
Well-coupled particles follow the gas motion throughout the disk, as small-particle drift timescales are longer than disk lifetimes.
The distribution skewed towards smaller particles increases in solid mass throughout the disk since there are more small-particles beyond $r_{\rm c}$ that are not accreted.
The bulk $\Sigma_{\rm s}$ of the distribution skewed towards larger particles is depleted throughout the inner and middle disk regions more than the fiducial case as there is more mass drifting inwards.

Figure \ref{fig:solid_particle_modeling}c also includes the analytic growth timescale approximation for particles in the Epstein regime from \cite{birnstiel_dust_2024}
\begin{equation} \label{eq:t_growth}
    t_{\rm growth} \simeq \frac{\Sigma_{\rm g}}{\Sigma_{\rm s}\;\Omega}
\end{equation}
in comparison with the drift timescale since collisional growth can transfer mass from different size bins, and is particularly efficient for smaller sizes \citep{stammler_redistribution_2017,powell_depletion_2022}.
While we do not directly include particle size evolution through collisional growth, it is important to note that particles are likely growing and drifting at similar rates, excluding growth barriers and disk evolution interior to the H$_2$O ice line.
We further discuss the potential impact of solid particle growth and destruction in Section \ref{ssec:mech_sizedist}.


\section{Ice Line Model} \label{sec:molecular_physics}

The evolution of gas and solid abundances of disk volatiles, along with their ice line locations, depends on both the physical structure of the disk, as discussed in the previous section, and on molecular properties and processes. 
Section~\ref{ssec:species} introduces the main carriers of volatile oxygen (O), carbon (C), and nitrogen (N) expected in disks, along with their assumed abundances. 
Section~\ref{ssec:ads_and_des} describes thermal adsorption and desorption on particle surfaces, each species’ adopted desorption parameters, and how the model tracks changes in the evolving species-dependent surface densities.

\subsection{Molecular Species}\label{ssec:species}

We focus on molecules that are proposed major carriers of volatile C, N, and O within disks, in order to gain an understanding of how the C/N/O gas and solid ratios may evolve during formation. 
We consider constraints from protostellar, disk, and cometary ices \citep{boogert_observations_2015,oberg_astrochemistry_2021,mcclure_ice_2023}, which in general agree about the nature of the most common C/N/O carriers, even though abundances can differ between sources by up to an order-of-magnitude.
We do not consider any chemical alterations, which is a reasonable starting point for the outer disk, since icy grains are mainly inherited from the interstellar cloud and do not undergo significant chemical processing \citep{bergner_ice_2021}. 
This assumption does not hold for the chemically reprocessed inner disk \citep{eistrup_molecular_2018,eistrup_chemical_2022,eistrup_chemical_2022-1}, and even in the outer disk, vertical turbulent mixing may lead to substantial chemical evolution \citep{bergner_evolutionary_2020,zhang_chemistry_2024}. 

\begin{deluxetable}{l|c|ccc}[!b]
\tablecolumns{5}
\tablecaption{Species Initial Abundances and Kinetic Desorption Properties \label{tab:material_props}}
\tablehead{ 
 \colhead{\bf{Species}} &
 \colhead{\bf{$\bf x_{\rm i} \times 10^{-4}$}} &
 \colhead{\bf{$E_{\rm des}$ [K]}} &
 \colhead{\bf{$\nu_{\rm f}$ [s$^{-1}$]}} &
 \colhead{}
}
\startdata
         H$_2$O           & 1     & 5705 & 4.96$\times 10^{15}$ & \tablenotemark{a} \\
         NH$_4^+$         & 0.1   & 8286 & 7.70$\times 10^{15}$ & \tablenotemark{b} \\
         NH$_3$           & 0.05  & 2965 & 2.10$\times 10^{12}$ & \tablenotemark{c} \\
         CO$_2$           & 0.3   & 3196 & 6.81$\times 10^{16}$ & \tablenotemark{a} \\
         CH$_4$           & 0.05  & 1232 & 5.43$\times 10^{13}$ & \tablenotemark{a} \\
         CO               & 0.2   & 1155 & 8.34$\times 10^{11}$ &  \tablenotemark{d} \\
         N$_2$            & 0.275 & 1034 & 7.89$\times 10^{11}$ &  \tablenotemark{d} \\
\enddata 
\tablenotetext{a}{\cite{minissale_thermal_2022}}
\tablenotetext{b}{\cite{kruczkiewicz_ammonia_2021}}
\tablenotetext{c}{\cite{martin-domenech_thermal_2014}}
\tablenotetext{d}{\cite{fayolle_n2_2016}}
\end{deluxetable}

The main carriers of O, C, and N in all observable phases of star and planet formation (beyond the H$_2$O ice line) are water (H$_2$O), carbon dioxide (CO$_2$), ammonia (NH$_3$), methane (CH$_4$), methanol (CH$_3$OH)\footnote{We do not include CH$_3$OH in our model since it is a minor contributor to the C and O budget, and has similar desorption properties to H$_2$O, entailing that it cannot be efficiently enhanced or depleted with respect to H$_2$O from combined dynamic and kinetic processes alone.}, carbon monoxide (CO), and theoretically, molecular nitrogen (N$_2$). Some nitrogen may also be locked up in ammonium (NH${_4^+}$) salts on grains, for example as ammonium formate (NH${_4^+}$COOH$^-$), produced in the reaction between NH$_3$ and HCOOH \citep{bergner_kinetics_2016}.
Together, these seven species span the range of volatile desorption kinetics present in disks beyond the H$_2$O ice line to the outer disk. 
The abundances we use to initialize our model are summarized in Table \ref{tab:material_props}. The H$_2$O ice abundance of $\sim10^{-4}$ comes from protostellar and molecular cloud ice observations \citep{boogert_observations_2015}. Recent disk observations suggests that in disks this value may be higher, but more data is needed to confirm this \citep{sturm_jwst_2023,bergner_jwst_2024}. For the remaining species we use low-mass protostellar ice abundances as our starting point \citep{boogert_observations_2015}. 
CO and CO$_2$ are then initially present at around 20\% and 30\% that of H$_2$O ice. Based on both disk and comet observations, these values are lower when considering the mature disk as a whole, but this could simply be due to sublimation in the warmer disk regions rather than chemistry \citep{mumma_chemical_2011,altwegg_cometary_2019, bergner_jwst_2024}. CH$_4$ and NH$_3$ ices range between 1-10\% in protostars, while comets are closer to $\sim1$\%, and no detections in disks have yet been published. We thus assume an abundance of $\sim$5\% with respect to H$_2$O for both species. The remaining nitrogen carriers in our model are NH$_4^+$ salt and N$_2$. Following the range for ice observations in the ISM from \cite{boogert_observations_2015}, we adopt an NH$_4^+$ abundance of 10\%. We assume that the remaining nitrogen is in N$_2$ with an abundance of 27.5\% with respect to H$_2$O to account for the solar system N abundance \citep{lodders_solar_2003,asplund_chemical_2009,asplund_chemical_2021}. 

\subsection{Adsorption and Desorption}\label{ssec:ads_and_des}

To calculate the division of volatiles between the solid and gas phases, we use the desorption and adsorption framework presented in \cite{price_ice-coated_2021}, based on \cite{hollenbach_water_2008}, which considers the changes in surface density for particles of each species based on the their desorption energies and attempt frequencies. The desorption rate per molecule is set by
\begin{equation}\label{eq:Rdes}
    R_{\rm des} = \nu_{f} e^{-E_{\rm des}/T}
\end{equation}
where $E_{\rm des}$ is the desorption energy in units of Kelvin, and $\nu_f$ is the attempt frequency. When calculating the desorption rate, we assume that the temperature of the particle is that of the disk midplane at that particle's disk position, $T=T(r,t)$. Species kinetic parameters are obtained from laboratory work, where data from temperature programmed desorption (TPD) experiments provide species-specific concurrent fits of $E_{\rm des}$ and $\nu_{\rm f}$~\citep{collings_laboratory_2004,cuppen_laboratory_2024}. 
For the purposes of this study, we choose values for desorption off of Amorphous Solid Water (ASW) when available, which may somewhat overestimate the binding energies for the most volatile species once H$_2$O becomes depleted. 
This should not impact the overall results, but could in practice lead to receding hypervolatile ice line locations with time, all else being equal. 

\begin{figure*}[!ht]
 \centering
 \includegraphics[width=8.6cm]{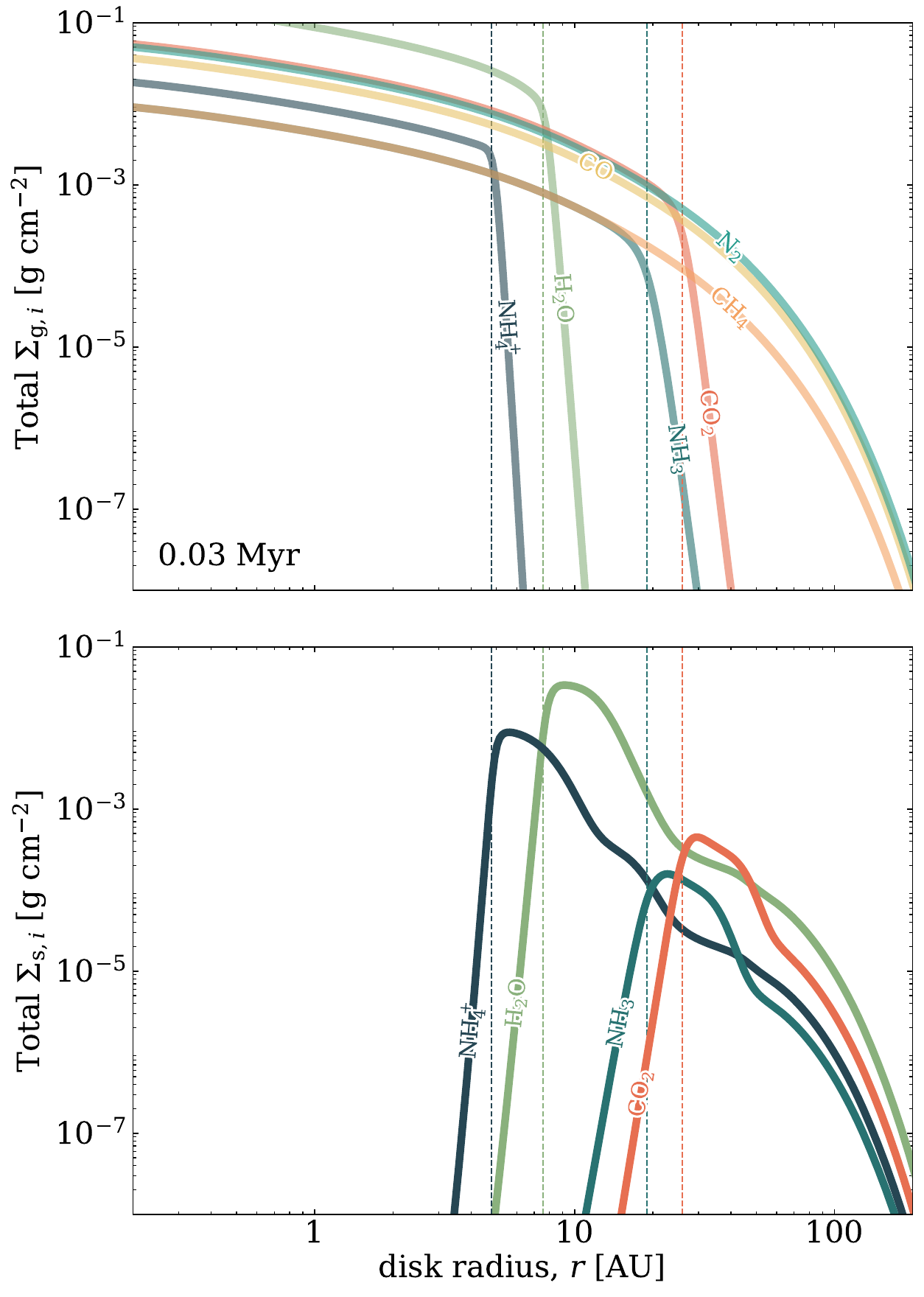}
 \includegraphics[width=8.7cm]{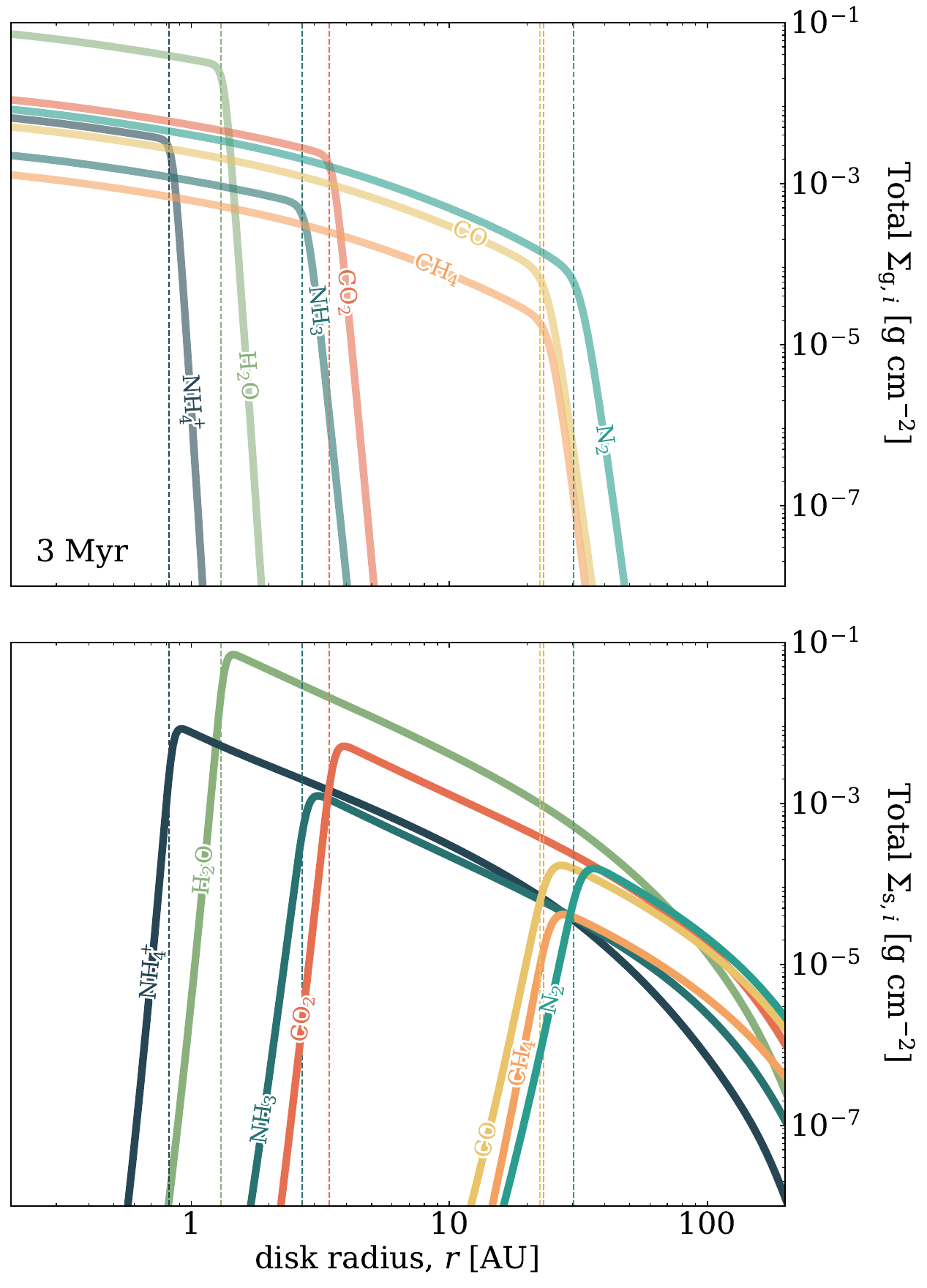}
 \caption{{\bf Compositional ice lines can be numerically approximated where the total solid surface density for each species, combined over all particle sizes, is that of the gas surface density.} The top and bottom panels are the gas and total solid surface densities across each species included in this work at 0.03 Myr (left) and 3 Myr (right). Dashed lines indicate the numerical ice lines, the radial intersection between the gaseous and solid surface densities. The colors correspond to species and will stay consistent across the remainder of the plots unless otherwise specified. }
 \label{fig:numerical_ice lines}
\end{figure*}  
Our choices for the desorption energies and associated attempt frequencies for each species are also summarized in Table \ref{tab:material_props}. 
For H$_2$O, CO$_2$, and CH$_4$, we adopt the recommended values from Table 2 of \cite{minissale_thermal_2022}, which correspond to the sub-monolayer regime on ASW. 
These values are developed through a critical review of experimental and theoretical data, and provide a consistent data set for astrochemical modeling. 
Since NH$_4^+$ is not included in \cite{minissale_thermal_2022}, we use the desorption parameters of the NH$_4^+$ salt, NH$_4^+$COOH$^-$, from the first-order Polanyi–Wigner fit in Table 3 of \cite{kruczkiewicz_ammonia_2021}, as it provides a chemically relevant proxy for N-bearing ionic species expected in processed icy mantles. 
NH$_3$ exhibits a broad range of desorption energies shaped by hydrogen bonding, and no ASW value is presented in \cite{minissale_thermal_2022}. We therefore adopt the multilayer pure ice values from the experiments of \cite{martin-domenech_thermal_2014}, and convert their reported pre-exponential factor into an attempt frequency assuming 1 monolayer $\sim$ $10^{15}$ molecules cm$^{-2}$. 
This approach may lead to desorption at somewhat higher temperatures than the values reported by \cite{minissale_thermal_2022}, but may be appropriate for NH$_3$-rich disk regions. 
For CO and N$_2$, \cite{minissale_thermal_2022} recommend values from \cite{smith_desorption_2016}, which are broadly consistent with other studies. However, we adopt the values from the detailed experiments of \cite{fayolle_n2_2016} on compact ASW (non-porous) at $\sim$1 ML coverage, instead.

Adsorption depends on the collision rate of the adsorbing gas molecule colliding with a particle 
\begin{equation}\label{eq:Rads}
    R_{\rm ads} = n_{\rm s}\sigma_{\rm s}v_{\rm th}
\end{equation}
where $v_{\rm th} = \sqrt{8k_{\rm B}T/\pi m_i}$ is the thermal velocity of the gas species, $\sigma_{\rm s}=\pi s^2$ is the cross sectional area of the solid particle, and $n_{\rm s}$ is the number density of solids of that species set by the molecule's initial abundance and solid surface density evolution. The volumetric gaseous source term for the numerical model is a combination of the rates of adsorption and desorption for each species, such that $s_{\rm g} = n_{\rm s}R_{des} - n_{\rm g}R_{ads}$. By vertically integrating and normalizing by the species mass, we find adsorption 
\begin{equation}
    S_{\rm ads} = \frac{\sigma_{\rm s}v_{\rm th}\Sigma_{\rm g}\Sigma_{\rm s}}{\sqrt{2\pi(1+\xi^2)}\: m_{\rm s}H_{\rm g}}
\end{equation}
and desorption 
\begin{equation}
    S_{\rm des} = R_{\rm des}\Sigma_{\rm s}
\end{equation}
surface sources, with units of g cm$^{-2}$ s$^{-1}$.
The vertical integration assumes a gaussian solid distribution where the scale height of the solid particles is some fraction of the gas scale height, $\xi=H_{\rm s}/H_{\rm g}$.
To imitate particle settling, we assume that marginally-coupled particles have $\xi=0.1$, while well-coupled particles have $\xi=1$.
These terms now reflect the change in surface density for the different species desorption and adsorption across each particle size bin. The contributing sources for gas and solid phases of the different species are then $S_{\rm g}=S_{\rm des}-S_{\rm ads}$ and $S_{\rm s}=S_{\rm ads}-S_{\rm des}$, and can be added to the surface density evolution equations (Eq. \ref{eq:diff_eq_gas} and \ref{eq:diff_eq_solid}). For a detailed derivation of the species source terms see \cite{price_ice-coated_2021}.

Ice lines are defined as the disk radii where the gas and solid surface densities of each species are equal. 
Figure~\ref{fig:numerical_ice lines} shows each species total surface densities at 0.03 and 3 Myr with separate panels for gases and bulk solids, and dashed vertical lines marking the ice lines. 
We note that when adding particle drift, these ice lines are expected to shift inwards for marginally-coupled particles due to similar drift and desorption timescales \citep{piso_co_2015}.
Based on the modeled ice line locations, we group species according to their relative volatilities. H$_2$O and NH$_4^+$ salt sublimate closest to the star and are classified as ``water-like.” 
NH$_3$ and CO$_2$ occupy intermediate regions and are referred to as ``mid-volatiles,” while the highly volatile species CH$_4$, CO, and N$_2$ sublimate farthest out and are grouped as ``hypervolatiles.”

\begin{figure*}[!p]
 \centering
 \includegraphics[width=\linewidth]{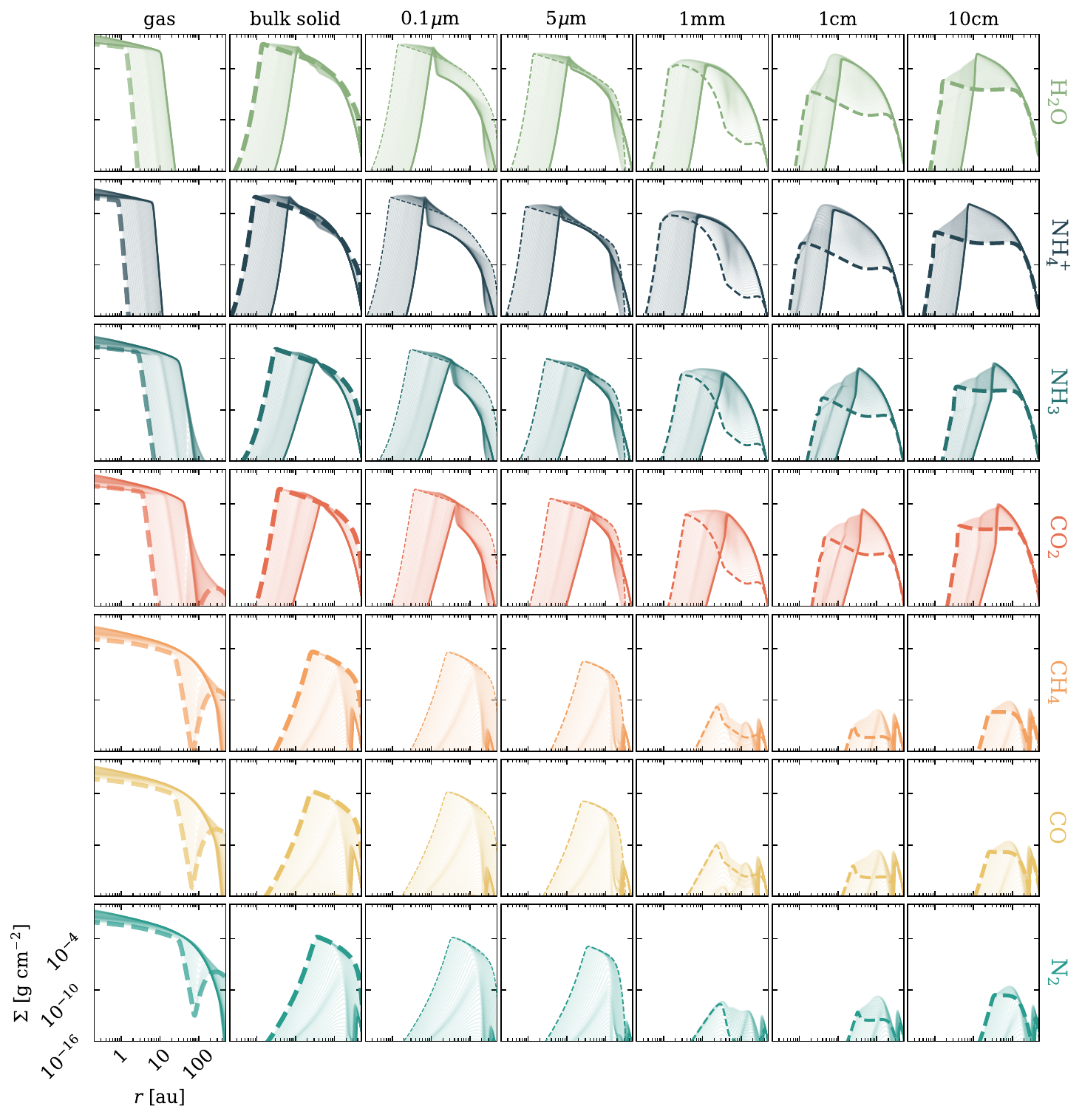}
 \caption{{\bf Volatile vapor ($\Sigma_{\rm g}$) and ice ($\Sigma_{\rm s}$) surface density evolution varies significantly across different species and particle sizes, indicative of interconnected thermal, physical, and molecular processes setting the distribution of planet forming materials.} Solid and dashed lines are times of 10 kyr and 3 Myr respectively, while in between timesteps are more transparent. Each color, and row, corresponds to a particular species such that the C-bearing molecules are warmer tones, N-bearing molecules are cooler blue tones, and H$_2$O is green. The rows from top to bottom are in order of desorption from the inner disk to the outer disk. The first column traces $\Sigma_{\rm g}$, while the remaining columns trace size-dependent $\Sigma_{\rm s}$ with increasing particle size, and 3 Myr line width, from left to right.}.
 \label{fig:sigmas_allcomps}
\end{figure*}


\section{Volatile Enhancements in the Fiducial Disk Model} \label{sec:results}

Our fiducial model is based on the evolution of a 0.5$M_{\odot}$ star surrounded by an actively accreting disk and a time-dependent stellar luminosity, $L(t)$, and mass accretion rate, $\dot{M}_{\star}(t)$, as described in Section~\ref{ssec:disk_temp}. We assume a constant turbulence parameter of $\alpha = 10^{-3}$, and a dust-to-gas mass ratio of $10^{-3}$. The initial distribution of particles is set by a collisional cascade, with mass fractions binned over five particle sizes accounting for the total solid component of the disk mass, $0.025 M_{\odot} \times 10^{-3}$ (see Section~\ref{ssec:size_dist}). The minimum particle size is set by the average ISM grain, $s_{\rm min} = 0.1$~\textmu m, while the largest particle size is $s_{\rm max} = 10$ cm, such that most of the mass is likely to be marginally-coupled to the gas throughout a large portion of the disk. We also assume that the disk inherited all of its ices in a cold start, where all the species start frozen out onto dust grains, motivated by \cite{bergner_ice_2021},  \cite{oberg_astrochemistry_2021}, and \cite{topchieva_ices_2024}\footnote{We do not consider mixed \citep{bergner_jwst_2024} or entrapped ices \citep{simon_entrapment_2023}, which would likely decrease the level of volatile/H$_2$O ice enhancement in the outer disk since a larger portion of the volatile budget will be trapped on drifting solids that are carried into the inner disk and towards the host star.}.
We explore varying volatile initial conditions and disk modeling choices in Section~\ref{sec:variations_from_fiducial}.

\subsection{Volatile Surface Density Evolution}

\begin{figure}[!hb]
    \centering
    \includegraphics[width=\linewidth]{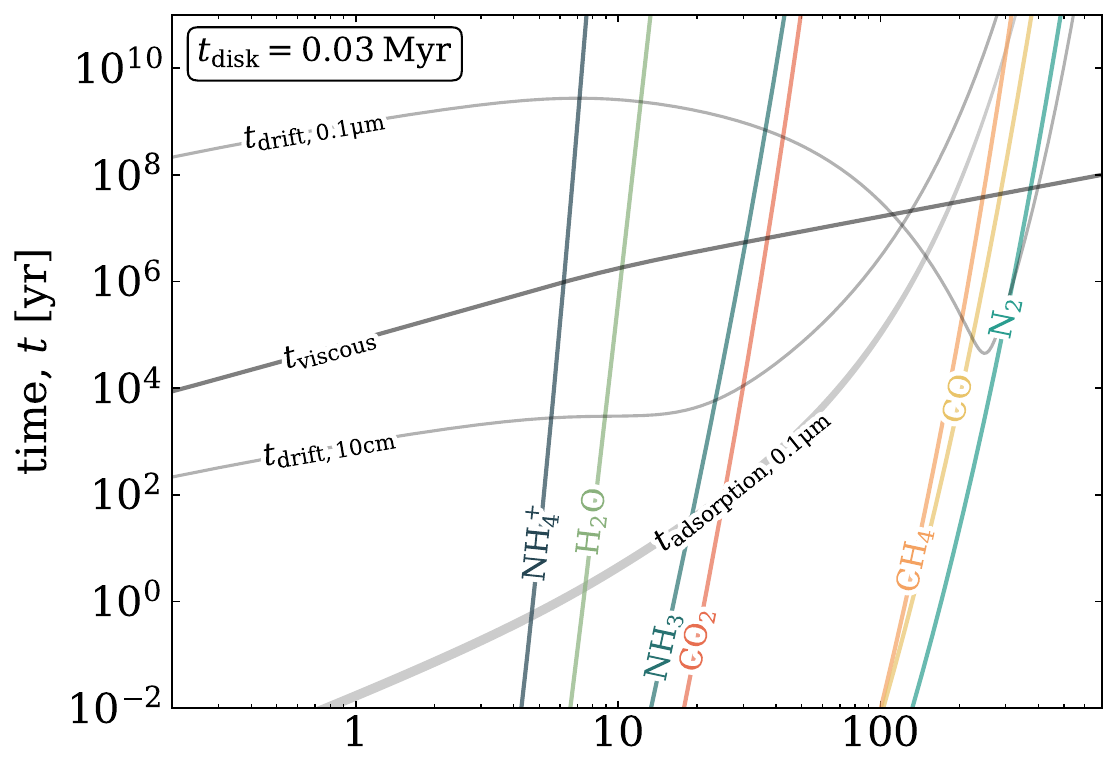}
    \includegraphics[width=\linewidth]{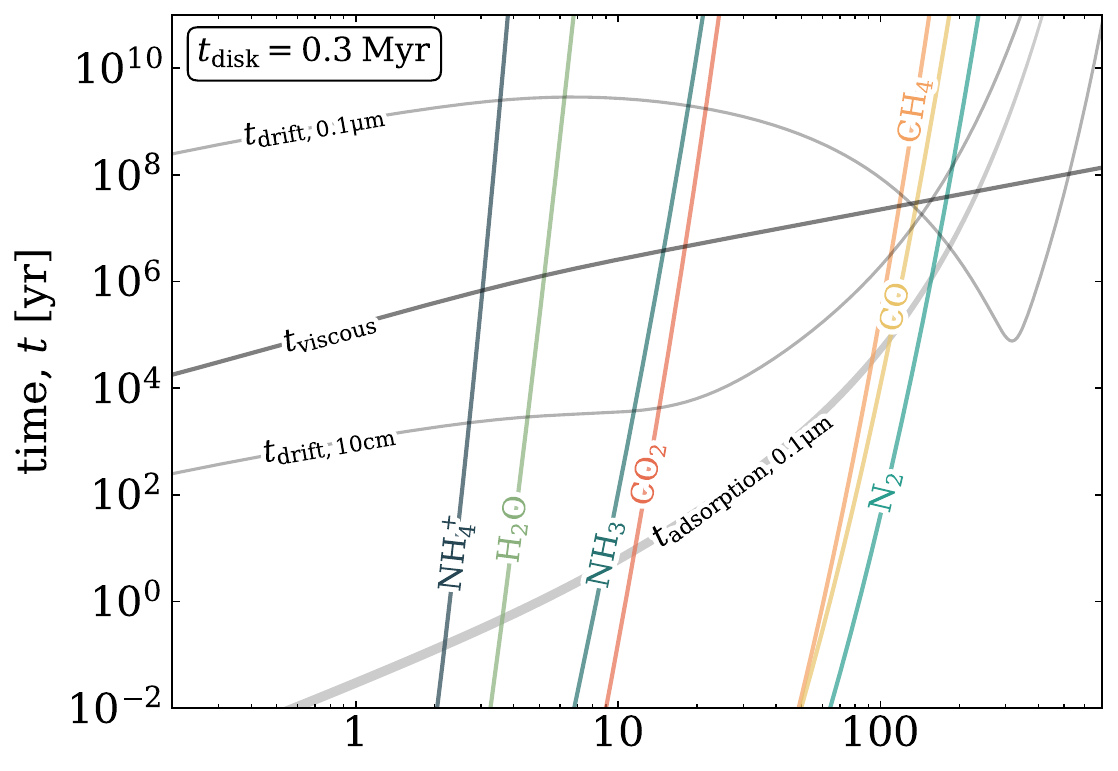}
    \includegraphics[width=\linewidth]{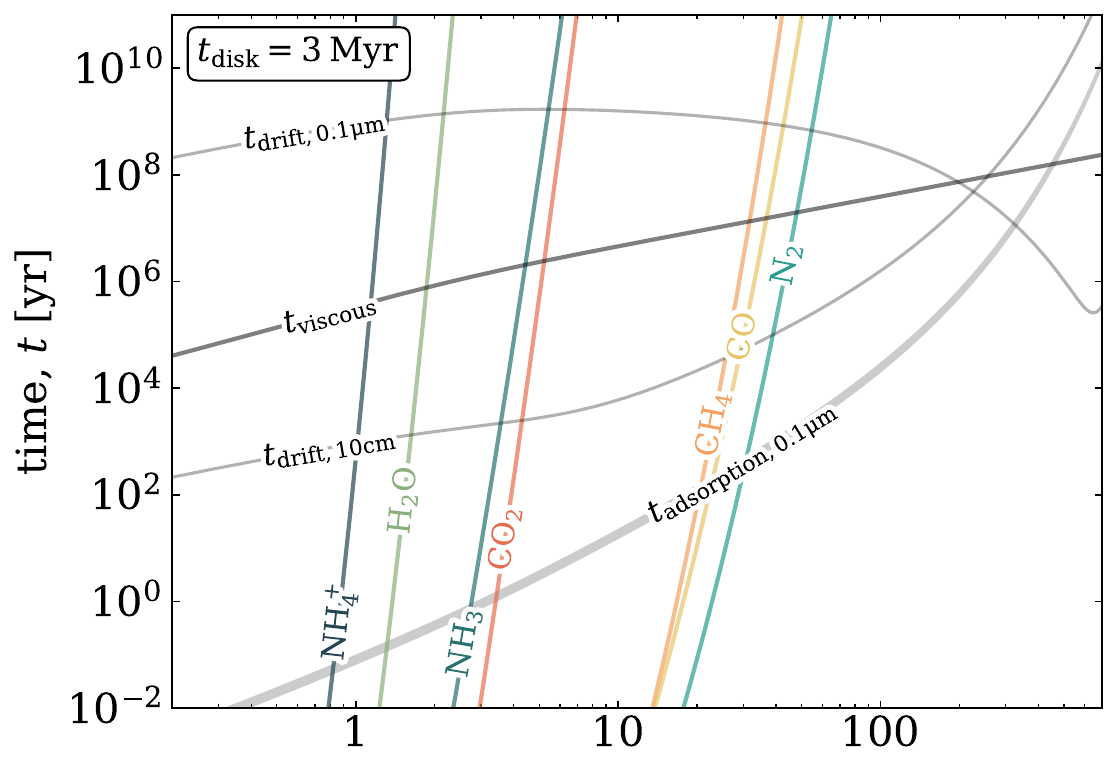}
    \caption{{\bf Analytic viscous (Eq. \ref{eq:t_viscous}), drift (Eq. \ref{eq:t_drift}), adsorption (inverse of Eq. \ref{eq:Rads}), and desorption (inverse of Eq. \ref{eq:Rdes}) timescales calculated using the fiducial model bulk gas surface density and temperature at disk times of 0.03, 0.3, and 3 Myr.} The drift and adsorption timescales assume a particle size of 0.1~{\textmu}m, tracing the dominant solid surface area. The species labeled lines correspond to  desorption timescales calculated using kinetic parameters from Table \ref{tab:material_props}, where intersections with the adsorption timescale indicate the existence of that species' ice line.} 
    \label{fig:timescales}
\end{figure}

Figure~\ref{fig:sigmas_allcomps} displays the radial and temporal evolution of the gas and solid, bulk and size-dependent, species surface densities for the fiducial disk model. 
These evolving profiles can be understood from the interplay of volatile adsorption and desorption, gas and small particle advection, and larger particle drift.
Analytic timescales for the relevant processes, calculated using the numerical bulk gas surface density and temperature at 0.03, 0.3, and 3 Myr are shown in Figure \ref{fig:timescales}, which are used below to explain the different evolutionary tracks of different volatiles.
The steep inner drop in species' bulk gas and solid surface densities in the first two columns of Figure~\ref{fig:sigmas_allcomps} mark each ice line and how they evolve in time due to the decreasing disk temperature.
These evolving numerical ice lines match well with the analytic ice lines, the intersections between the desorption and adsorption timescales in Figure~\ref{fig:timescales}. 
The rapidly changing desorption timescales (each defined as the inverse of the desorption rate per molecule; see Eq.~\ref{eq:Rdes}), reflect a strong dependence on species-specific kinetic parameters and on the disk temperature evolution. 
The adsorption timescale (defined as the inverse of the adsorption rate per molecule; see Eq.~\ref{eq:Rads}), shown in gray for the complete species range, depends on the available surface area, which is primarily the population of 0.1 {\textmu}m particles that are well-coupled to the evolving bulk gas. 

For a static, steady-state disk model, volatiles are expected to be almost completely in the solid phase beyond their respective ice lines, which is seen here only for the water-like species. 
For hypervolatiles, the initial disk model is too warm for any freeze-out, and ice lines only appear with time as the disk cools (top panel of Figure \ref{fig:timescales}, $t_{\rm disk}=0.03$~Myr), but even as the ice lines appear (middle panel of Figure \ref{fig:timescales}, $t_{\rm disk}=0.3$~Myr), the hypervolatile gas is not completely depleted. 
Instead, starting around 0.1 Myr, hypervolatile ice build-up depends on the adsorption timescale, which is comparable to the disk lifetime in the outermost disk regions (Eq.~\ref{eq:t_viscous}). 
Because of these long adsorption timescales, hypervolatile and mid-volatile species that diffuse outwards due to concentration gradients or that are viscously advected outward together with the bulk gas, will only over time adsorb onto outer disk solid particles, resulting in an initial build-up of vapor for all these species around 100~au in Figure \ref{fig:sigmas_allcomps}. Outward advection is only efficient beyond the critical radius, and hence we do not see a similar effect for water-like species, whose ice lines are far interior to the critical radius.  

Finally, solid surface densities are shaped by a combination of inwards drift of larger icy particles, outward advection of small icy particles, and adsorption of volatile vapors onto these particles, further increasing solid surface densities. Marginally-coupled solid particles (1 mm to 10 cm) drift inward rapidly on $\sim$kyr timescales (Eq.~\ref{eq:t_drift}) throughout most of the disk due to gas-induced drag, depleting the outer disk of solid mass while enriching the region interior the species' ice line in vapor through desorption. 
This is best seen for the water-like volatiles, whose decreasing solid column densities, especially in the outer disk, are largely explained by this process. 
By contrast, the overall solid column densities of mid- and hypervolatiles increase over time in the outer disk regions as a result of efficient adsorption of initially gas-phase volatiles (only the case for hypervolatiles), adsorption of outwardly diffused and advected hypervolatile and mid-volatile vapors, and outward advected small, well-coupled, icy particles, as seen in Figure~\ref{fig:bulksurfdens}. 
The latter process both moves the volatile solid peaks further out in the disk, and provides much of the surface area on which the gas-phase vapors can adsorb throughout the later stages of the disk evolution.
Together these processes result in a substantial buildup of hypervolatile and mid-volatile ice in the outer disk, which ultimately leads to significant relative enhancements of mid- and hypervolatile ices as compared to H$_2$O ice, a key result explored in the following subsection.

\subsection{Volatile Ice Relative Enhancements}

\begin{figure}[!t]
 \centering
 \includegraphics[width=\linewidth]{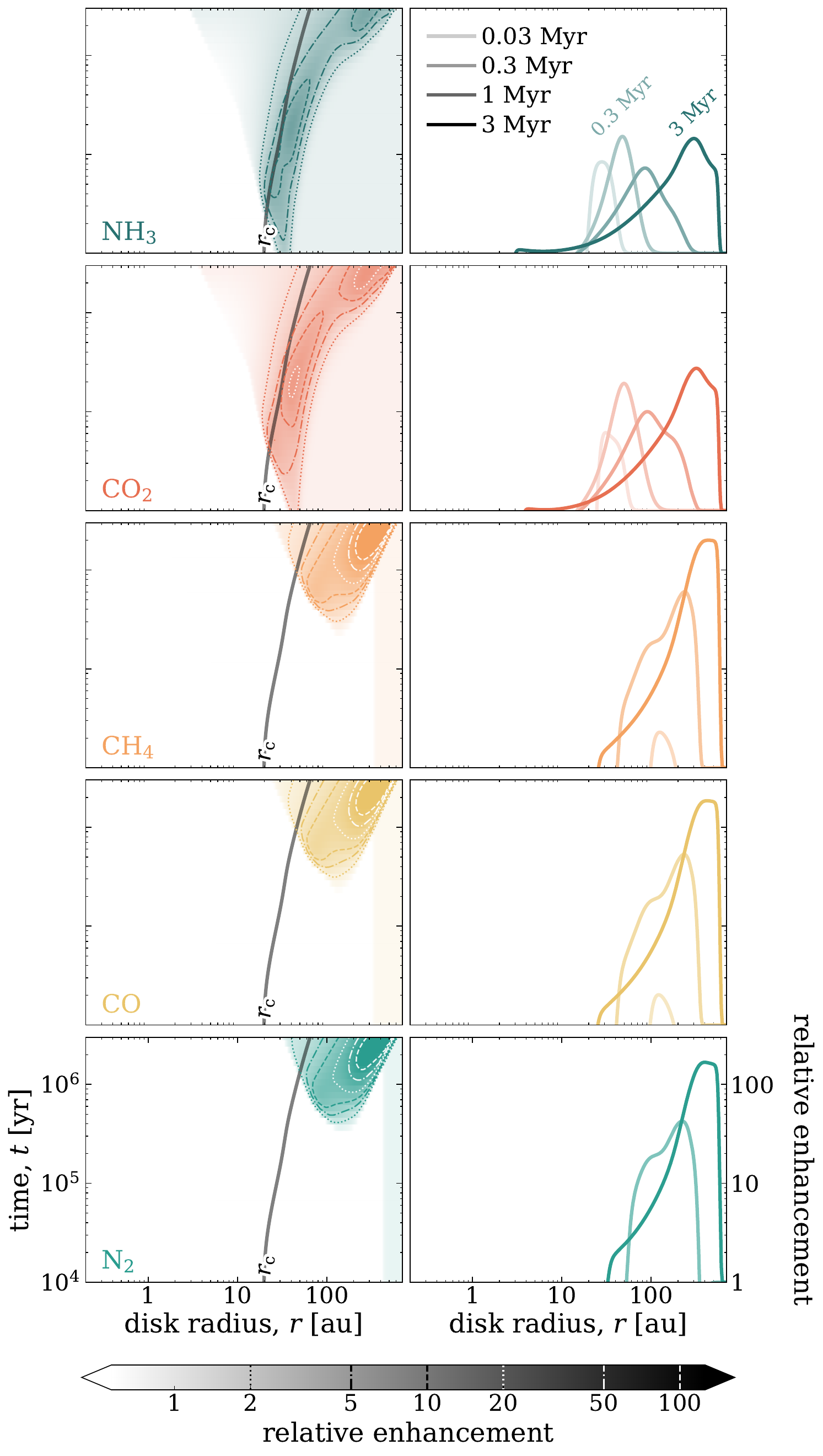}
 \caption{{\bf Hypervolatile and mid-volatile ices can over time become $\sim10-100\times$ enhanced beyond their ice lines as compared to H$_2$O ice, where the emerging enhanced region is advected according to the evolving critical radius.} The left column is the radial--temporal relative enhancement evolution as defined by Eq.~\ref{eq:enhancement}, bounded by the species ice line where the color first appears. Contours of 2, 5, 10, 20, 50, and 100$\times$ enhancements are indicated with corresponding styles as in the color bar. The evolving critical radius, $r_{\rm c}$, fitted iteratively with $\Sigma_{\rm c}$ using Eq. \ref{eq:self_similar}, is included for reference to the bulk gas evolution as a thick gray labeled line. The right column compares the enhancements at 0.03, 0.3, 1, and 3~Myr.}
 \label{fig:rel_enh}
\end{figure}

As a result of the disk volatile evolution described above, certain disk regions may become relatively enhanced over time in one volatile ice as compared to H$_2$O ice, quantified as 
\begin{equation}\label{eq:enhancement}
    \frac{\Sigma_{\rm s,i}(r,t)\: /\: \Sigma_{\rm s,H_2O}(r,t)}{x_{\rm i}\:/\:x_{\rm H_2O}} > 1
\end{equation}
where initial abundances, $x_{\rm i}$, can be found for species $i$ in Table \ref{tab:material_props}, and $\Sigma_{\rm s,i}$ is the species bulk solid surface density, summed over all particle sizes.  
Figure~\ref{fig:rel_enh} provides the relative icy enhancement evolution in time and space (left column) and radial enhancement profiles at a few different times (right column) for NH$_3$, CO$_2$, CH$_4$, CO, and N$_2$\footnote{We do not include NH$_4^+$ salt in the relative enhancement analysis since it evolves very similarly to H$_2$O due to similar ice line locations.}. 
Relative icy enhancements emerge for all species, but the degree of enhancement, as well as the time evolution, varies substantially between the mid-volatile and hypervolatile species.
Generally, relative icy volatile enhancements with respect to H$_2$O since initial are due to (1) the depletion and drying out of H$_2$O ice from the outer disk through inward drifting, marginally-coupled large particles; (2) the desorption of volatiles at their ice lines followed by viscous diffusion and advection of coupled gases and solids around the critical disk radius \citep[similar to the the ``vertical cold finger effect'' highlighted in][]{meijerink_radiative_2009}; and (3) the build-up of ice on small, well-coupled particle surfaces through increasingly more efficient (re)adsorption. 

Mid-volatile species develop $\sim$20–30$\times$ enhancements relative to H$_2$O ice beyond their ice lines as early as 0.03 Myr, with peak enhancement regions tracing the region just beyond $r_{\rm c}$ throughout the disk evolution, as seen in the second panel of Figure~\ref{fig:rel_enh}. 
The maximum enhancement is limited by the loss of some mid-volatile mass through the initial inward drift of icy particles. 
Hypervolatile species do not exhibit substantial enhancement until after $\sim$0.5 Myr since, at earlier times, the ice lines do not exist (i.e., no intersection of the hypervolatile desorption and adsorption timescales  in the top panel of Figure \ref{fig:timescales}). 
At later disk times, efficient adsorption coupled with outwards advection shapes the hypervolatile build-up on the solids reaching peak enhancements of $\gtrsim100\times$; similar to the mid-volatiles, the peak enhancement region depends on the evolving critical radius, beyond which effective outwards advection is possible.  
The larger enhancements can be understood when considering that  there is no significant loss of hypervolatiles in the outer disk, either through inward advection of hypervolatile-rich gas since the bulk of the vapors always reside far beyond $r_{\rm c}$, or through inward pebble drift, since these species only become frozen out after most pebble drift has occurred.

\begin{figure} [!t]
    \centering
    \includegraphics[width=1\linewidth]{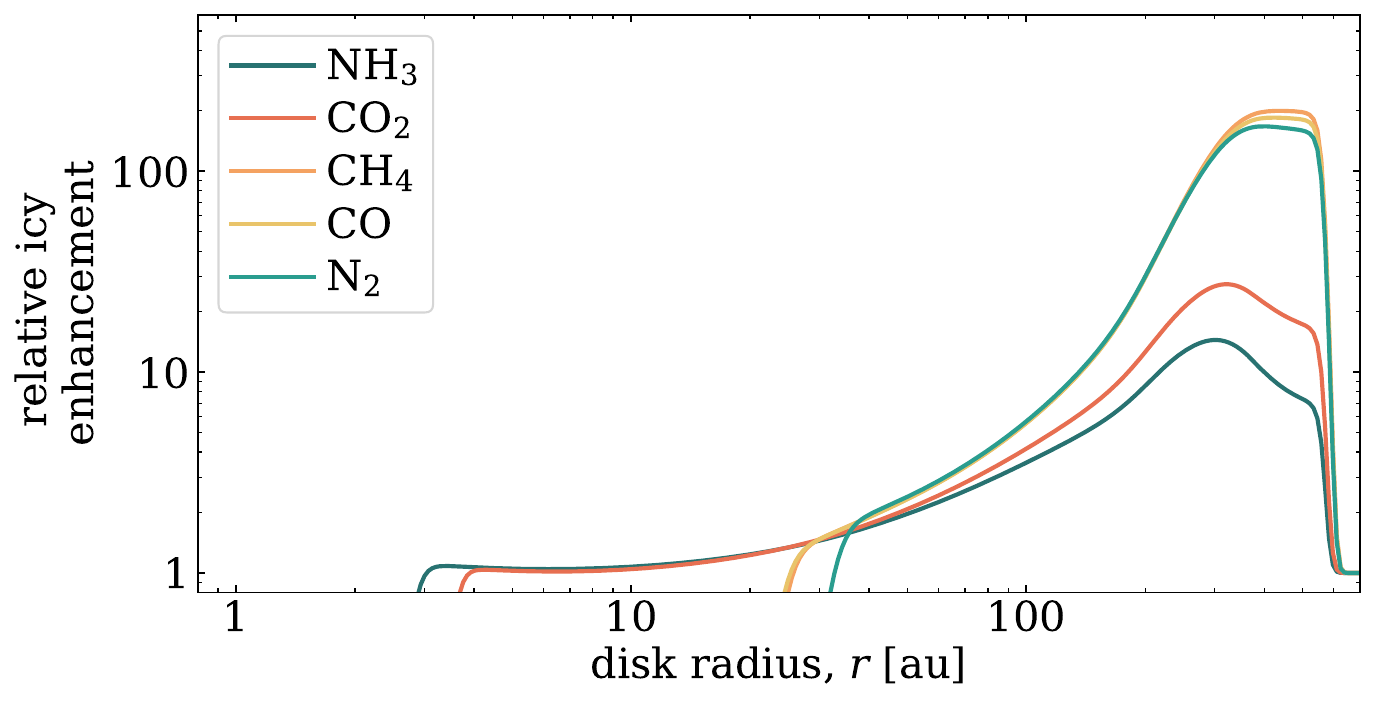}
    \caption{Relative enhancement at 3 Myr of various volatile ices to H$_2$O ice as compared with initial abundances. All species become relatively enhanced beyond $\sim$20~au for the fiducial disk model of an actively evolving disk where all species are inherited in their solid phase.}
    \label{fig:3Myr_relenh}
\end{figure}

While the enhancement levels differ, the final radial enhancement profiles all reach a similar shape, summarized in Figure~\ref{fig:3Myr_relenh}.
The final enhancement peak location corresponds to the advection wave following the critical disk radius evolution, and the overall volatile enhancement reflects the interplay of the different enhancement processes discussed above, where at early disk times, volatile enhancement is governed by fast inward drift of water-rich ice particles, and at late times, it is dominated by vapor and small icy particle redistribution and efficient (re)adsorption for species whose ice line is near or beyond the critical disk radius at any point in the disk's lifetime. 
 
\subsection{Comparison with previous dynamical disk models}

We can summarize our results as follows: for species whose $r_{i,\:\rm ice}(t)\gtrsim r_{c}(t)$, our fiducial model shows that the combination of drift, desorption, diffusion, advection, and (re)adsorption onto well-coupled particles, leads to development of relative icy volatile enhancements with respect to H$_2$O beyond their ice lines.
While inward drift of larger, marginally-coupled icy particles effectively removes H$_2$O vapor from the outer disk before it can freeze out, mid-volatile and hypervolatile vapors and small icy particles are advected outwards on viscous diffusion timescales, such that icy enhancements follow the $r_{\rm c}(t)$ outward advection wave.
Hypervolatile ices are enhanced by $\sim100\times$, and mid-volatiles by $\sim10\times$, with the lower enhancement explained by an initial $r_{\rm ice}(t) \lesssim r_{\rm c}(t)$, resulting in some inward loss of CO$_2$ and NH$_3$. 
Ice-enhanced regions are therefore not only set by the species ice line, as highlighted in \cite{price_ice-coated_2021}, who used a precursor to the this model to find large enhancements in CO/H$_2$O ice surface densities, but also by the evolving critical disk radius, which determines how vapor, small particles, and volatile-enhanced regions will be advected.

Gas and solid enhancements around species ice lines have been observed previously in models that couple sublimation of inward-drifting H$_2$O ice \citep{booth_chemical_2017,schneider_how_2021}. 
Such models show volatile enrichment in the gas-phase interior to the H$_2$O ice line \citep{stevenson_rapid_1988,cuzzi_material_2004}, as well as outwardly-diffusing H$_2$O vapor that condenses onto particles just beyond the H$_2$O ice line \citep{ciesla_evolution_2006,ros_ice_2013}. 
Similar effects are seen in the 2D radial--vertical models of \citet{krijt_transport_2018,krijt_co_2020}, which simulate CO drift, desorption, vertical settling, and some chemical processing, and are therefore more comparable to our outer disk results. 
They too find substantial volatile gas-phase enhancement relative to the bulk gas interior to the ice line at the midplane, in this case the CO ice line. Such localized enhancements are not apparent in Figure \ref{fig:sigmas_allcomps}, and on closer inspection show that we only see such vapor enhancements inside the respective ice lines at early times, due to a combination of drift slow-down over time, and efficient turbulent diffusion, which spreads out the desorbing vapors. The magnitude of this effect is hence sensitive to the details of the disk model assumptions.

The models most directly comparable to the one presented here are 2D radial--azimuthal hydrodynamic models of gravito-viscous disks by \citet{molyarova_gravitoviscous_2021,molyarova_co_2025} which include CO, CO$_2$, CH$_4$, and H$_2$O, though we note that the simulations of \cite{molyarova_co_2025} have $t_{\rm max}=0.5$~Myr, and later time icy enhancement in the outer disk from outward advection was hence not included in their study. 
Similar to this work, they find that the volatile solids can build up in the outer disk with respect to H$_2$O.
It is quite remarkable that our 1D, lower-mass, semi-analytic, disk model produces comparable gas and solid enhancement trends across species. It suggests that the results emerging from our fiducial model are at least qualitatively robust to the particular model set-up as well as the modeling methods.


\section{Enhancement Dependencies on Disk Parameters} \label{sec:variations_from_fiducial}

Based on the discussion in the previous section, we should expect both the magnitude and location of the peak in relative volatile icy enhancements ($\sim200-300$~au at 3 Myr in Figure \ref{fig:3Myr_relenh}),  to be shaped by modeling choices regarding the disk, solid particle, and gas initial conditions and dynamics. 
To assess their relative influence on the resulting volatile enhancements, we vary the parameters governing viscous evolution and solid transport in Section~\ref{ssec:modelparamdepend}, particle size distribution in Section~\ref{ssec:mech_sizedist}, disk thermal evolution in Section~\ref{ssec:thermal_depend}, initial volatile phase in Section~\ref{ssec:initdepend_disc}, and critical disk radius in Section~\ref{ssec:rcrit_disc}, where in each case we show the impact on CO (hypervolatile) and CO$_2$ (mid-volatile) icy enhancements relative to H$_2$O ice.

\subsection{Turbulent viscosity, particle drift efficiency, and dust-to-gas mass ratio }\label{ssec:modelparamdepend}

\begin{figure}[!t]
    \centering
    \includegraphics[width=\linewidth]{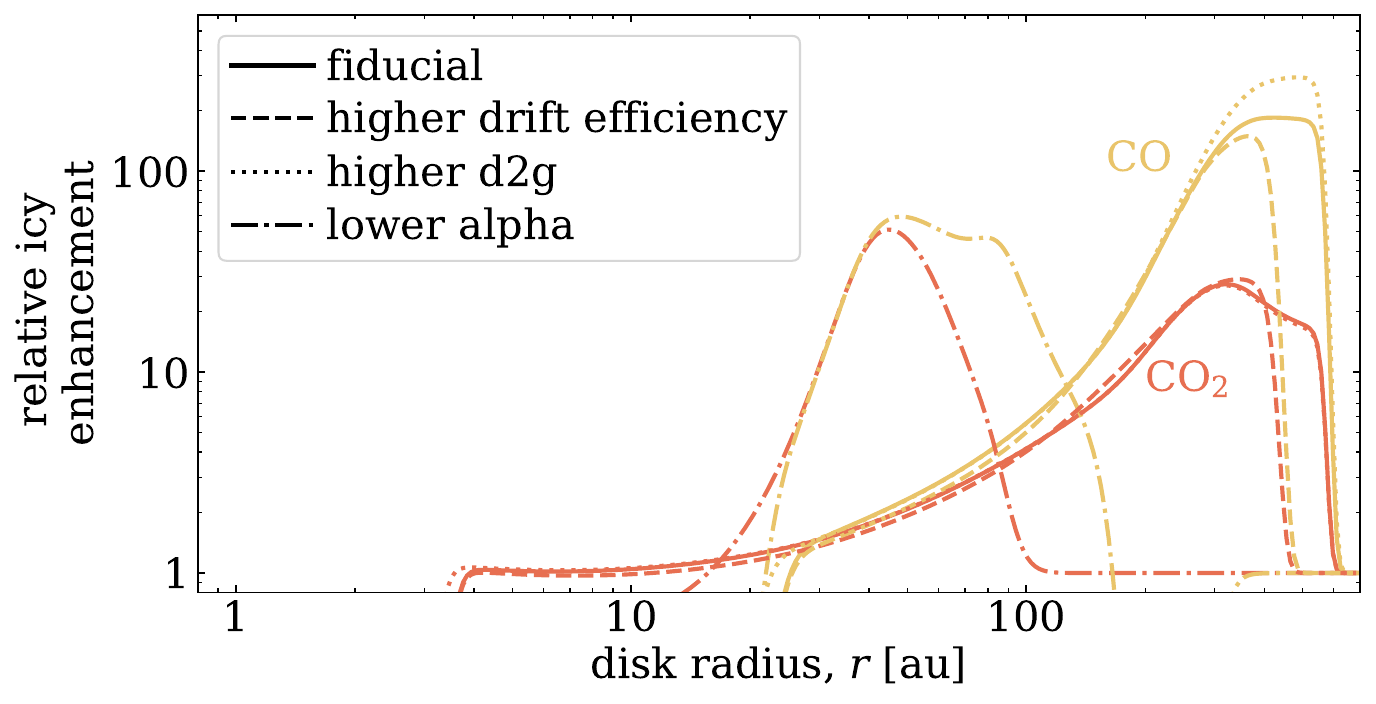}
    \caption{{\bf Relative icy enhancements occur regardless of variation to physical disk parameters.} Relative enhancement of CO$_2$ and CO ice to H$_2$O ice as compared with initial abundances at 3 Myr assuming different efficiencies for the physical dependencies. The solid line is the fiducial model, where the drift efficiency is $f_{\rm d}=0.1$, dust-to-gas ratio is $10^{-3}$, and turbulence parameter is $\alpha=10^{-3}$. The dashed line is for a model with higher drift efficiency, $f_{\rm d}=0.5$. The dotted line is for a disk with an order-of-magnitude higher dust-to-gas mass ratio of $10^{-2}$. The dotted-dashed line is for a disk with an order-of-magnitude lower turbulence of $\alpha=10^{-4}$.}
    \label{fig:3Myr_relenh_physmod}
\end{figure}

We test the sensitivity of volatile ice enhancements as compared with H$_2$O ice across a set of parameters governing the physical timescales in the disk: turbulence strength, particle drift efficiency, and dust-to-gas mass ratio. 
Figure~\ref{fig:3Myr_relenh_physmod} compares the 3 Myr fiducial CO$_2$ and CO relative icy enhancements (solid line) with one that assumes an order-of-magnitude lower turbulence strength (dash-dotted line), one that has higher radial drift efficiency for marginally-coupled particles (dashed line), and one that has an order-of-magnitude higher dust-to-gas ratio (dotted line). 
Lowering the turbulence parameter to $\alpha = 10^{-4}$ increases the viscous timescale relative to the fiducial case of $\alpha = 10^{-3}$, reducing the efficiency of viscous diffusion, and as a result, volatile enhanced icy particles remain more concentrated near the initial $r_{\rm c} = 20$~au. 
The 3 Myr peak icy enhancement for hypervolatile CO decreases to $\sim40 - 50\times$ from $>100\times$ in the fiducial model.

The icy particle drift efficiency sets how quickly the outer disk is depleted of radially inward-drifting, marginally-coupled particles. When the particle drift efficiency is increased from the fiducial value of $f_d = 0.1$ to 0.5 \citep[an approximation for particles growing and drifting on similar timescales;][]{birnstiel_gas-_2010}, the enhancement peaks become sharper and reduced by $\sim30\%$. 
The overall enhancement reduction is due to relatively more loss of volatile species from the outer disk; all ice surface densities decrease in this model, but the decrease is larger for more volatile species. 
In other words, when drift is slower, it is more selective in desiccating the outer disk of water-like species, while an overall faster drift means that drift is important at a larger range of disk times, and hence is less discriminatory between different kinds of volatiles. 

Disks with a higher dust-to-gas mass ratio, increasing from $10^{-3}$ to $10^{-2}$, raises the peak CO icy enhancement by $\sim2\times$, while CO$_2$ enhancements remain largely unchanged.
This difference may be understood by considering that for hypervolatiles, whether the dust-to-gas ratio increases by $10\times$ through more solid particles and/or a reduction in gas density, there will be more vapor leftover for efficient late time adsorption, resulting in the $2\times$ increase in relative icy enhancement for CO to H$_2$O as compared to that of the fiducial model.
Whereas the mid-volatile species' enhancements are regulated by early time loss of volatile material through drifting and desorbing ices, with less material leftover for later time adsorption, which should not change the evolution between the mid-volatiles and water significantly with a uniform change in the dust-to-gas ratio.

\subsection{Particle size distributions}\label{ssec:mech_sizedist}

\begin{figure}[!t]
    \centering
    \includegraphics[width=1\linewidth]{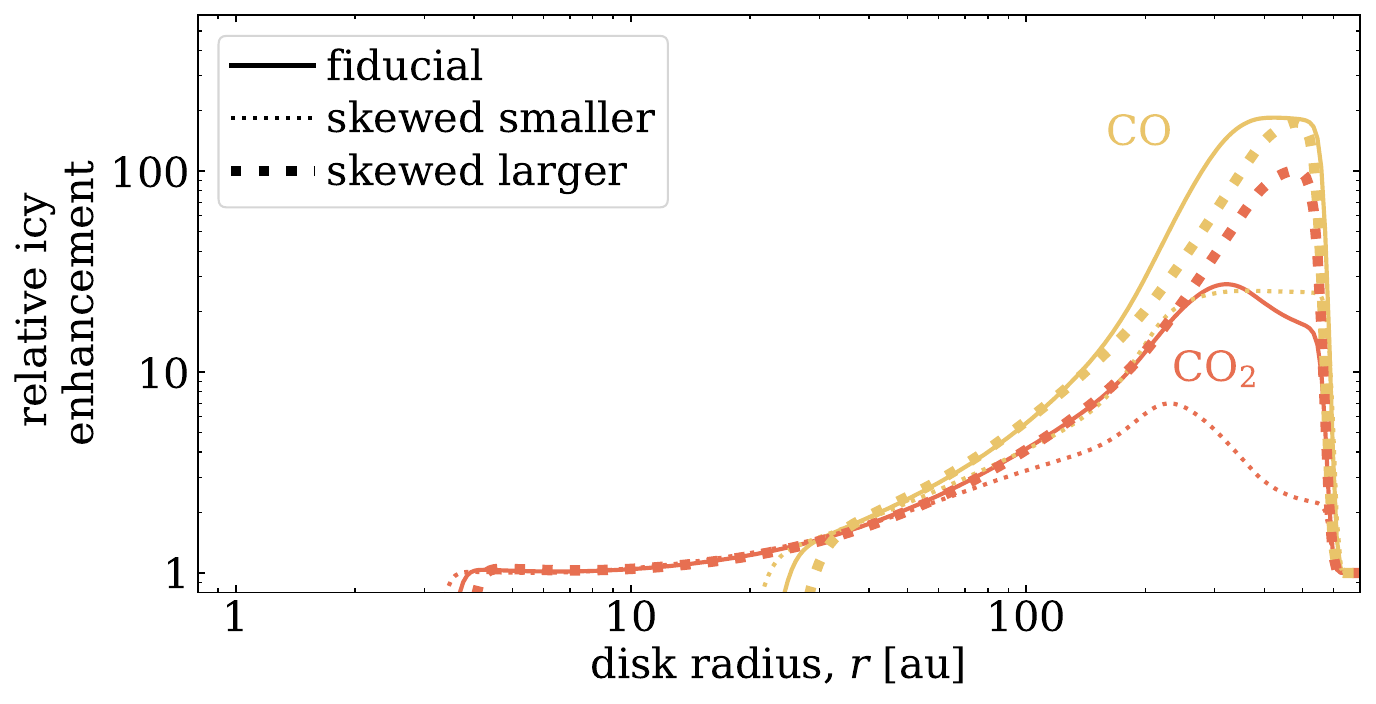}
    \caption{{\bf The degree of relative enhancement matches between the mid-volatile and hypervolatile species when the distribution is skewed towards larger particle sizes, but shows a decrease for distributions skewed towards smaller particle sizes.} Relative enhancement of CO$_2$ and CO ice to H$_2$O ice at 3 Myr since initial, comparing the fiducial (solid) collisional cascade model with distributions that are skewed towards larger (dotted-dashed) or smaller (dotted) solid particles by 5\%.}
    \label{fig:3Myr_relenh_sizedist}
\end{figure}

Figure~\ref{fig:3Myr_relenh_sizedist} displays the 3 Myr CO$_2$ and CO icy relative enhancements compared to H$_2$O ice for particle size distributions that are skewed from the fiducial solid mass distribution by 5\% towards smaller (dotted line) or larger (dash-dotted line) particles. The mass distribution that is skewed towards smaller particles, imitating a scenario in which collisional fragmentation is more efficient than particle growth, results in weaker enhancements by $\sim80\%$.
Although there is more surface area for adsorption in the skewed smaller case, there is also less mass drifting in on icy particles and hence less drying out of the outer disk.
Even when there is less inward-drifting solid material feeding the vapor content for later-time adsorption, both mid- and hypervolatiles still become about an order-of-magnitude enhanced in the ice as compared to H$_2$O ice.

When the distribution is skewed towards larger particles, imitating a scenario where particle growth is more efficient than, or is unimpeded by, collisional fragmentation and bouncing barriers \citep[i.e., as in the species-dependent growth, drift, sublimation, fragmentation model of icy particles in][]{yunerman_pathway_2024}, all species become relatively enhanced in volatile ice as compared to H$_2$O ice by $\sim50\times$.
Interestingly, while the relative hypervolatile icy enhancement decreases by $\sim0.5\times$ as compared to the fiducial case, the mid-volatile enhancement increases by $\sim2\times$. 
The increase in CO$_2$ enhancement is likely due to more CO$_2$ mass drifting and sublimating as compared with the fiducial case, allowing for more CO$_2$ vapor used to build up the relative icy enhancement in the outer disk, even if there is less surface area among the small particles for adsorption when the mass distribution is skewed towards large particles.
The reduction of CO ice enhancement is less intuitive, but is likely due to a loss of surface area available for adsorption in the outer disk, hence extending adsorption timescales beyond the lifetime of the disk. 
We note that while particle size distributions affect the precise abundances and enhancement profiles, order-of-magnitude enhancements of volatiles with respect to H$_2$O ice appears robust to precise assumptions about particle size distributions, and hence the underlying particle growth and fragmentation physics.

One possible caveat to these results is that our model does not take into account detailed solid particle size evolution and the impact that may have on the radial size distribution over the disk lifetime. 
Solid particle size evolution is classically set by collisional growth and fragmentation.
These processes, in addition to particle drift, act to reset the maximum particle size radially over time, while redistributing mass between size bins \citep[][and references therein]{birnstiel_dust_2024}. 
The balance between these processes in shaping size distributions strongly depends on radial distance, where particle growth and drift are important at all radii, while fragmentation may become increasingly more important at smaller radii.
We use the particle size evolution model in the review by \citet{birnstiel_dust_2024} to illustrate how these processes may change icy relative enhancements in our model.
They find that in the outer disk ($r=100$~au), micron-sized particles can grow through collisions by roughly 1-2 orders-of-magnitude in size within 0.1-1 Myr before they are drift-limited.
In our model, all species become relatively enhanced at these radii on Myr timescales because of long freeze-out and advection times (Figures \ref{fig:solid_particle_modeling} and \ref{fig:rel_enh}).
This implies that outer disk icy volatile to H$_2$O relative enhancements should remain fairly unchanged when considering growth of small particles. 
At smaller disk radii, the relative enhancement may be more impacted by particle size evolution.
Particles in the middle disk ($r=30$~au) grow quicker, such that the maximum particle size of the distribution increases from 10 microns to 1 cm within 0.1 Myr, but then decreases to less than 1 mm due to the drift limit. 
However, mid-volatile species icy relative enhancements in Figure \ref{fig:rel_enh} develop as early as 10 kyr, and will therefore be impacted by particle size evolution processes in addition to drift, desorption, advection, and adsorption.
In the very inner disk ($r=3$~au), the particle size distribution changes dramatically due to very fast growth, drift, and fragmentation. 
Under most circumstances including particle size evolution at these radii would therefore be very important \citep{nietiadi_collisions_2020,rozner_aeolian-erosion_2020,sirono_elasticity_2021,yunerman_pathway_2024}, but in our model we only look at enhancements that happen at 10-100s of au and should not be impacted by particle size distribution evolution of the inner disk.

A second difference between our model and those with growth, is that the maximum particle size will be different at different radii, while we initialize our model with the same distribution throughout the disk. The main discrepancy between our model and those with detailed size evolution, is that in the outer disk, drift prevents growth to above 100 microns. 
In reality, we may also be underestimating how quickly the outer disk can be dried out because of significant solid mass in larger particles that may drift more slowly.
Bringing it all together, the icy relative enhancements we find in our model at larger radii, $\sim100$'s of au, are likely robust to particle size evolution processes, but at smaller radii, $\sim10$'s of au, our model may be overestimating the degree of enhancement.

\subsection{Thermal Evolution}\label{ssec:thermal_depend}

\begin{figure}[!t]
    \centering
    \includegraphics[width=\linewidth]{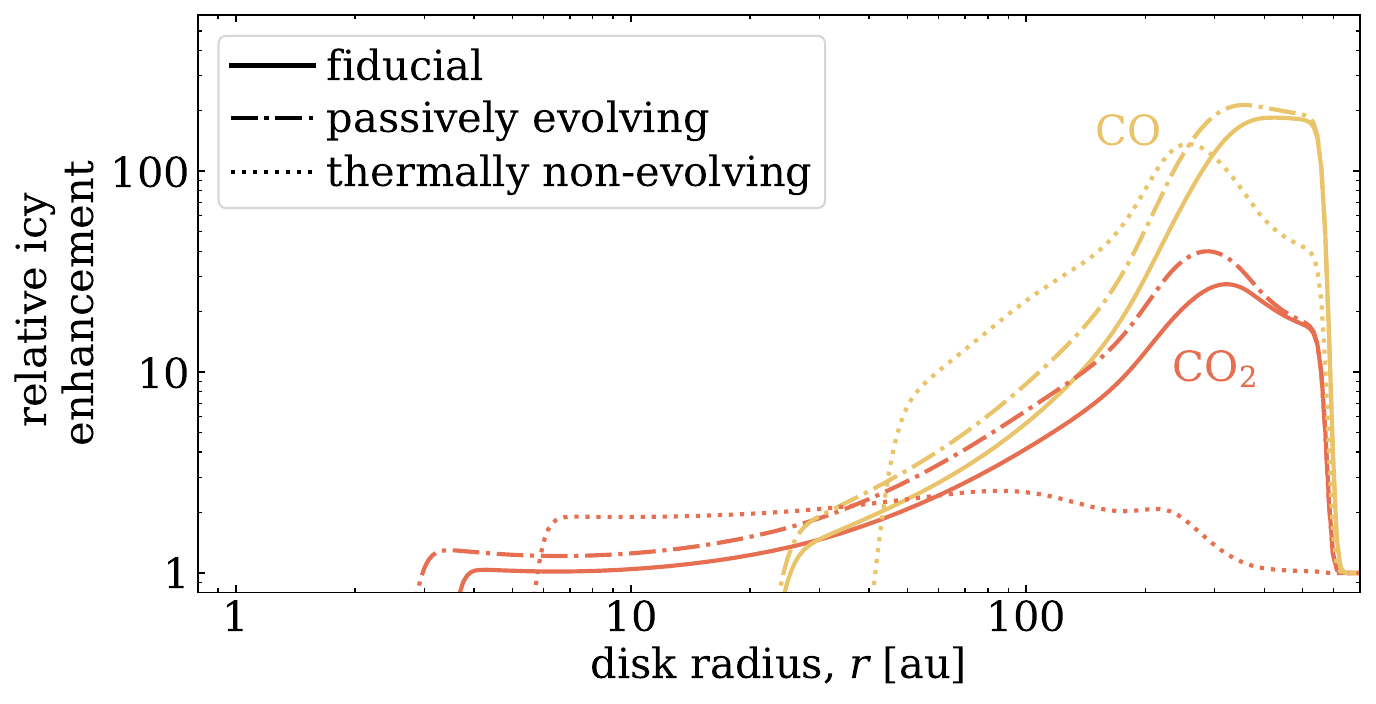}
    \caption{{\bf Relative icy enhancement between mid-volatile and hypervolatile species is sensitive to disk thermal evolution.} Relative enhancement of CO$_2$ and CO ice to H$_2$O ice as compared with initial abundances at 3 Myr across differing disk temperature scenarios. The fiducial model (solid), an actively evolving disk, is compared to one that is passively evolving (dotted-dashed) and one that is active, but non-evolving in time (thermally non-evolving, dotted). }
    \label{fig:3Myr_relenh_temp}
\end{figure}

Different assumptions about the disk thermal evolution (or lack thereof) have profound impacts on ice line locations through time, and in this subsection, we investigate how icy volatile enhancements differ between the fiducial actively evolving disk, a thermally non-evolving disk, and a passively evolving disk. 
Figure~\ref{fig:3Myr_relenh_temp} compares the 3 Myr relative enhancements for CO$_2$ and CO for the three scenarios.
The thermally non-evolving case assumes an active disk with constant values for the stellar luminosity and mass accretion rate for a $0.5{\rm M_{\odot}}$ star at 1 Myr, taking $L_{\star} = 0.65$ L$_{\odot}$ and $\dot{M} = 10^{-8}$ M$_{\odot}$ yr$^{-1}$ \citep[i.e. that of an average T Tauri star, ][]{hartmann_accretion_1998}. 
The passively evolving case uses an evolving stellar luminosity but excludes the active temperature component completely, resulting in a thermally evolving cooler disk.

All scenarios result in some hypervolatile and mid-volatile ice enhancements beyond their respective ice lines, and the passively evolving case is nearly identical to that of the actively evolving one, with the only difference appearing in the inner disk ice where the CO$_2$ ice line is shifted $\sim0.5$~au inwards.
The thermally non-evolving case provides an isolated view of the volatile ice enhancement evolution as a result of viscous diffusion, coupled particle drift, and molecular physics alone, i.e. without the added impact of inward-sweeping ice lines.
The result differs both in the location and magnitude of the peak hypervolatile and mid-volatile ice enhancements. 
Compared to the evolving disks, the enhancements happen at smaller disk radii and stall out around factors of $\sim2\times$ for mid-volatiles and $\sim20\times$ for hypervolatiles.
While the drift timescales are the same for similar sized particles of CO$_2$ and H$_2$O in the thermally non-evolving case, CO$_2$’s adsorption and desorption timescales are still longer than that for H$_2$O leading to the roughly $\sim2\times$ increase uniformly for CO$_2$ enhancements outwards of the CO$_2$ ice line. 
This model shows that even without ice lines sweeping inwards through the disk  \citep[as in the thermally non-evolving case here and the ``fixed $T$'' case of][]{price_ice-coated_2021}, enhancements can be substantial, especially for hypervolatile species.

\subsection{Volatile Phase Initial Conditions}\label{ssec:initdepend_disc}

\begin{figure}[!t]
    \centering
    \includegraphics[width=\linewidth]{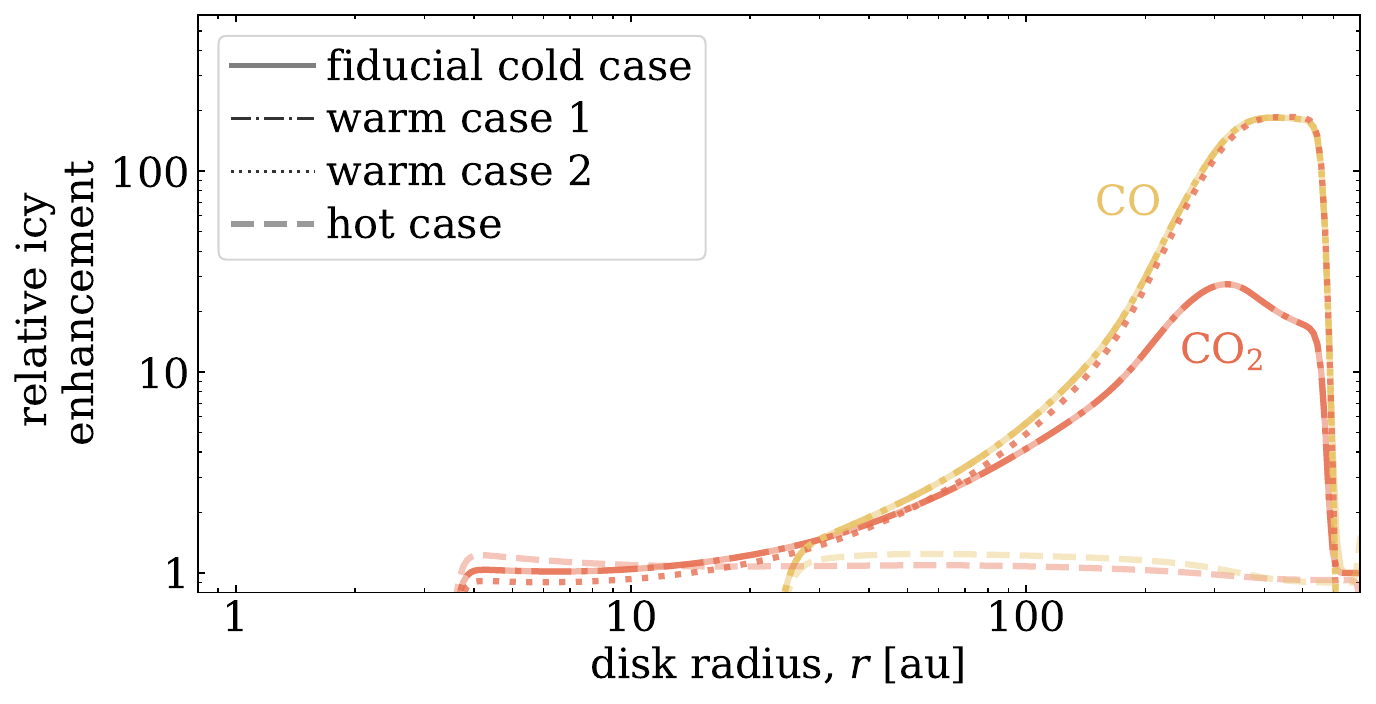}
    \caption{{\bf As long as H$_2$O ice is inherited, all species become significantly relatively enhanced as compared to H$_2$O ice.} Relative enhancement of CO$_2$ and CO ice to H$_2$O ice as compared with initial abundances at 3 Myr are shown for a variety of different maximum disk temperatures reached during the disk formation stage. In the fiducial cold case (solid), all species are initially frozen out with everything starting in the solid phase. Warm case 1 (dotted-dashed) has hypervolatile abundances fully initialized in the gas phase and warm case 2 (dotted) also has mid-volatile species begin in the gas phase. In the hot case (dashed), all species, including H$_2$O and NH$_4^+$ salt, are initialized in the gas phase.}
    \label{fig:3Myr_relenh_initphase}
\end{figure}

In the fiducial cold case, we assume that all species are inherited from the molecular cloud in the solid phase during disk formation, implying that the forming disk did not heat up above hypervolatile sublimation temperatures. 
This choice was motivated by theoretical models and observations tracing the volatile compositions, primarily in the deuterium-to-hydrogen (D/H) ratio from protostellar clouds to circumstellar disks to comets, which all suggest that H$_2$O ice likely survives the disk formation processes \citep[e.g.,][]{visser_chemical_2009,visser_chemical_2011,hincelin_survival_2013,cleeves_multiple_2016,bergner_ice_2021}.
The fates of more volatile ices are less clear, and simulations suggest that these may be subject to at least some sublimation and reprocessing during the disk formation stage \citep[e.g.][]{visser_chemical_2009}.

We run three additional models to account for other possible scenarios of inheritance versus reset, shown in Figure~\ref{fig:3Myr_relenh_initphase}. 
Here, we do not mean a chemical reset, but simply a scenario where some or all interstellar ices have sublimated into the gas phase before our simulations begin. 
Intermediary warm case 1 assumes that the disk formation stage exceeded hypervolatile sublimation temperatures, such that hypervolatiles begin in the gas phase, while warm case 2 assumes that only H$_2$O and NH$_4^{+}$ salts are inherited as ices and all other species begin in the gas phase. 
The hot case assumes that the material accreting onto the disk exceeded sublimation temperatures of H$_2$O and NH$_4^{+}$ salts upon disk formation, such that all species begin in the gas phase. 

Volatile enhancement in warm case 1 is identical to the cold case.
This can be explained by 1) mid volatiles sharing the same initial phase in both scenarios and 2) hypervolatiles effectively beginning in the same phase in both scenarios due to fast initial hypervolatile sublimation in the cold case model.
Hypervolatile enhancements in warm case 2 are identical as in the previous cases, while mid-volatiles' icy enhancement increases to become as enhanced as the hypervolatiles.
In this scenario, mid-volatiles are not lost at early times due to inward drift and desorption since the mass is initially in the vapors and must freeze-out first onto the well-coupled particles over adsorption timescales.
There is little to no relative enhancement of any species relative to H$_2$O ice in the hot case, since all solids, including H$_2$O ice, must first build-up on Myr adsorption timescales. 
This results in minimal H$_2$O desiccation from the outer disk, and hence
no formation of more volatile-dominated solids. 
In summary, as long as H$_2$O ice is preserved during disk formation, and begins frozen-out while other species may be in their vapor phase, we should expect substantial $10$-$100\times$ mid -volatile and hypervolatile icy relative enhancements in the evolving outer disk.

\subsection{Critical disk radius}\label{ssec:rcrit_disc}
\begin{figure*}[!t]
    \centering

    \includegraphics[width=0.483\linewidth]{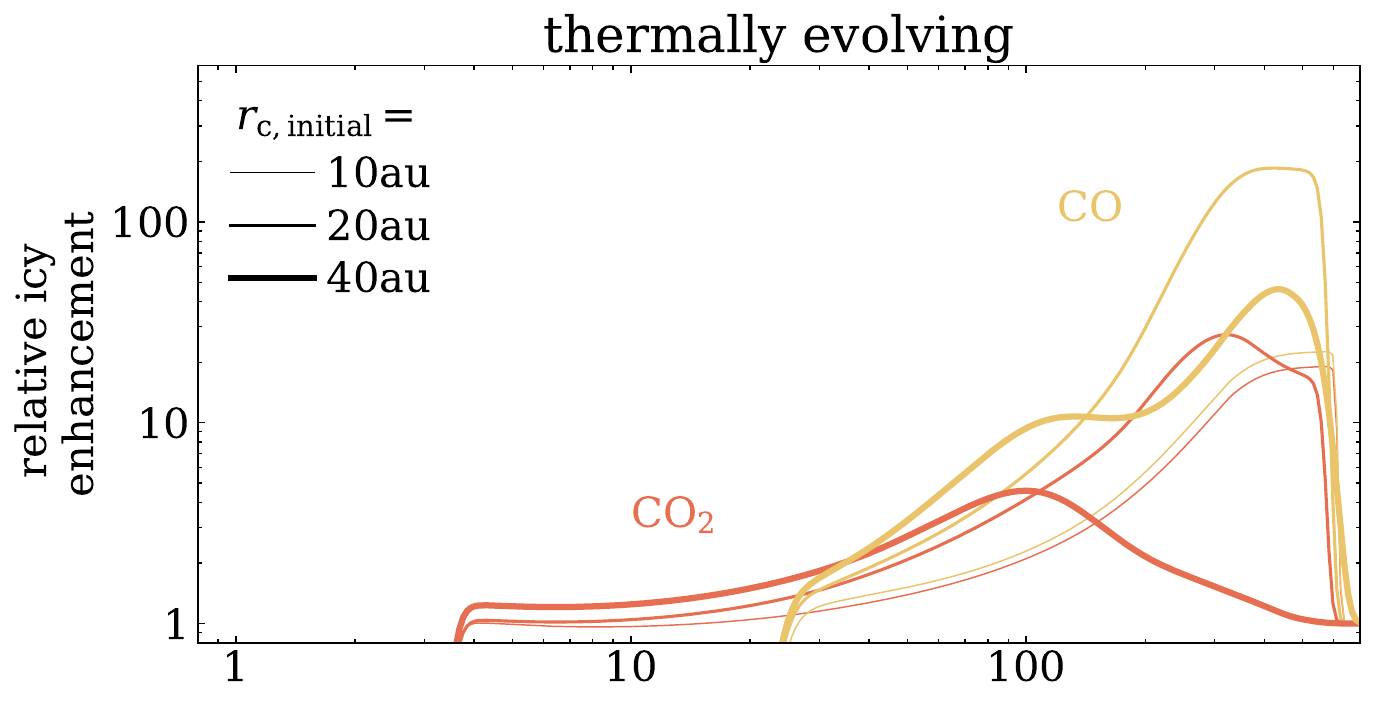}
    \hfill
    \includegraphics[width=0.483\linewidth]{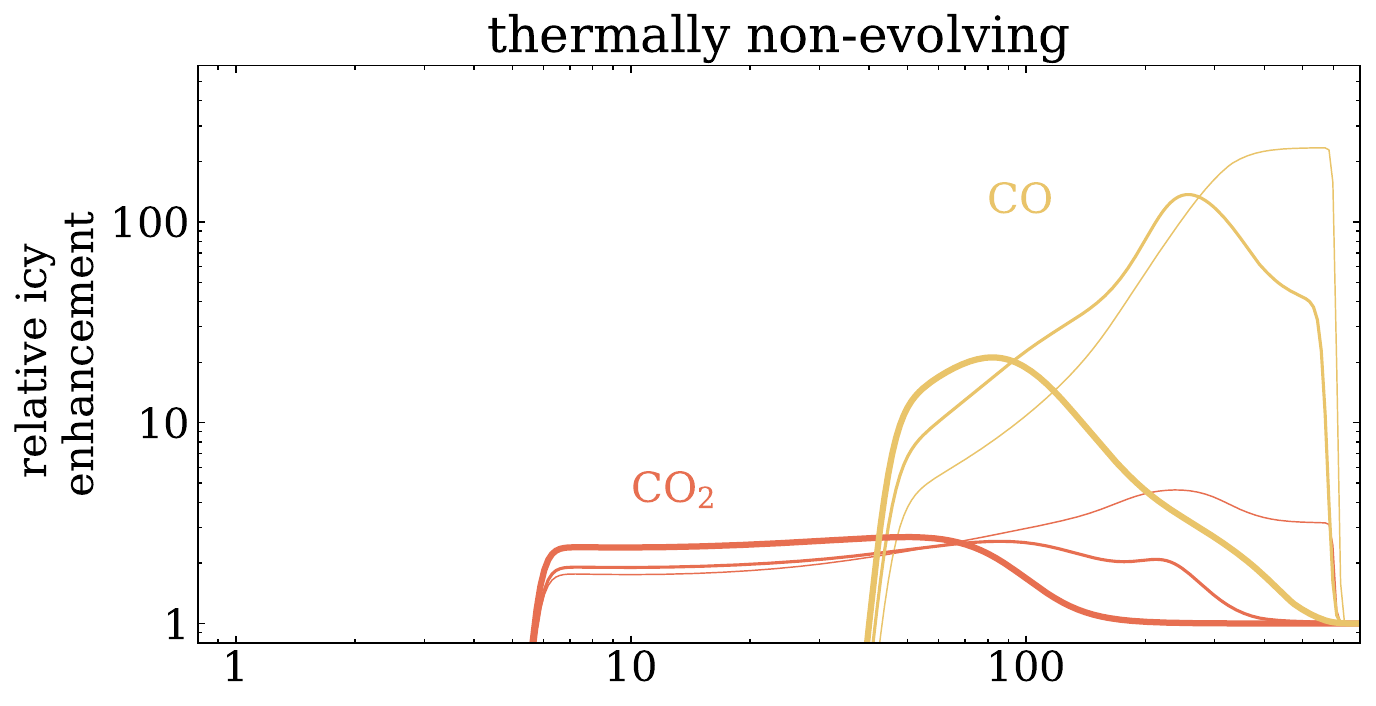}


    \includegraphics[width=0.475\linewidth]{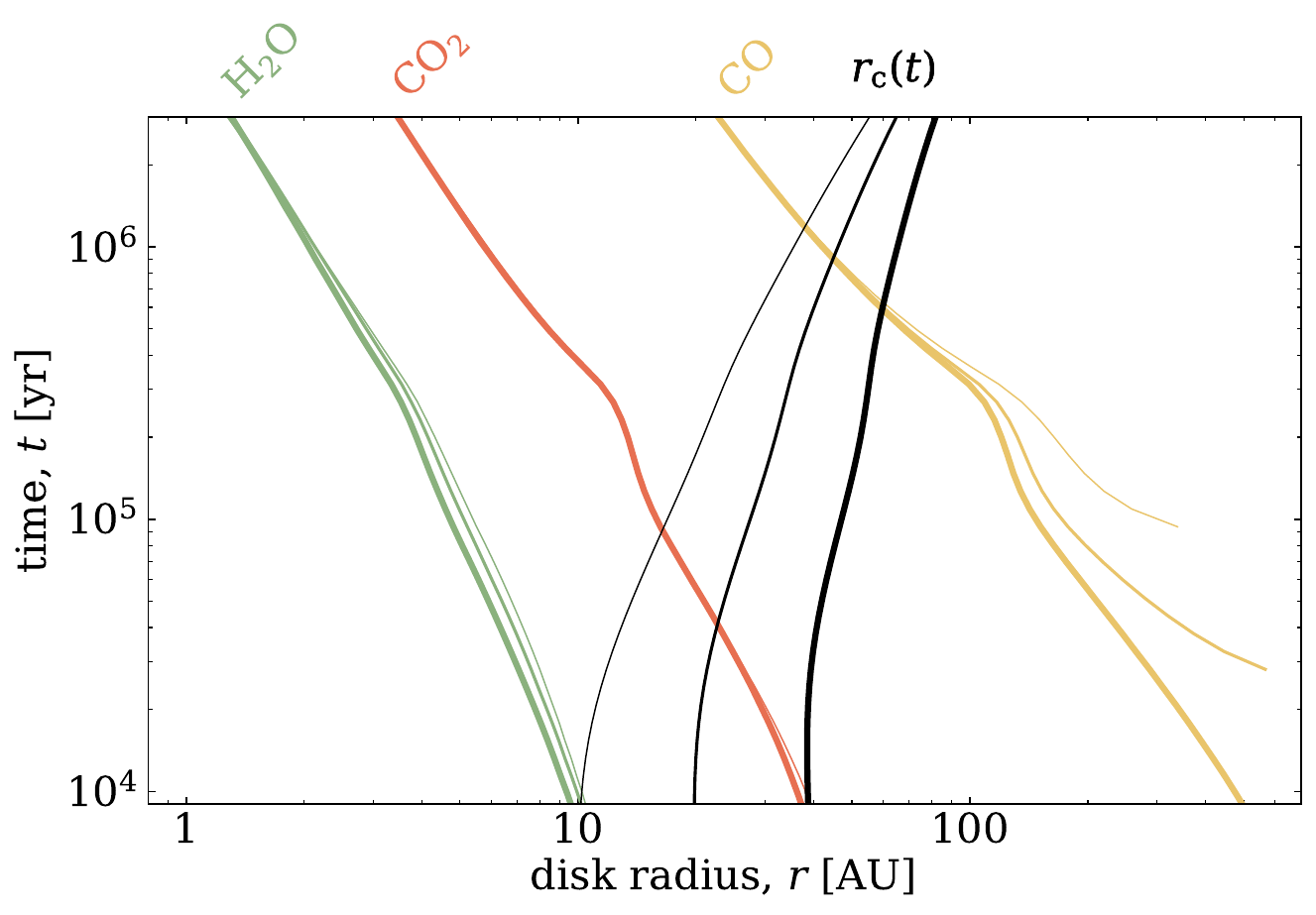}
    \hfill
    \includegraphics[width=0.475\linewidth]{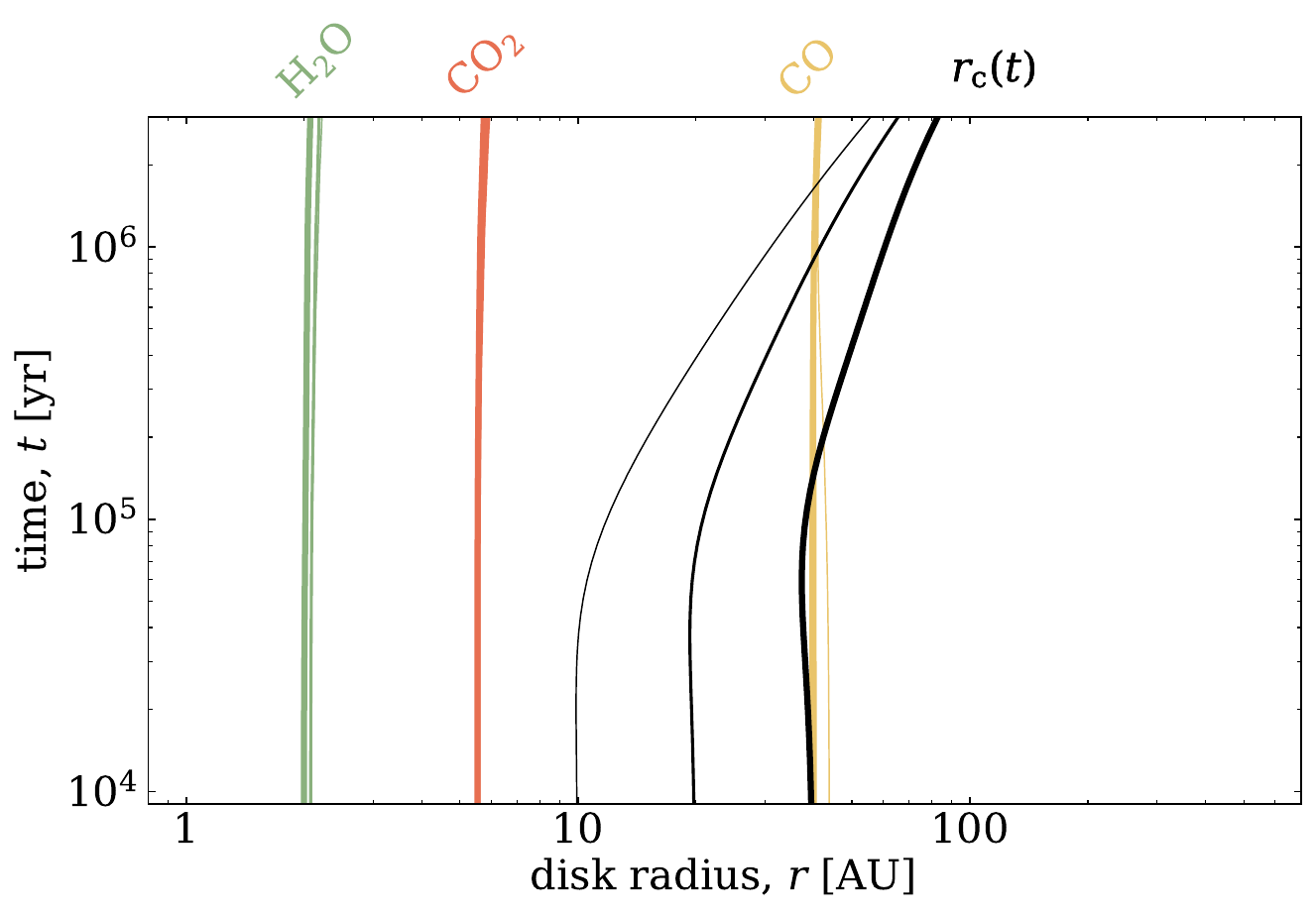}

    \caption{{\bf Relative icy enhancements occur when the species ice line is near, beyond, or evolves through the critical disk radius during the disk lifetime.} The top row shows relative enhancements of CO$_2$ and CO ice to H$_2$O ice, as compared with initial abundances at 3 Myr, for varying critical disk radii ($r_{\rm c}$). The three line thicknesses correspond to different models assuming initial $r_{\rm c}$ of 10, 20, and 40~au. For the same corresponding line thickness, in the bottom row, we display the evolution of $r_{\rm c}$, fitted iteratively with $\Sigma_{\rm c}$ using Eq. \ref{eq:self_similar}, and the H$_2$O, CO$_2$, and CO ice lines. The left column is for the fiducial, actively evolving disk, while the right column is for a thermally non-evolving disk, assuming $T(r,t=1$ Myr$)$.}
    \label{fig:3Myr_relenh_rcrit_combined_evolution_rcrit_depend}
\end{figure*}

Here we examine the dependence of the degree of relative icy enhancements on the assumed initial critical disk radius. 
As turbulent viscous diffusion becomes relevant ($t>10^4$~yr for this model based on Figure \ref{fig:timescales}), desorbing ices enriching the gas of volatiles interior to their ice lines may then be diffused outwards over the species ice line.
Vapors may also be advected outwards together with bulk gas depending on the relative position of the ice line, $r_{\rm ice}$, to the critical disk radius, $r_{\rm c}$, which both evolve during the disk lifetime\footnote{If we were to include angular momentum transport through disk winds, as in \cite{chambers_analytic_2019}, we would expect icy enhancements of lesser degree but still concentrated near the critical radius. Here $r_{\rm c}$ may not evolve outwards as much over the disk lifetime as compared to a disk driven by turbulent viscosity alone.}.
Figure~\ref{fig:3Myr_relenh_rcrit_combined_evolution_rcrit_depend} displays the 3 Myr relative icy enhancements for CO$_2$ and CO ice with respect to H$_2$O ice in the top row for initial $r_{\rm c} \in \{10,20,40\}$~au, while the bottom row displays the species ice line and critical disk radius evolution. 
We provide the thermally evolving disk in the left column and thermally non-evolving disk in the right column to isolate the critical radius and thermal evolution dependence. 

In the fiducial, thermally evolving case, reducing the critical radius ($r_{\rm c,0} = 10$~au) reduces the relative icy enhancements to $\sim10\times$ for hypervolatiles and mid-volatiles alike.
This can be understood when considering that, for most of the disk's lifetime, there is less efficient outwards advection across the species' icelines since, especially at early times, there is a large distance between $r_{\rm c}$ and $r_{\rm ice}$.
When $r_{\rm c,0} = 40$~au, the mid-volatile ices are less enhanced, which is due to the fact that beyond 10~kyrs, the critical radius is exterior to the ice line, and hence advection no longer contributes to mid-volatile enhancement. 
Outward diffusion is still active and is responsible for the lower level enhancements seen in Figure~\ref{fig:3Myr_relenh_rcrit_combined_evolution_rcrit_depend}.
By contrast, the hypervolatile ice enhancement is substantial and bimodal, with a maximum $\sim 40\times$ peak at $\sim400$~au, while a secondary $\sim 10\times$ peak occurs at $\sim 100$~au. 
The primary peak is associated with later time outward advection, while the secondary peak location is set by diffusion-driven enhancement at earlier times.

In the thermally non-evolving case, the balance between diffusion and advection driven ice enhancement shifts because the ice lines are constant and therefore located at smaller radii in the young disk. 
For CO$_2$, low levels of enhancement can be explained by diffusion when $r_{\rm c}=20-40$~au, while there is a small additional contribution from advection for the smallest critical radius, which is initially located only a few au exterior to the CO$_2$ ice line. 
For CO, diffusion dominates enhancement (up to a factor of $10\times$) for $r_{\rm c}=40$~au, while advection results in enhancements of $>100\times$ when $r_{\rm c}=10$~au.
Smaller $r_{\rm c}$ in the thermally non-evolving case leads to larger enhancements for CO since there is more material available beyond $r_{\rm c}$ later on due to the CO ice line existing at earlier times.
Fiducial $r_{\rm c}=20$~au shows a mixed behavior resulting in a double peak, similar to the case of $r_{\rm c}=40$~au in a thermally evolving disk.
 
To summarize, for mid-volatiles, $\gtrsim10\times$ relative icy enhancements can only occur if a substantial portion of volatile-rich gas can be advected outward and then efficiently readsorb on Myr timescales in the outer disk. 
This places a boundary on significant relative icy enhancements occurring when $r_{\rm c} \lesssim r_{\rm ice}$. 
Hypervolatile ices are more robust to relative icy enhancements, both because diffusion alone can produce $10\times$ enhancements, and because advection and readsorption are more likely to play a role across a larger parameter space; i.e. it is more likely that $r_{\rm c} \lesssim r_{\rm ice}$ for the majority of the disk lifetime for hypervolatiles.
By comparing the outer disk mid-volatile and hypervolatile ice abundances with comet compositions, we may be able to constrain the protosolar nebula's $r_{\rm c}(t)$. 
For example, a comet enhanced in CO and not in CO$_2$, may suggest a large $r_{\rm c}$. 
More generally, the relative enhancements of mid-volatile and hypervolatile species should encode important information about the chemo-dynamical evolution of the birth disk.


\section{Final Ice Abundances and Implications for Comet Compositions} \label{sec:comets}

\begin{figure*}[!t]
 \centering
 \includegraphics[width=17cm]{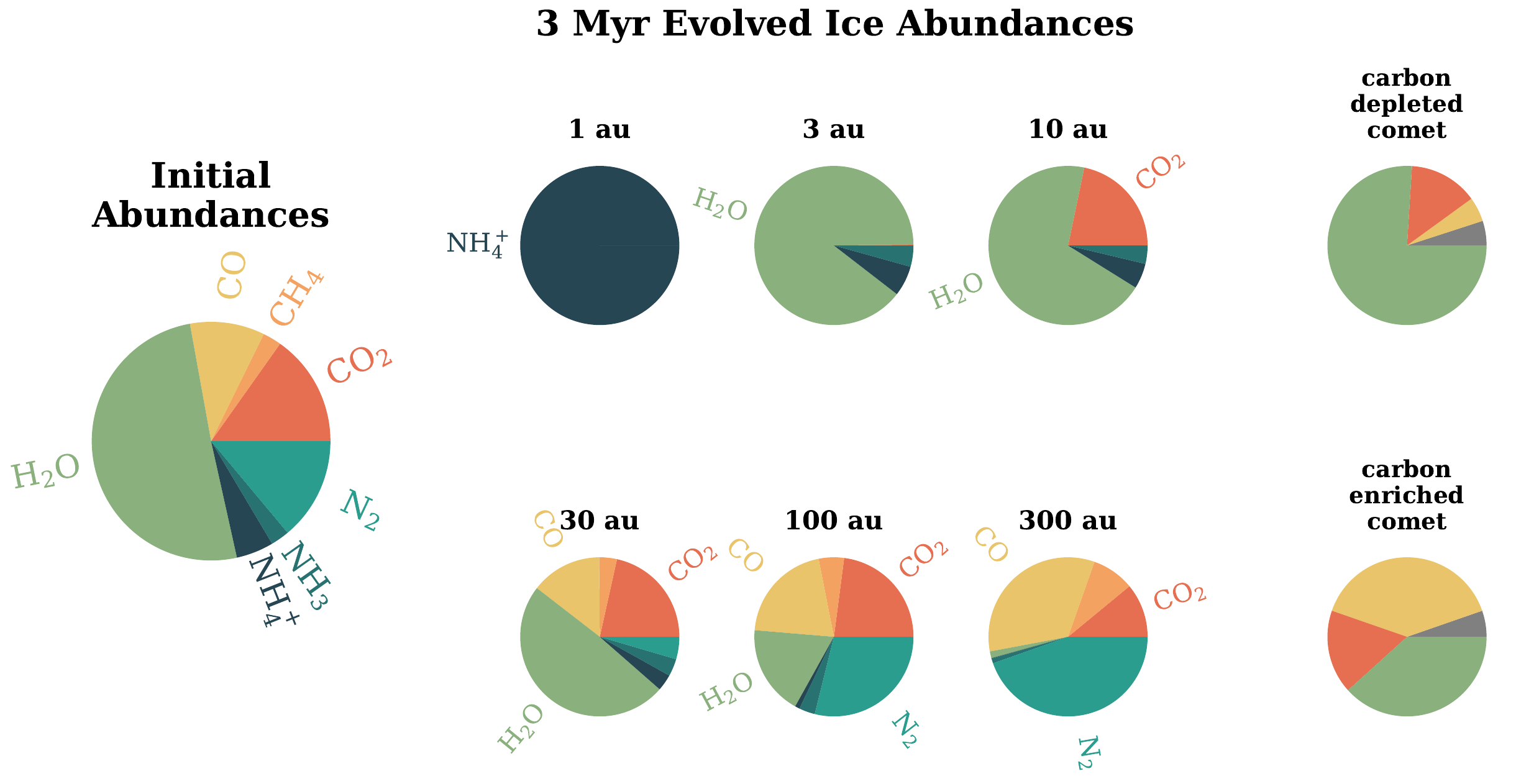}
 \caption{{\bf Outer disk regions become increasingly H$_2$O ice poor as compared to dominating N$_2$, CO, and CO$_2$ ices over a disk's lifetime.}
 The leftmost pie displays the assumed initial abundances across all species that arrive in the disk frozen-out on solid particles. 
 Individual species making up the total volatile ice composition are labeled and color-coded. 
 Radial ice abundances from 1 to 300~au are displayed at 3 Myr in the six small pies to the right of the initial abundances. 
 Species labels are only included in the radial 3 Myr pies if their ice abundance makes up at least 10\% of the total ice composition. Displayed to the right of the 3 Myr radial ice composition pies are those of representative solar system carbon-depleted and carbon-enriched comets adapted from \cite{jewitt_interstellar_2023}, which use the average of measured H$_2$O, CO$_2$, and CO ice production rates from a number of solar system and interstellar comets for the carbon-depleted comet, detailed in \cite{seligman_volatile_2022}, and comet C/2006 W3 (Christensen) from  \cite{ootsubo_akari_2012} for the carbon-enriched case. Refractory species are not considered in these pies, only volatile species.}
 \label{fig:icecomps}
\end{figure*}

Redistribution of volatiles through coupled drift, diffusion, desorption, adsorption, and advection mechanisms leads to the development of significant relative icy enhancements of mid- and hypervolatile ices as compared to H$_2$O ice, and sets the available material and ice content for planets, moons, and comets forming in the mid-to-outer disk. 
Final 3 Myr ice compositions in our fiducial disk model, displayed in Figure~\ref{fig:icecomps} for several disk radii, are notably marked by an innermost region (1 au) enriched in ammonium salts, an inner disk (1-10~au) that is H$_2$O ice-rich, a middle disk region that is primarily H$_2$O ice but also has significant fractions of CO$_2$ and CO ices, and an outer disk ($\gtrsim$100~au) dominated by N$_2$ and CO ices and almost devoid of H$_2$O ice. 

We can compare our 3 Myr disk ice compositions with those of the representative ``carbon depleted'' and ``carbon enriched'' comets in Figure~\ref{fig:icecomps} \citep[as defined in][]{seligman_volatile_2022,jewitt_interstellar_2023}. 
The majority of measured comets fall into the ``carbon-depleted'' or water-rich category \citep{mumma_chemical_2011,ahearn_cometary_2012,cochran_composition_2015,altwegg_cometary_2019}, while the paradigmatic ``carbon-enriched'' comet is that of C/2006 W3 Christensen \citep{ootsubo_akari_2012}. Generally, what makes both W3 and other ``carbon-enriched" comets rich in carbon is that they  contain substantial amounts of mid- and hypervolatile species, many of which contain carbon \citep{mckay_evolution_2015,mckay_evolution_2018,altwegg_cometary_2019,weissman_origin_2020,pinto_survey_2022}. 
Previous studies have concluded that the carbon-depleted comets likely formed not too far beyond the H$_2$O ice line,
from particles rich in H$_2$O ice, and with modest abundances of more volatile species present through entrapment or freeze-out beyond their native ice lines \citep{bockelee-morvan_composition_2017,lippi_investigation_2021,blum_formation_2022,willacy_comets_2022}. Comparing their typical composition to our fiducial model (Figure \ref{fig:icecomps}) shows a reasonable agreement around 10 au, where the presence of CO$_2$ ice is accounted for by CO$_2$ condensation just outside of the CO$_2$ ice line in the mature disk.

The carbon-enriched comet contains similar parts CO and H$_2$O ices, and around $15\%$ in CO$_2$ ice, which resembles our 3 Myr $\sim30$~au ice composition in our model, except that we see somewhat less CO enhancement and somewhat more CO$_2$ enhancement. 
We only obtain the `correct' proportion between CO and CO$_2$ ice at much larger distances, but there, water ice is completely gone. Perhaps this mismatch suggests that the Solar Nebula is more similar to one of our model variations that resulted in overall more CO enhancement and/or less CO$_2$ enhancement in the outer disk.
Finally, the growing subset of comets that have more CO ice than H$_2$O ice likely formed not just beyond the CO ice line \citep{eistrup_cometary_2019,price_ice-coated_2021,mckay_quantifying_2021,malamud_are_2022,biver_chemical_2024}, but far enough out in the disk to produce a large advection-driven hypervolatile enhancement. 
Comet C/2016 R2 (PANSTARRS) is one of the most extreme examples, exhibiting CO and N$_2$ as its dominant volatiles, with smaller fractions of CO$_2$ and CH$_4$, and almost no H$_2$O \citep{biver_extraordinary_2018,mckay_peculiar_2019,altwegg_evidence_2020}. 
Comets C/2021 A1 (Leonard) and C/2023 A3 (Tsuchinshan–ATLAS) also have high CO/H$_2$O production rate ratios, though theirs are less extreme than that of C/2016 R2 \citep{pinto_survey_2022,lippi_chemical_2024,biver_chemical_2024}.
The relatively small number of carbon-dominated comets may suggest that in the solar system, there was little solid material beyond ~30 au; alternatively, comets that formed in the outermost disk may have been preferentially scattered and lost, or they are rarely observed for other reasons. 
Still, the fact that these hypervolatile rich comets do exist suggests that we can exclude a solar nebula where all icy grains were completely sublimated prior to disk formation (see Section~\ref{ssec:initdepend_disc}).

The number of interstellar comets is still small, but we may expect their compositions to only partially overlap with Solar System comets, since the likelihood of comets escaping a planetary system may depend on their formation radii. 
We speculate that planetesimals that form at large disk radii in the cold, hypervolatile ice-rich regions are more likely to be dynamically susceptible to planetary system ejection via gravitational encounters, and should hence be over-represented in the interstellar comet population \citep{hahn_orbital_1999,malamud_are_2022,jewitt_interstellar_2023}. 
This hypothesis will require more observations to verify, but in the meantime, we note that the interstellar comet 2I/Borisov showed strong CO outgassing and a high CO/H$_2$O ratio \citep{bodewits_carbon_2020,cordiner_unusually_2020}, while 1I/’Oumuamua, experienced a substantial acceleration, possibly driven by the release of trapped H$_2$ and/or N$_2$ \citep{seligman_volatile_2022,bergner_acceleration_2023}. 
Recently-observed 3I/ATLAS \citep{jewitt_hubble_2025}, likely older \citep{taylor_kinematic_2025} and unrelated to 1I and 2I \citep{hopkins_different_2025}, will continue to provide more context to icy planetesimal formation as its volatile content is revealed over the next months. 
We expect that more interstellar objects will be observed and measured in the future, given that the estimated number density of such objects is $\sim0.1$~au$^{-3}$ \citep{marceta_synthetic_2023,jewitt_interstellar_2023}; our model-based expectation is that they on average be more rich in hypervolatile ices, particularly CO and N$_2$, and relatively depleted in H$_2$O ice compared to the typical Solar System comet. 


\section{C/N/O Ratios and Implications for Gas Giant Compositions} \label{sec:gas_giants}

\begin{figure}[!t]
 \centering
 \includegraphics[width=8cm]{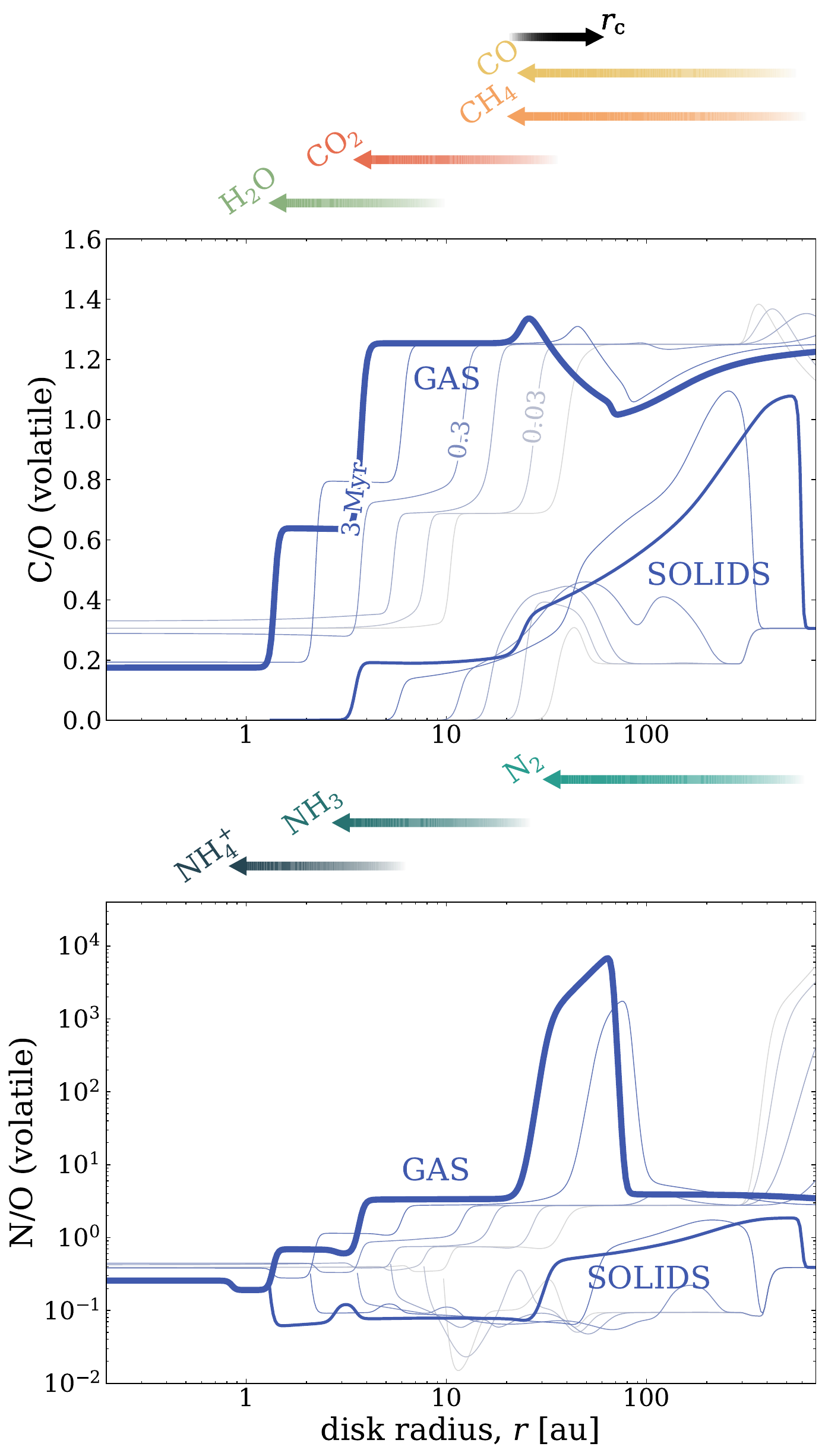}
 \caption{{\bf Volatile C/N/O abundance ratios evolve significantly over a disk's lifetime as a result of coupled drift, desorption, advection, and adsorption.} Fiducial radially and temporally evolving C/O and N/O elemental abundance ratios are displayed in the top and bottom panels, respectively. Ratios are split into their gas and solid components indicated as the two thickest lines for 3 Myr, where the gas is the thicker of the two. Previous timesteps are included as increasingly less colorful thin lines, where 0.03, 0.3, and 3 Myr times are labeled. ``GAS'' AND ``SOLIDS'' labels differentiate between the ratios in the gas and those in the total solids. Evolving ratios are compared with evolving ice lines displayed for each species as colorful arrows where the tail starts at the first instance where species desorption and adsorption timescales are the same and ends at the final 3 Myr ice line radius. The full critical disk radius evolution is also displayed as an arrow at the top of the figure labeled $r_{\rm c}$. }
 \label{fig:elem_ratios}
\end{figure}

\begin{figure*}[p]
 \centering
 \includegraphics[width=17.5cm]{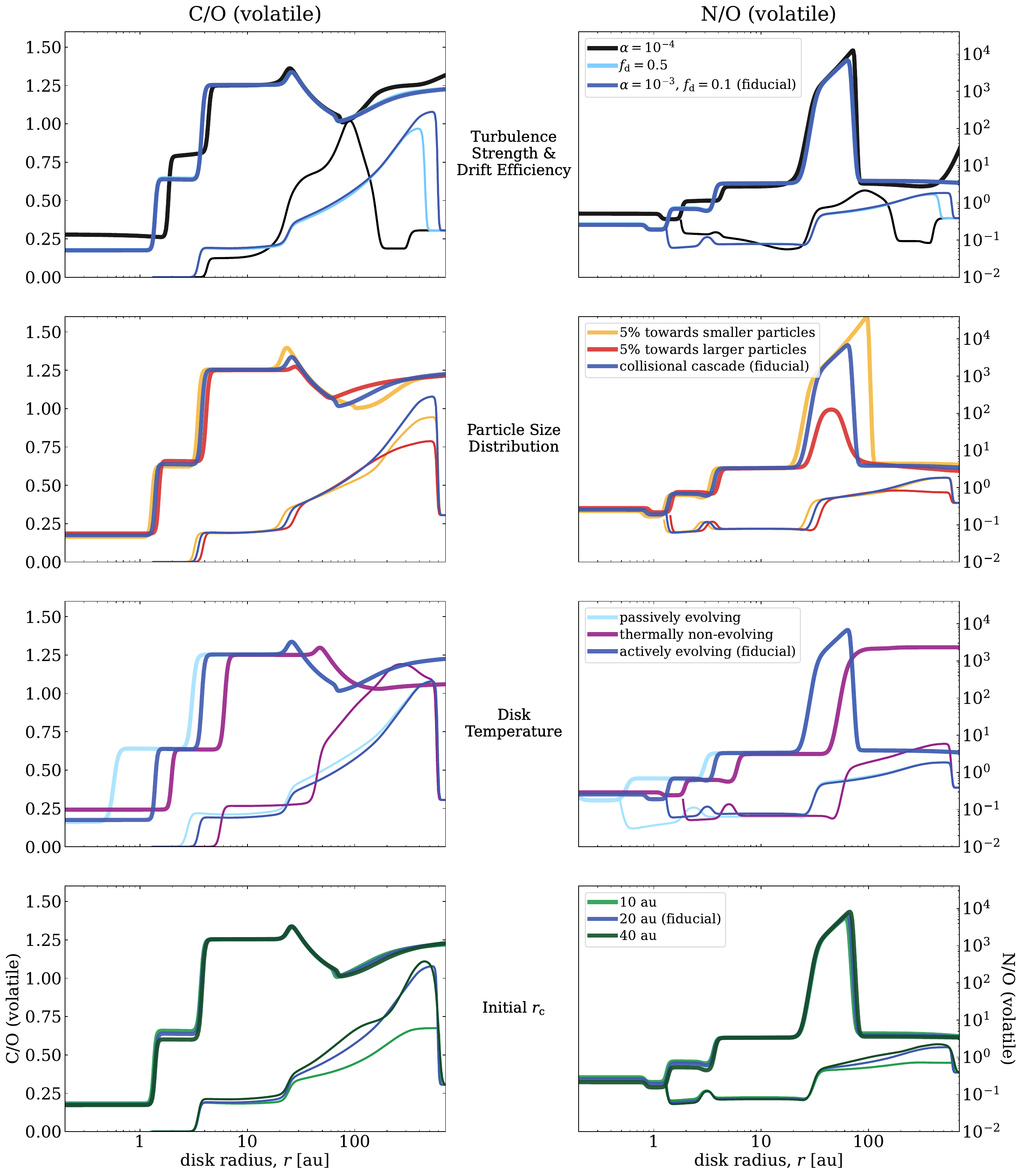}
 \caption{{\bf Expected C/N/O ratios are sensitive to model assumptions.} The fiducial 3 Myr C/O (left column) and N/O (right column) ratios across the disk (blue), as first shown in Figure~\ref{fig:elem_ratios}, are compared with those of varying initial conditions for each row. First are the physical disk parameters, namely turbulence strength and particle drift efficiency, followed by particle size distribution effects when skewing towards smaller or larger particles. Next are disk temperature effects where the fiducial actively evolving disk is compared with one that is static and one that is passively evolving. Last are the resulting C/N/O ratios assuming different initial critical disk radii, $r_{\rm c}$.}
 \label{fig:elemratio_depends}
\end{figure*}

Dynamical disk processes redistributing volatile-rich gases and ices should also regulate the evolving elemental C/N/O ratios available for gas giant atmosphere formation. 
Taking into account a combination of gas accretion and pollution by icy pebbles and planetesimals, elemental ratios, such as C/O, C/H, and O/H, may be used to distinguish the gas giant formation location \citep{oberg_effects_2011,turrini_tracing_2021,bitsch_how_2022}.
While directly linking planetary elemental ratios to disk location is complicated by the interdependent processes that change the physical and chemical makeup of the disk and planet in time \citep{molliere_interpreting_2022,knierim_constraining_2022,knierim_unraveling_2025}, the distribution of volatiles in the disk is still what sets the reservoir for planetary compositions \citep{mah_close-ice_2023,penzlin_bowie-align_2024}. 
We further discuss several of these complications in the context of our results for the C/O and N/O ratios at the end of this section.
Figure~\ref{fig:elem_ratios} compares the evolving fiducial gas and solid volatile C/O and N/O ratios across the disk at different time steps. 
The evolving species' ice lines and critical disk radius are also shown as arrows beginning and ending from initial radii (when $t_{\rm des} = t_{\rm ads}$ for species ice lines and $r_{\rm c,0} = 20$~au) to final, 3 Myr radii.
We do not include or consider the C, N, and O abundances locked up in refractory grains\footnote{See recent work by \cite{houge_burned_2025} focusing on modeling the distribution of refractory organics in the inner disk.} and instead focus solely on the volatile material. 
At early times, our fiducial C/O and N/O gas and solid ratios resemble those of static disk models that consider desorption and adsorption, except for a small inward displacement of ice lines due to particle drift \citep{piso_co_2015}. 
Gas C/O and N/O ratios are $<1$ interior to the CO$_2$ ice line, while exterior to it C/O $\sim 1.2$ and N/O $>3$.
Solid C/O is small throughout the disk and even in the outer disk it is $<0.2$, while the solid N/O is even lower. 
With time, the ice lines shift radially inwards, similar to \cite{miley_impact_2021}, but otherwise the impact of dynamics on the gas-phase elemental ratios are modest. 
Especially in the inner disk, these changes are smaller compared to what is seen in many previous disk dynamical studies that similarly include drift, desorption, and adsorption.
While we do see some depletion of inner disk gas C/O, models by \citet{booth_chemical_2017}, \citet{booth_planet-forming_2019}, \citet{mah_close-ice_2023}, \citet{lienert_changing_2025}, \citet{williams_locked_2025}, and \citet{houge_burned_2025}, find that the inner disk gas C/O first becomes substantially depleted, more so than in Figure \ref{fig:elem_ratios}, and in some models at later times, the gas C/O then increases to above initial levels.
The earlier depletion is caused by inward drift and sublimation of H$_2$O ice, which we also see to some degree.
The magnitude of this depletion is set by the balance between the inward pebble flux and the diffusive and advective removal of the water vapor, where the latter appears more efficient in our model as compared to some earlier works. 
The later time increase in inner disk gas C/O is determined by the inward flux of carbon-rich vapor. 
This can be due to both diffusion and advection, and in our disk model, most carbon-rich vapor is advected outwards due to a smaller assumed critical disk radius.

The solid elemental ratios are strongly impacted by dynamics after $\sim$0.5~Myr, where in the outer disk, C/O and N/O solid ratios increase significantly due to the mid-volatile and hypervolatile icy enhancements building up along $r_{\rm c}(t)$.
Most previous models do not see such continuous large enhancements in the outer disk. 
Instead, models that include diffusion and readsorption find large and sharp enhancements in solid C/O ratios, just beyond abundant carbon species ice lines \citep{booth_planet-forming_2019, schneider_how_2021, williams_locked_2025}.  
The magnitude of these changes are similar to those seen before $\sim 1$ Myr for our fiducial solid C/O in Figure \ref{fig:elem_ratios}, but with the difference that our enhancements are smoother.
This difference between our and previous results are due to efficient spreading of materials in models that have a smaller $r_{\rm c}$ and/or are more efficient at diffusing material beyond the relevant ice lines \citep{booth_chemical_2017,houge_burned_2025,molyarova_co_2025}.
The enhancement in solid C/O that we find in the outer disk after 3 Myr is much larger as compared with other models, increasing to near unity, because these relative icy enhancement regions continue to be fed more efficiently in time for more volatile species as $r_{\rm c}$ moves outward.
Models with larger $r_{\rm c}$, further than species ice lines, will not diffuse out materials significantly, such that shorter timescale processes, like particle size evolution, and their impact on relative abundances, may persist without being smoothed out, perhaps resembling those of previous studies.
Generally, the solid C/O and N/O increases we find are a result of the combined effects of depletion of H$_2$O and outwards advection and efficient readsorption of mid- and hypervolatile vapors onto the small, well-coupled solid particles that remain in the outer disk.
In well-desiccated disks, the solid C/O ratio depends mainly on the relative enhancements of CO and CO$_2$ in the outer disk, while the N/O ratio depends on the relative enhancements of N$_2$ and NH$_3$ versus CO and CO$_2$. 
In disks where H$_2$O desiccation is less efficient, the amount of icy H$_2$O particles also needs to be taken into account when calculating the elemental solid ratios.

Figure~\ref{fig:elemratio_depends} displays the variation in C/N/O ratios at 3 Myr, across the model-dependent parameter space introduced in Section \ref{sec:variations_from_fiducial}, as compared to our fiducial model; below we note some of the most important effects.
When particle drift is more efficient, the C/O enhancements occur somewhat closer to the CO and N$_2$ ice lines and are slightly depressed, while the gas-phase ratios change little.
When turbulent viscous diffusion is less efficient, solid enhancements are not advected outwards as quickly, which causes a small decrease and inward shift of the super-solar solid C/O and N/O ratios.
Skewing the mass distribution towards larger particles depresses the C/O and N/O ratios because of an increase in CO$_2$ enhancement, and a decrease in hypervolatile enhancement as discussed in Section \ref{sec:variations_from_fiducial}. We also note that the ice line locations appear closer to the star as more solid mass is skewed towards larger particles, which is consistent with the models of \cite{piso_co_2015}, where larger particles dominate.
Skewing the solid mass distribution towards smaller particles has almost no impact on elemental ratios.
Passive disks are cooler, and as such, their inner disk ice lines, particularly those of H$_2$O and NH$_4^+$ salt, and to some degree, those of NH$_3$ and CO$_2$, occur at smaller radii, which is reflected in the inner disk ($r\lesssim10$~au) gas and solid C/O and N/O ratios. 
When the disk is thermally non-evolving, mid-volatile ices are lost to drift and desorption at early times, decreasing the CO$_2$ solid abundance, and thus increasing the solid C/O and N/O ratios by factors of $2-5\times$ in much of the outer disk. Finally, changing the initial critical disk radius from 20~au in the fiducial model to 10 or 40~au has arguably the largest impact in the predicted C/O and N/O ratios.
First, changing $r_{c,0}$ from 10 to 20, to 40~au, pushes the effective H$_2$O ice line location outwards due to less effective inwards gas advection, which shift the gas-phase C/O and N/O curves inwards as well. 
Perhaps more importantly, the solid C/O and N/O peaks in the outer disk decrease substantially with decreasing $r_{\rm c}$ due to an increasing relative enhancement of CO$_2$ in the outer disk when $r_{\rm c}$ is smaller. 
It is interesting to note that changing $r_{\rm c}$ and changing the skewing of the particle size distribution have similar effects on the elemental gas and solid ratios, particularly seen in gas N/O. When decreasing $r_{\rm c}$ and when skewing the distribution towards large particles, the outer disk solid and gas C/O and N/O ratios decrease since both of these scenarios have a larger influx of H$_2$O at earlier times, increasing the denominator of the gas elemental ratios.

Taken together, this parameter study shows that, in the majority of disk models, the outermost disk regions have super-solar C/O and N/O ratios in {\it both} the solid and gas-phase. 
Planets accreting their atmospheres from these enhanced regions should form C/O- and N/O-rich, regardless of whether their metals originate primarily from the gas or from polluting pebbles and vapors. 
The magnitude of the enhancement does depend on the particulars of the model assumptions, however, and should be most evident in directly imaged gas giants at large separations (100~au or more).
One model assumption that may make a difference, for example, is if the angular momentum is not transported through turbulence, but instead through disk photoevaportive or MHD-driven winds, then the outer disk may not be as enhanced as material would not be advected outwards and the solid C/O and N/O ratios will be smaller \citep{chambers_analytic_2019,tabone_secular_2022,lienert_changing_2025, tabone_alma_2025}.
Another disk process that may change elemental ratios is disk chemistry, but absent of substantial vertical mixing the impact on the midplane composition should be minimal
\citep{walsh_chemical_2012,eistrup_molecular_2018,notsu_composition_2020,eistrup_chemical_2022,eistrup_chemical_2022-1}. 
Finally, much uncertainty in disk elemental ratios remains in how the solid particle size evolution is accounted for as particles can grow through collisions and surface energy effects (sintering, condensation, etc.) or get destroyed by fragmentation, erosion, and sublimation \citep{ros_effect_2019,rozner_aeolian-erosion_2020,schneider_how_2021,estrada_global_2022,powell_depletion_2022,bitsch_how_2022}. 
Solid particle structure, composition, and volatile entrapment may further change the solid volatile disk distribution \citep{xenos_how_2025,houge_burned_2025,williams_locked_2025}.

Even with an established disk composition, there are additional complications between linking  elemental ratios between disks and planetary atmospheres. 
First, a planet can accrete material at multiple times and different disk locations \citep{danti_composition_2023,guzman_franco_how_2026}. 
For example, \citet{mah_close-ice_2023} proposed that around M-dwarfs planets  accreting at early times are expected to have lower C/O ratios as compared to late accretors.
Another example concerns planet migration;
\citet{penzlin_bowie-align_2024} find that misaligned planets formed through high-eccentricity migration have a higher C/O as compared to aligned planets that may have formed through disk migration. 
Furthermore, the atmosphere may chemically evolve depending on the planetary interior structure and efficiency of convective mixing, changing the atmospheric composition from initial \citep{knierim_constraining_2022,knierim_convective_2024,knierim_unraveling_2025}. 
All these complications aside, planets forming and accreting at large disk radii, should still be impacted by outwardly advecting and increasing large enhancements in disk C/O and N/O ratios, both in the gas and solid phases.


\section{Summary and Conclusions}\label{sec:summary_conc}

We calculate radially and temporally evolving gaseous and solid surface densities for a range of volatiles in a viscous disk considering particle drift, gas diffusion, advection, species-dependent adsorption and desorption, and evolving midplane temperature profiles, while tracing five particle sizes (0.1 {\textmu}m, 5 {\textmu}m, 1 mm, 1 cm, 10 cm) and seven disk species (NH$_4^+$ salt, H$_2$O, NH$_3$, CO$_2$, CH$_4$, CO, and N$_2$, in order of ice line location in our model). 
For an actively evolving disk around a low-mass star (0.5 M$_{\odot}$), we find that:
\begin{enumerate}
    \item All mid- and hypervolatile species become relatively enhanced in the ice phase as compared to H$_2$O after 3 Myr, unless H$_2$O is initiated in the gas phase. 
    \item Hypervolatiles (CH$_4$, CO, and N$_2$) become up to $\sim100\times$ enhanced compared to initial abundances with respect to H$_2$O, while mid-volatiles (NH$_3$ and CO$_2$) can become enhanced by $\sim2-50\times$, where the latter number depends on the assumed thermal evolution, solid particle evolution, and initial disk structure.
    \item Volatile enhancement beyond a given ice line is set at early times ($t\lesssim0.5$~Myr) by particle drift and desorption, desiccating the outer disk, followed by mid-volatile and hypervolatile diffusion and advection across the ice line, and deposition onto remaining well-coupled particles. Enhancement regions are then advected outwards and further enhanced at later disk times ($t\gtrsim0.5$Myr), following the critical disk radius evolution, with the largest enhancements often seen in the outermost disk regions ($r\gtrsim100$~au).
    \item While the inner disk remains H$_2$O ice-rich ($r\lesssim30$~au), N$_2$, CO, and CO$_2$ ices come to dominate the outer disk, which has implications for comets formed in our solar system and beyond.
    \item The mid-volatile and hypervolatile enhancements in the outer disk cause solid C/O and N/O ratios beyond the hypervolatile ice lines to increase above unity in most disk models, resulting in extrasolar ratios in both gas and solids at large disk radii. 
\end{enumerate}

In conclusion, we find that the outer disk midplane becomes increasingly rich in hypervolatile and mid-volatile ices, with the enhancement peak following the outwards advection wave set by turbulent viscous diffusion. Qualitatively, this result appears robust to the specific model choices. The impacts of many model parameters are still left to explore, however, including vertical mixing and diffusion, hypervolatile entrapment in less volatile ices, gas and ice chemistry, and particle growth and fragmentation. Of these entrapment may have the largest effect based on a recent study \citep{williams_locked_2025}, and we plan to examine its role within the context of our model in a forthcoming study.

\begin{acknowledgments}

E.S.Y. thanks Dr. Ruth A. Murray-Clay for insightful discussions on disk processes, coding, and life, Dr. Elettra L. Piacentino for countless hours explaining chemistry and molecular physics, Vedant Chandra for figuring out particle size distributions with me, and Marissa Maney for fruitful discussions in the office. We also thank Drs. Sean M. Andrews, David J. Wilner, James Owen, Jenny K. Calahan, Kristina Monsch, Matthew J. Holman, Robin Wordsworth, and Dimitar D. Sasselov.
E.M.P. gratefully acknowledges the generous support of the Heising-Simons Foundation through the 51 Pegasi b Postdoctoral Fellowship during the majority of this work.
This work is also supported from a Simons Foundation grant (grant No. 686302 - K.I.Ö.). We thank the anonymous referee for providing a detailed review which has greatly improved this manuscript.
\end{acknowledgments}

We acknowledge the use of OpenAI's \href{https://openai.com/chatgpt/overview/}{ChatGPT} in the preparation of this manuscript, specifically in refining figure design and optimizing code snippets, as well as refining the text for clarity, grammar, and narrative structure. 

\software{
        \texttt{fruitypebbles} and \href{https://github.com/emprice/benzaiten}{\texttt{benzaiten}} \citep{price_ice-coated_2021},
        \href{https://waps.cfa.harvard.edu/MIST/index.html}{\texttt{MIST}} \citep{dotter_mesa_2016,choi_mesa_2016},
        \href{https://github.com/astropy/astropy}{\texttt{astropy}} \citep{astropy_collaboration_astropy_2022}, 
        \href{https://github.com/scipy/scipy}{\texttt{scipy}} \citep{virtanen_scipy_2020},
        \href{https://github.com/numpy/numpy}{\texttt{numpy}} \citep{harris_array_2020},
        \href{https://github.com/pandas-dev/pandas}{\texttt{pandas}} \citep{mckinney_data_2010},
        \href{https://github.com/matplotlib/matplotlib}{\texttt{matplotlib}} \citep{hunter_matplotlib_2007},
        \href{https://github.com/cphyc/matplotlib-label-lines}{\texttt{matplotlib-label-lines}} \citep{cadiou_matplotlib_2022} 
    }

\bibliography{bibliography}{}

@article{powell_new_2019,
	title = {New {Constraints} {From} {Dust} {Lines} {On} {The} {Surface} {Densities} {Of} {Protoplanetary} {Disks}},
	volume = {878},
	issn = {1538-4357},
	url = {http://arxiv.org/abs/1905.03252},
	doi = {10.3847/1538-4357/ab20ce},
	number = {2},
	urldate = {2019-07-03},
	journal = {ApJ},
	author = {Powell, Diana and Murray-Clay, Ruth and Pérez, Laura M. and Schlichting, Hilke E. and Rosenthal, Mickey},
	month = jun,
	year = {2019},
	note = {arXiv: 1905.03252},
	pages = {116},
}

@article{weidenschilling_aerodynamics_1977,
	title = {Aerodynamics of solid bodies in the solar nebula},
	volume = {180},
	issn = {0035-8711},
	url = {https://academic.oup.com/mnras/article/180/2/57/1034183},
	doi = {10.1093/mnras/180.2.57},
	language = {en},
	number = {2},
	urldate = {2019-07-03},
	journal = {Mon Not R Astron Soc},
	author = {Weidenschilling, S. J.},
	month = sep,
	year = {1977},
	pages = {57--70},
}

@article{ros_ice_2013,
	title = {Ice condensation as a planet formation mechanism},
	volume = {552},
	url = {https://ui.adsabs.harvard.edu/abs/2013A%26A...552A.137R/abstract},
	doi = {10.1051/0004-6361/201220536},
	language = {en},
	urldate = {2019-07-03},
	journal = {Astronomy and Astrophysics},
	author = {Ros, K. and Johansen, A.},
	month = apr,
	year = {2013},
	pages = {A137},
}

@article{birnstiel_simple_2012,
	title = {A simple model for the evolution of the dust population in protoplanetary disks},
	volume = {539},
	url = {https://ui.adsabs.harvard.edu/abs/2012A%26A...539A.148B/abstract},
	doi = {10.1051/0004-6361/201118136},
	language = {en},
	urldate = {2019-07-03},
	journal = {Astronomy and Astrophysics},
	author = {Birnstiel, T. and Klahr, H. and Ercolano, B.},
	month = mar,
	year = {2012},
	pages = {A148},
}

@article{boogert_observations_2015,
	title = {Observations of the {Icy} {Universe}},
	volume = {53},
	issn = {0066-4146, 1545-4282},
	url = {http://arxiv.org/abs/1501.05317},
	doi = {10.1146/annurev-astro-082214-122348},
	number = {1},
	urldate = {2022-06-28},
	journal = {Annu. Rev. Astron. Astrophys.},
	author = {Boogert, Adwin and Gerakines, Perry and Whittet, Douglas},
	month = aug,
	year = {2015},
	note = {arXiv:1501.05317 [astro-ph]},
	pages = {541--581},
}

@article{altwegg_cometary_2019,
	title = {Cometary {Chemistry} and the {Origin} of {Icy} {Solar} {System} {Bodies}: {The} {View} {After} \textit{{Rosetta}}},
	volume = {57},
	issn = {0066-4146, 1545-4282},
	shorttitle = {Cometary {Chemistry} and the {Origin} of {Icy} {Solar} {System} {Bodies}},
	url = {https://www.annualreviews.org/doi/10.1146/annurev-astro-091918-104409},
	doi = {10.1146/annurev-astro-091918-104409},
	language = {en},
	number = {1},
	urldate = {2022-06-28},
	journal = {Annu. Rev. Astron. Astrophys.},
	author = {Altwegg, Kathrin and Balsiger, Hans and Fuselier, Stephen A.},
	month = aug,
	year = {2019},
	pages = {113--155},
}

@article{altwegg_evidence_2020,
	title = {Evidence of ammonium salts in comet {67P} as explanation for the nitrogen depletion in cometary comae},
	volume = {4},
	issn = {2397-3366},
	url = {http://www.nature.com/articles/s41550-019-0991-9},
	doi = {10.1038/s41550-019-0991-9},
	language = {en},
	number = {5},
	urldate = {2022-06-28},
	journal = {Nat Astron},
	author = {Altwegg, Kathrin and Balsiger, Hans and Hänni, Nora and Rubin, Martin and Schuhmann, Markus and Schroeder, Isaac and Sémon, Thierry and Wampfler, Susanne and Berthelier, Jean-Jacques and Briois, Christelle and Combi, Mike and Gombosi, Tamas I. and Cottin, Hervé and De Keyser, Johan and Dhooghe, Frederik and Fiethe, Björn and Fuselier, Steven A.},
	month = may,
	year = {2020},
	pages = {533--540},
}

@article{bergner_kinetics_2016,
	title = {Kinetics and mechanisms of the acid-base reaction between {NH}\$\_3\$ and {HCOOH} in interstellar ice analogs},
	volume = {829},
	issn = {1538-4357},
	url = {http://arxiv.org/abs/1608.00010},
	doi = {10.3847/0004-637X/829/2/85},
	number = {2},
	urldate = {2022-06-28},
	journal = {ApJ},
	author = {Bergner, Jennifer B. and Oberg, Karin I. and Rajappan, Mahesh and Fayolle, Edith C.},
	month = sep,
	year = {2016},
	note = {arXiv:1608.00010 [astro-ph, physics:physics]},
	pages = {85},
}

@article{ros_effect_2019,
	title = {Effect of nucleation on icy pebble growth in protoplanetary discs},
	volume = {629},
	issn = {0004-6361},
	url = {https://ui.adsabs.harvard.edu/abs/2019A&A...629A..65R/abstract},
	doi = {10.1051/0004-6361/201834331},
	language = {en},
	urldate = {2022-05-15},
	journal = {Astronomy and Astrophysics},
	author = {Ros, Katrin and Johansen, Anders and Riipinen, Ilona and Schlesinger, Daniel},
	month = sep,
	year = {2019},
	pages = {A65},
}

@article{siess_internet_2000,
	title = {An internet server for pre-main sequence tracks of low- and intermediate-mass stars},
	volume = {358},
	issn = {0004-6361},
	url = {https://ui.adsabs.harvard.edu/abs/2000A&A...358..593S},
	urldate = {2022-05-15},
	journal = {Astronomy and Astrophysics},
	author = {Siess, L. and Dufour, E. and Forestini, M.},
	month = jun,
	year = {2000},
	note = {ADS Bibcode: 2000A\&A...358..593S},
	pages = {593--599},
}

@article{oberg_molecules_2021,
	title = {Molecules with {ALMA} at {Planet}-forming {Scales} ({MAPS}) {I}: {Program} {Overview} and {Highlights}},
	volume = {257},
	issn = {0067-0049, 1538-4365},
	shorttitle = {Molecules with {ALMA} at {Planet}-forming {Scales} ({MAPS}) {I}},
	url = {http://arxiv.org/abs/2109.06268},
	doi = {10.3847/1538-4365/ac1432},
	number = {1},
	urldate = {2022-05-12},
	journal = {ApJS},
	author = {Oberg, Karin I. and Guzman, Viviana V. and Walsh, Catherine and Aikawa, Yuri and Bergin, Edwin A. and Law, Charles J. and Loomis, Ryan A. and Alarcon, Felipe and Andrews, Sean M. and Bae, Jaehan and Bergner, Jennifer B. and Boehler, Yann and Booth, Alice S. and Bosman, Arthur D. and Calahan, Jenny K. and Cataldi, Gianni and Cleeves, L. Ilsedore and Czekala, Ian and Furuya, Kenji and Huang, Jane and Ilee, John D. and Kurtovic, Nicolas T. and Gal, Romane Le and Liu, Yao and Long, Feng and Menard, Francois and Nomura, Hideko and Perez, Laura M. and Qi, Chunhua and Schwarz, Kamber R. and Sierra, Anibal and Teague, Richard and Tsukagoshi, Takashi and Yamato, Yoshihide and Hoff, Merel L. R. van 't and Waggoner, Abygail R. and Wilner, David J. and Zhang, Ke},
	month = nov,
	year = {2021},
	note = {arXiv: 2109.06268},
	pages = {1},
}

@article{andrews_disk_2018,
	title = {The {Disk} {Substructures} at {High} {Angular} {Resolution} {Project} ({DSHARP}): {I}. {Motivation}, {Sample}, {Calibration}, and {Overview}},
	volume = {869},
	issn = {2041-8213},
	shorttitle = {The {Disk} {Substructures} at {High} {Angular} {Resolution} {Project} ({DSHARP})},
	url = {http://arxiv.org/abs/1812.04040},
	doi = {10.3847/2041-8213/aaf741},
	number = {2},
	urldate = {2022-05-12},
	journal = {ApJ},
	author = {Andrews, Sean M. and Huang, Jane and Pérez, Laura M. and Isella, Andrea and Dullemond, Cornelis P. and Kurtovic, Nicolás T. and Guzmán, Viviana V. and Carpenter, John M. and Wilner, David J. and Zhang, Shangjia and Zhu, Zhaohuan and Birnstiel, Tilman and Bai, Xue-Ning and Benisty, Myriam and Hughes, A. Meredith and Öberg, Karin I. and Ricci, Luca},
	month = dec,
	year = {2018},
	note = {arXiv: 1812.04040},
	pages = {L41},
}

@article{birnstiel_gas-_2010,
	title = {Gas- and dust evolution in protoplanetary disks},
	volume = {513},
	issn = {0004-6361, 1432-0746},
	url = {http://www.aanda.org/10.1051/0004-6361/200913731},
	doi = {10.1051/0004-6361/200913731},
	urldate = {2022-04-28},
	journal = {A\&A},
	author = {Birnstiel, T. and Dullemond, C. P. and Brauer, F.},
	month = apr,
	year = {2010},
	pages = {A79},
}

@article{piso_co_2015,
	title = {C/{O} {AND} {SNOWLINE} {LOCATIONS} {IN} {PROTOPLANETARY} {DISKS}: {THE} {EFFECT} {OF} {RADIAL} {DRIFT} {AND} {VISCOUS} {GAS} {ACCRETION}},
	volume = {815},
	issn = {1538-4357},
	shorttitle = {C/{O} {AND} {SNOWLINE} {LOCATIONS} {IN} {PROTOPLANETARY} {DISKS}},
	url = {https://iopscience.iop.org/article/10.1088/0004-637X/815/2/109},
	doi = {10.1088/0004-637X/815/2/109},
	language = {en},
	number = {2},
	urldate = {2022-04-26},
	journal = {ApJ},
	author = {Piso, Ana-Maria A. and Öberg, Karin I. and Birnstiel, Tilman and Murray-Clay, Ruth A.},
	month = dec,
	year = {2015},
	pages = {109},
}

@article{price_ice-coated_2021,
	title = {Ice-coated {Pebble} {Drift} as a {Possible} {Explanation} for {Peculiar} {Cometary} {CO}/{H2O} {Ratios}},
	volume = {913},
	issn = {0004-637X},
	url = {https://ui.adsabs.harvard.edu/abs/2021ApJ...913....9P},
	doi = {10.3847/1538-4357/abf041},
	urldate = {2022-04-25},
	journal = {The Astrophysical Journal},
	author = {Price, Ellen M. and Cleeves, L. Ilsedore and Bodewits, Dennis and Öberg, Karin I.},
	month = may,
	year = {2021},
	note = {ADS Bibcode: 2021ApJ...913....9P},
	pages = {9},
}

@article{nietiadi_collisions_2020,
	title = {Collisions between ice-covered silica grains: {An} atomistic study},
	volume = {352},
	issn = {0019-1035},
	shorttitle = {Collisions between ice-covered silica grains},
	url = {https://ui.adsabs.harvard.edu/abs/2020Icar..35213996N},
	doi = {10.1016/j.icarus.2020.113996},
	urldate = {2022-04-08},
	journal = {Icarus},
	author = {Nietiadi, Maureen L. and Rosandi, Yudi and Urbassek, Herbert M.},
	month = dec,
	year = {2020},
	note = {ADS Bibcode: 2020Icar..35213996N},
	pages = {113996},
}

@article{andrews_observations_2020,
	title = {Observations of {Protoplanetary} {Disk} {Structures}},
	volume = {58},
	url = {https://doi.org/10.1146/annurev-astro-031220-010302},
	doi = {10.1146/annurev-astro-031220-010302},
	number = {1},
	urldate = {2022-03-24},
	journal = {Annual Review of Astronomy and Astrophysics},
	author = {Andrews, Sean M.},
	year = {2020},
	note = {\_eprint: https://doi.org/10.1146/annurev-astro-031220-010302},
	pages = {483--528},
}

@article{shakura_black_1973,
	title = {Black holes in binary systems. {Observational} appearance.},
	volume = {24},
	issn = {0004-6361},
	url = {https://ui.adsabs.harvard.edu/abs/1973A%26A....24..337S/abstract},
	language = {en},
	urldate = {2022-03-24},
	journal = {Astronomy and Astrophysics, Vol. 24, p. 337 - 355},
	author = {Shakura, N. I. and Sunyaev, R. A.},
	year = {1973},
	pages = {337},
}

@article{lodders_solar_2003,
	title = {Solar {System} {Abundances} and {Condensation} {Temperatures} of the {Elements}},
	volume = {591},
	issn = {0004-637X},
	url = {https://ui.adsabs.harvard.edu/abs/2003ApJ...591.1220L},
	doi = {10.1086/375492},
	urldate = {2021-12-15},
	journal = {The Astrophysical Journal},
	author = {Lodders, Katharina},
	month = jul,
	year = {2003},
	note = {ADS Bibcode: 2003ApJ...591.1220L},
	pages = {1220--1247},
}

@article{hollenbach_water_2008,
	title = {{WATER}, {O2}, {AND} {ICE} {IN} {MOLECULAR} {CLOUDS}},
	volume = {690},
	issn = {0004-637X},
	url = {https://doi.org/10.1088/0004-637x/690/2/1497},
	doi = {10.1088/0004-637X/690/2/1497},
	language = {en},
	number = {2},
	urldate = {2021-05-29},
	journal = {ApJ},
	publisher = {American Astronomical Society},
	author = {Hollenbach, David and Kaufman, Michael J. and Bergin, Edwin A. and Melnick, Gary J.},
	month = dec,
	year = {2008},
	pages = {1497--1521},
}

@article{oberg_effects_2011,
	title = {The {Effects} of {Snowlines} on {C}/{O} in {Planetary} {Atmospheres}},
	volume = {743},
	issn = {0004-637X},
	url = {http://adsabs.harvard.edu/abs/2011ApJ...743L..16O},
	doi = {10.1088/2041-8205/743/1/L16},
	urldate = {2021-05-29},
	journal = {The Astrophysical Journal Letters},
	author = {Öberg, Karin I. and Murray-Clay, Ruth and Bergin, Edwin A.},
	month = dec,
	year = {2011},
	pages = {L16},
}

@article{bergner_evolutionary_2020,
	title = {An {Evolutionary} {Study} of {Volatile} {Chemistry} in {Protoplanetary} {Disks}},
	volume = {898},
	url = {http://adsabs.harvard.edu/abs/2020ApJ...898...97B},
	doi = {10.3847/1538-4357/ab9e71},
	urldate = {2020-10-21},
	journal = {The Astrophysical Journal},
	author = {Bergner, Jennifer B. and Öberg, Karin I. and Bergin, Edwin A. and Andrews, Sean M. and Blake, Geoffrey A. and Carpenter, John M. and Cleeves, L. Ilsedore and Guzmán, Viviana V. and Huang, Jane and Jørgensen, Jes K. and Qi, Chunhua and Schwarz, Kamber R. and Williams, Jonathan P. and Wilner, David J.},
	month = aug,
	year = {2020},
	pages = {97},
}

@article{krijt_transport_2018,
	title = {Transport of {CO} in {Protoplanetary} {Disks}: {Consequences} of {Pebble} {Formation}, {Settling}, and {Radial} {Drift}},
	volume = {864},
	issn = {1538-4357},
	shorttitle = {Transport of {CO} in {Protoplanetary} {Disks}},
	url = {http://arxiv.org/abs/1808.01840},
	doi = {10.3847/1538-4357/aad69b},
	number = {1},
	urldate = {2020-09-25},
	journal = {ApJ},
	author = {Krijt, Sebastiaan and Schwarz, Kamber R. and Bergin, Edwin A. and Ciesla, Fred J.},
	month = aug,
	year = {2018},
	note = {arXiv: 1808.01840},
	pages = {78},
}

@article{rozner_aeolian-erosion_2020,
	title = {The aeolian-erosion barrier for the growth of metre-size objects in protoplanetary discs},
	volume = {496},
	url = {http://adsabs.harvard.edu/abs/2020MNRAS.496.4827R},
	doi = {10.1093/mnras/staa1864},
	urldate = {2020-09-25},
	journal = {Monthly Notices of the Royal Astronomical Society},
	author = {Rozner, Mor and Grishin, Evgeni and Perets, Hagai B.},
	month = jun,
	year = {2020},
	pages = {4827--4835},
}

@article{perets_wind-shearing_2011,
	title = {Wind-shearing in gaseous protoplanetary disks and the evolution of binary planetesimals},
	volume = {733},
	issn = {0004-637X, 1538-4357},
	url = {http://arxiv.org/abs/1103.1629},
	doi = {10.1088/0004-637X/733/1/56},
	number = {1},
	urldate = {2020-02-18},
	journal = {ApJ},
	author = {Perets, Hagai B. and Murray-Clay, Ruth A.},
	month = may,
	year = {2011},
	note = {arXiv: 1103.1629},
	pages = {56},
}

@article{hartmann_accretion_1998,
	title = {Accretion and the {Evolution} of {T} {Tauri} {Disks}},
	volume = {495},
	issn = {0004-637X},
	url = {http://adsabs.harvard.edu/abs/1998ApJ...495..385H},
	doi = {10.1086/305277},
	urldate = {2019-08-05},
	journal = {The Astrophysical Journal},
	author = {Hartmann, Lee and Calvet, Nuria and Gullbring, Erik and D'Alessio, Paola},
	month = mar,
	year = {1998},
	pages = {385--400},
}

@article{lynden-bell_evolution_1974,
	title = {The evolution of viscous discs and the origin of the nebular variables.},
	volume = {168},
	issn = {0035-8711},
	url = {http://adsabs.harvard.edu/abs/1974MNRAS.168..603L},
	doi = {10.1093/mnras/168.3.603},
	urldate = {2019-08-05},
	journal = {Monthly Notices of the Royal Astronomical Society},
	author = {Lynden-Bell, D. and Pringle, J. E.},
	month = sep,
	year = {1974},
	pages = {603--637},
}

@article{sirono_elasticity_2021,
	title = {Elasticity of a {Sintered} {Contact} between {Dust} {Grains}},
	volume = {911},
	issn = {0004-637X},
	url = {https://ui.adsabs.harvard.edu/abs/2021ApJ...911..114S},
	doi = {10.3847/1538-4357/abec7c},
	urldate = {2023-06-09},
	journal = {The Astrophysical Journal},
	author = {Sirono, Sin-iti and Kudo, Daiki},
	month = apr,
	year = {2021},
	note = {ADS Bibcode: 2021ApJ...911..114S},
	pages = {114},
}

@article{dohnanyi_collisional_1969,
	title = {Collisional model of asteroids and their debris},
	volume = {74},
	issn = {2156-2202},
	url = {https://onlinelibrary.wiley.com/doi/abs/10.1029/JB074i010p02531},
	doi = {10.1029/JB074i010p02531},
	language = {en},
	number = {10},
	urldate = {2023-05-30},
	journal = {Journal of Geophysical Research (1896-1977)},
	author = {Dohnanyi, J. S.},
	year = {1969},
	note = {\_eprint: https://agupubs.onlinelibrary.wiley.com/doi/pdf/10.1029/JB074i010p02531},
	pages = {2531--2554},
}

@article{tanaka_steady-state_1996,
	title = {Steady-{State} {Size} {Distribution} for the {Self}-{Similar} {Collision} {Cascade}},
	volume = {123},
	issn = {0019-1035},
	url = {https://www.sciencedirect.com/science/article/pii/S0019103596901700},
	doi = {10.1006/icar.1996.0170},
	language = {en},
	number = {2},
	urldate = {2023-05-30},
	journal = {Icarus},
	author = {Tanaka, Hidekazu and Inaba, Satoshi and Nakazawa, Kiyoshi},
	month = oct,
	year = {1996},
	pages = {450--455},
}

@article{cleeves_multiple_2016,
	title = {{MULTIPLE} {CARBON} {MONOXIDE} {SNOW} {LINES} {IN} {DISKS} {SCULPTED} {BY} {RADIAL} {DRIFT}},
	volume = {816},
	issn = {2041-8205},
	url = {https://dx.doi.org/10.3847/2041-8205/816/2/L21},
	doi = {10.3847/2041-8205/816/2/L21},
	language = {en},
	number = {2},
	urldate = {2023-03-07},
	journal = {ApJL},
	publisher = {The American Astronomical Society},
	author = {Cleeves, L. Ilsedore},
	month = jan,
	year = {2016},
	pages = {L21},
}

@article{smith_desorption_2016,
	title = {Desorption {Kinetics} of {Ar}, {Kr}, {Xe}, {N2}, {O2}, {CO}, {Methane}, {Ethane}, and {Propane} from {Graphene} and {Amorphous} {Solid} {Water} {Surfaces}},
	volume = {120},
	issn = {1520-6106},
	url = {https://doi.org/10.1021/acs.jpcb.5b10033},
	doi = {10.1021/acs.jpcb.5b10033},
	number = {8},
	urldate = {2023-02-01},
	journal = {J. Phys. Chem. B},
	publisher = {American Chemical Society},
	author = {Smith, R. Scott and May, R. Alan and Kay, Bruce D.},
	month = mar,
	year = {2016},
	pages = {1979--1987},
}

@article{kruczkiewicz_ammonia_2021,
	title = {Ammonia snow lines and ammonium salts desorption},
	volume = {652},
	issn = {0004-6361, 1432-0746},
	url = {https://www.aanda.org/10.1051/0004-6361/202140579},
	doi = {10.1051/0004-6361/202140579},
	language = {en},
	urldate = {2023-01-26},
	journal = {A\&A},
	author = {Kruczkiewicz, F. and Vitorino, J. and Congiu, E. and Theulé, P. and Dulieu, F.},
	month = aug,
	year = {2021},
	pages = {A29},
}

@article{martin-domenech_thermal_2014,
	title = {Thermal desorption of circumstellar and cometary ice analogs},
	volume = {564},
	issn = {0004-6361, 1432-0746},
	url = {http://www.aanda.org/10.1051/0004-6361/201322824},
	doi = {10.1051/0004-6361/201322824},
	urldate = {2023-01-24},
	journal = {A\&A},
	author = {Martín-Doménech, R. and Muñoz Caro, G. M. and Bueno, J. and Goesmann, F.},
	month = apr,
	year = {2014},
	pages = {A8},
}

@article{minissale_thermal_2022,
	title = {Thermal {Desorption} of {Interstellar} {Ices}: {A} {Review} on the {Controlling} {Parameters} and {Their} {Implications} from {Snowlines} to {Chemical} {Complexity}},
	volume = {6},
	shorttitle = {Thermal {Desorption} of {Interstellar} {Ices}},
	url = {https://ui.adsabs.harvard.edu/abs/2022ESC.....6..597M},
	doi = {10.1021/acsearthspacechem.1c00357},
	urldate = {2022-12-14},
	journal = {ACS Earth and Space Chemistry},
	author = {Minissale, Marco and Aikawa, Yuri and Bergin, Edwin and Bertin, Mathieu and Brown, Wendy A. and Cazaux, Stephanie and Charnley, Steven B. and Coutens, Audrey and Cuppen, Herma M. and Guzman, Victoria and Linnartz, Harold and McCoustra, Martin R. S. and Rimola, Albert and Schrauwen, Johanna G. M. and Toubin, Celine and Ugliengo, Piero and Watanabe, Naoki and Wakelam, Valentine and Dulieu, Francois},
	month = mar,
	year = {2022},
	note = {ADS Bibcode: 2022ESC.....6..597M},
	pages = {597--630},
}

@article{fayolle_n2_2016,
	title = {N2 and {CO} {Desorption} {Energies} from {Water} {Ice}},
	volume = {816},
	issn = {0004-637X},
	url = {https://ui.adsabs.harvard.edu/abs/2016ApJ...816L..28F},
	doi = {10.3847/2041-8205/816/2/L28},
	urldate = {2022-12-14},
	journal = {The Astrophysical Journal},
	author = {Fayolle, Edith C. and Balfe, Jodi and Loomis, Ryan and Bergner, Jennifer and Graninger, Dawn and Rajappan, Mahesh and Öberg, Karin I.},
	month = jan,
	year = {2016},
	note = {ADS Bibcode: 2016ApJ...816L..28F},
	pages = {L28},
}

@article{bohlin_survey_1978,
	title = {A survey of interstellar {H} {I} from {Lalpha} absorption measurements. {II}.},
	volume = {224},
	issn = {0004-637X},
	url = {https://ui.adsabs.harvard.edu/abs/1978ApJ...224..132B},
	doi = {10.1086/156357},
	urldate = {2022-11-22},
	journal = {The Astrophysical Journal},
	author = {Bohlin, R. C. and Savage, B. D. and Drake, J. F.},
	month = aug,
	year = {1978},
	note = {ADS Bibcode: 1978ApJ...224..132B},
	pages = {132--142},
}

@article{garaud_effect_2007,
	title = {The {Effect} of {Internal} {Dissipation} and {Surface} {Irradiation} on the {Structure} of {Disks} and the {Location} of the {Snow} {Line} around {Sun}-like {Stars}},
	volume = {654},
	issn = {0004-637X},
	url = {https://ui.adsabs.harvard.edu/abs/2007ApJ...654..606G},
	doi = {10.1086/509041},
	urldate = {2022-11-17},
	journal = {The Astrophysical Journal},
	author = {Garaud, P. and Lin, D. N. C.},
	month = jan,
	year = {2007},
	note = {ADS Bibcode: 2007ApJ...654..606G},
	pages = {606--624},
}

@article{rosenthal_how_2020,
	title = {How {Flow} {Isolation} {May} {Set} the {Mass} {Scale} for {Super}-{Earth} {Planets}},
	volume = {898},
	issn = {0004-637X},
	url = {https://ui.adsabs.harvard.edu/abs/2020ApJ...898..108R},
	doi = {10.3847/1538-4357/ab9eb2},
	urldate = {2022-11-17},
	journal = {The Astrophysical Journal},
	author = {Rosenthal, M. M. and Murray-Clay, R. A.},
	month = aug,
	year = {2020},
	note = {ADS Bibcode: 2020ApJ...898..108R},
	pages = {108},
}

@article{powell_depletion_2022,
	title = {Depletion of gaseous {CO} in protoplanetary disks by surface-energy-regulated ice formation},
	issn = {2397-3366},
	url = {https://ui.adsabs.harvard.edu/abs/2022NatAs.tmp..184P},
	doi = {10.1038/s41550-022-01741-9},
	urldate = {2022-10-05},
	journal = {Nature Astronomy},
	author = {Powell, Diana and Gao, Peter and Murray-Clay, Ruth and Zhang, Xi},
	month = aug,
	year = {2022},
	note = {ADS Bibcode: 2022NatAs.tmp..184P},
}

@article{hartmann_accretion_2016,
	title = {Accretion onto {Pre}-{Main}-{Sequence} {Stars}},
	volume = {54},
	issn = {0066-4146},
	url = {https://ui.adsabs.harvard.edu/abs/2016ARA&A..54..135H},
	doi = {10.1146/annurev-astro-081915-023347},
	urldate = {2024-02-17},
	journal = {Annual Review of Astronomy and Astrophysics},
	author = {Hartmann, Lee and Herczeg, Gregory and Calvet, Nuria},
	month = sep,
	year = {2016},
	note = {ADS Bibcode: 2016ARA\&A..54..135H},
	pages = {135--180},
}

@article{yunerman_pathway_2024,
	title = {A {Pathway} for {Collisional} {Planetesimal} {Growth} in the {Ice}-dominant {Regions} of {Protoplanetary} {Disks}},
	volume = {961},
	issn = {0004-637X},
	url = {https://ui.adsabs.harvard.edu/abs/2024ApJ...961...33Y},
	doi = {10.3847/1538-4357/ad05b9},
	urldate = {2024-02-25},
	journal = {The Astrophysical Journal},
	author = {Yunerman, Elizabeth and Powell, Diana and Murray-Clay, Ruth},
	month = jan,
	year = {2024},
	note = {ADS Bibcode: 2024ApJ...961...33Y},
	pages = {33},
}

@article{asplund_chemical_2009,
	title = {The {Chemical} {Composition} of the {Sun}},
	volume = {47},
	issn = {0066-4146},
	url = {https://ui.adsabs.harvard.edu/abs/2009ARA&A..47..481A},
	doi = {10.1146/annurev.astro.46.060407.145222},
	urldate = {2024-02-29},
	journal = {Annual Review of Astronomy and Astrophysics},
	author = {Asplund, Martin and Grevesse, Nicolas and Sauval, A. Jacques and Scott, Pat},
	month = sep,
	year = {2009},
	note = {ADS Bibcode: 2009ARA\&A..47..481A},
	pages = {481--522},
}

@article{fischer_accretion_2023,
	title = {Accretion {Variability} as a {Guide} to {Stellar} {Mass} {Assembly}},
	volume = {534},
	url = {https://ui.adsabs.harvard.edu/abs/2023ASPC..534..355F},
	doi = {10.48550/arXiv.2203.11257},
	urldate = {2024-03-31},
	author = {Fischer, W. J. and Hillenbrand, L. A. and Herczeg, G. J. and Johnstone, D. and Kospal, A. and Dunham, M. M.},
	month = jul,
	year = {2023},
	note = {Conference Name: Protostars and Planets VII
Place: eprint: arXiv:2203.11257
ADS Bibcode: 2023ASPC..534..355F},
	pages = {355},
}

@article{lee_young_2020,
	title = {Young {Faithful}: {The} {Eruptions} of {EC} 53 as {It} {Cycles} through {Filling} and {Draining} the {Inner} {Disk}},
	volume = {903},
	issn = {0004-637X},
	shorttitle = {Young {Faithful}},
	url = {https://ui.adsabs.harvard.edu/abs/2020ApJ...903....5L},
	doi = {10.3847/1538-4357/abb6fe},
	urldate = {2024-03-31},
	journal = {The Astrophysical Journal},
	author = {Lee, Yong-Hee and Johnstone, Doug and Lee, Jeong-Eun and Herczeg, Gregory and Mairs, Steve and Varricatt, Watson and Hodapp, Klaus W. and Naylor, Tim and Peña, Carlos Contreras and Baek, Giseon and Haas, Martin and Chini, Rolf and {JCMT Transient Team}},
	month = nov,
	year = {2020},
	note = {ADS Bibcode: 2020ApJ...903....5L},
	pages = {5},
}

@article{fischer_herschel_2017,
	title = {The {Herschel} {Orion} {Protostar} {Survey}: {Luminosity} and {Envelope} {Evolution}},
	volume = {840},
	issn = {0004-637X},
	shorttitle = {The {Herschel} {Orion} {Protostar} {Survey}},
	url = {https://ui.adsabs.harvard.edu/abs/2017ApJ...840...69F},
	doi = {10.3847/1538-4357/aa6d69},
	urldate = {2024-03-31},
	journal = {The Astrophysical Journal},
	author = {Fischer, William J. and Megeath, S. Thomas and Furlan, Elise and Ali, Babar and Stutz, Amelia M. and Tobin, John J. and Osorio, Mayra and Stanke, Thomas and Manoj, P. and Poteet, Charles A. and Booker, Joseph J. and Hartmann, Lee and Wilson, Thomas L. and Myers, Philip C. and Watson, Dan M.},
	month = may,
	year = {2017},
	note = {ADS Bibcode: 2017ApJ...840...69F},
	pages = {69},
}

@article{dotter_mesa_2016,
	title = {{MESA} {Isochrones} and {Stellar} {Tracks} ({MIST}) 0: {Methods} for the {Construction} of {Stellar} {Isochrones}},
	volume = {222},
	issn = {0067-0049},
	shorttitle = {{MESA} {Isochrones} and {Stellar} {Tracks} ({MIST}) 0},
	url = {https://ui.adsabs.harvard.edu/abs/2016ApJS..222....8D},
	doi = {10.3847/0067-0049/222/1/8},
	urldate = {2024-03-31},
	journal = {The Astrophysical Journal Supplement Series},
	author = {Dotter, Aaron},
	month = jan,
	year = {2016},
	note = {ADS Bibcode: 2016ApJS..222....8D},
	pages = {8},
}

@article{choi_mesa_2016,
	title = {Mesa {Isochrones} and {Stellar} {Tracks} ({MIST}). {I}. {Solar}-scaled {Models}},
	volume = {823},
	issn = {0004-637X},
	url = {https://ui.adsabs.harvard.edu/abs/2016ApJ...823..102C},
	doi = {10.3847/0004-637X/823/2/102},
	urldate = {2024-03-31},
	journal = {The Astrophysical Journal},
	author = {Choi, Jieun and Dotter, Aaron and Conroy, Charlie and Cantiello, Matteo and Paxton, Bill and Johnson, Benjamin D.},
	month = jun,
	year = {2016},
	note = {ADS Bibcode: 2016ApJ...823..102C},
	pages = {102},
}

@article{paxton_modules_2011,
	title = {Modules for {Experiments} in {Stellar} {Astrophysics} ({MESA})},
	volume = {192},
	issn = {0067-0049},
	url = {https://ui.adsabs.harvard.edu/abs/2011ApJS..192....3P},
	doi = {10.1088/0067-0049/192/1/3},
	urldate = {2024-03-31},
	journal = {The Astrophysical Journal Supplement Series},
	author = {Paxton, Bill and Bildsten, Lars and Dotter, Aaron and Herwig, Falk and Lesaffre, Pierre and Timmes, Frank},
	month = jan,
	year = {2011},
	note = {ADS Bibcode: 2011ApJS..192....3P},
	pages = {3},
}

@article{paxton_modules_2013,
	title = {Modules for {Experiments} in {Stellar} {Astrophysics} ({MESA}): {Planets}, {Oscillations}, {Rotation}, and {Massive} {Stars}},
	volume = {208},
	issn = {0067-0049},
	shorttitle = {Modules for {Experiments} in {Stellar} {Astrophysics} ({MESA})},
	url = {https://ui.adsabs.harvard.edu/abs/2013ApJS..208....4P},
	doi = {10.1088/0067-0049/208/1/4},
	urldate = {2024-03-31},
	journal = {The Astrophysical Journal Supplement Series},
	author = {Paxton, Bill and Cantiello, Matteo and Arras, Phil and Bildsten, Lars and Brown, Edward F. and Dotter, Aaron and Mankovich, Christopher and Montgomery, M. H. and Stello, Dennis and Timmes, F. X. and Townsend, Richard},
	month = sep,
	year = {2013},
	note = {ADS Bibcode: 2013ApJS..208....4P},
	pages = {4},
}

@article{paxton_modules_2015,
	title = {Modules for {Experiments} in {Stellar} {Astrophysics} ({MESA}): {Binaries}, {Pulsations}, and {Explosions}},
	volume = {220},
	issn = {0067-0049},
	shorttitle = {Modules for {Experiments} in {Stellar} {Astrophysics} ({MESA})},
	url = {https://ui.adsabs.harvard.edu/abs/2015ApJS..220...15P},
	doi = {10.1088/0067-0049/220/1/15},
	urldate = {2024-03-31},
	journal = {The Astrophysical Journal Supplement Series},
	author = {Paxton, Bill and Marchant, Pablo and Schwab, Josiah and Bauer, Evan B. and Bildsten, Lars and Cantiello, Matteo and Dessart, Luc and Farmer, R. and Hu, H. and Langer, N. and Townsend, R. H. D. and Townsley, Dean M. and Timmes, F. X.},
	month = sep,
	year = {2015},
	note = {ADS Bibcode: 2015ApJS..220...15P},
	pages = {15},
}

@article{testi_protoplanetary_2022,
	title = {The protoplanetary disk population in the ρ-{Ophiuchi} region {L1688} and the time evolution of {Class} {II} {YSOs}},
	volume = {663},
	issn = {0004-6361},
	url = {https://ui.adsabs.harvard.edu/abs/2022A&A...663A..98T},
	doi = {10.1051/0004-6361/202141380},
	urldate = {2024-03-31},
	journal = {Astronomy and Astrophysics},
	author = {Testi, L. and Natta, A. and Manara, C. F. and de Gregorio Monsalvo, I. and Lodato, G. and Lopez, C. and Muzic, K. and Pascucci, I. and Sanchis, E. and Miranda, A. Santamaria and Scholz, A. and De Simone, M. and Williams, J. P.},
	month = jul,
	year = {2022},
	note = {ADS Bibcode: 2022A\&A...663A..98T},
	pages = {A98},
}

@article{piso_minimum_2015,
	title = {Minimum {Core} {Masses} for {Giant} {Planet} {Formation} with {Realistic} {Equations} of {State} and {Opacities}},
	volume = {800},
	issn = {0004-637X},
	url = {https://ui.adsabs.harvard.edu/abs/2015ApJ...800...82P},
	doi = {10.1088/0004-637X/800/2/82},
	urldate = {2024-03-31},
	journal = {The Astrophysical Journal},
	author = {Piso, Ana-Maria A. and Youdin, Andrew N. and Murray-Clay, Ruth A.},
	month = feb,
	year = {2015},
	note = {ADS Bibcode: 2015ApJ...800...82P},
	pages = {82},
}

@article{grant_dotm_2023,
	title = {The {\textbackslash}dot\{{M}\} -{M} disk {Relationship} for {Herbig} {Ae}/{Be} {Stars}: {A} {Lifetime} {Problem} for {Disks} with {Low} {Masses}?},
	volume = {166},
	issn = {0004-6256},
	shorttitle = {The {\textbackslash}dot\{{M}\} -{M} disk {Relationship} for {Herbig} {Ae}/{Be} {Stars}},
	url = {https://ui.adsabs.harvard.edu/abs/2023AJ....166..147G},
	doi = {10.3847/1538-3881/acf128},
	urldate = {2024-03-31},
	journal = {The Astronomical Journal},
	author = {Grant, Sierra L. and Stapper, Lucas M. and Hogerheijde, Michiel R. and van Dishoeck, Ewine F. and Brittain, Sean and Vioque, Miguel},
	month = oct,
	year = {2023},
	note = {ADS Bibcode: 2023AJ....166..147G},
	pages = {147},
}

@article{oberg_protoplanetary_2023,
	title = {Protoplanetary {Disk} {Chemistry}},
	volume = {61},
	issn = {0066-4146},
	url = {https://ui.adsabs.harvard.edu/abs/2023ARA&A..61..287O},
	doi = {10.1146/annurev-astro-022823-040820},
	urldate = {2024-03-31},
	journal = {Annual Review of Astronomy and Astrophysics},
	author = {Öberg, Karin I. and Facchini, Stefano and Anderson, Dana E.},
	month = aug,
	year = {2023},
	note = {ADS Bibcode: 2023ARA\&A..61..287O},
	pages = {287--328},
}

@article{mumma_chemical_2011,
	title = {The {Chemical} {Composition} of {Comets}—{Emerging} {Taxonomies} and {Natal} {Heritage}},
	volume = {49},
	issn = {0066-4146},
	url = {https://ui.adsabs.harvard.edu/abs/2011ARA&A..49..471M},
	doi = {10.1146/annurev-astro-081309-130811},
	urldate = {2024-03-31},
	journal = {Annual Review of Astronomy and Astrophysics},
	author = {Mumma, Michael J. and Charnley, Steven B.},
	month = sep,
	year = {2011},
	note = {ADS Bibcode: 2011ARA\&A..49..471M},
	pages = {471--524},
}

@article{flaherty_turbulence_2018,
	title = {Turbulence in the {TW} {Hya} {Disk}},
	volume = {856},
	issn = {0004-637X},
	url = {https://ui.adsabs.harvard.edu/abs/2018ApJ...856..117F},
	doi = {10.3847/1538-4357/aab615},
	urldate = {2024-03-31},
	journal = {The Astrophysical Journal},
	author = {Flaherty, Kevin M. and Hughes, A. Meredith and Teague, Richard and Simon, Jacob B. and Andrews, Sean M. and Wilner, David J.},
	month = apr,
	year = {2018},
	note = {ADS Bibcode: 2018ApJ...856..117F},
	pages = {117},
}

@article{rosotti_empirical_2023,
	title = {Empirical constraints on turbulence in proto-planetary discs},
	volume = {96},
	issn = {1387-6473},
	url = {https://ui.adsabs.harvard.edu/abs/2023NewAR..9601674R},
	doi = {10.1016/j.newar.2023.101674},
	urldate = {2024-03-31},
	journal = {New Astronomy Reviews},
	author = {Rosotti, Giovanni P.},
	month = jun,
	year = {2023},
	note = {ADS Bibcode: 2023NewAR..9601674R},
	pages = {101674},
}

@article{jiang_grain-size_2024,
	title = {Grain-size measurements in protoplanetary disks indicate fragile pebbles and low turbulence},
	volume = {682},
	issn = {0004-6361},
	url = {https://ui.adsabs.harvard.edu/abs/2024A&A...682A..32J},
	doi = {10.1051/0004-6361/202348271},
	urldate = {2024-03-31},
	journal = {Astronomy and Astrophysics},
	author = {Jiang, Haochang and Macías, Enrique and Guerra-Alvarado, Osmar M. and Carrasco-González, Carlos},
	month = feb,
	year = {2024},
	note = {ADS Bibcode: 2024A\&A...682A..32J},
	pages = {A32},
}

@article{bergner_acceleration_2023,
	title = {Acceleration of {1I}/`{Oumuamua} from radiolytically produced {H2} in {H2O} ice},
	volume = {615},
	issn = {0028-0836},
	url = {https://ui.adsabs.harvard.edu/abs/2023Natur.615..610B},
	doi = {10.1038/s41586-022-05687-w},
	urldate = {2024-04-26},
	journal = {Nature},
	author = {Bergner, Jennifer B. and Seligman, Darryl Z.},
	month = mar,
	year = {2023},
	note = {ADS Bibcode: 2023Natur.615..610B},
	pages = {610--613},
}

@article{meijerink_radiative_2009,
	title = {Radiative {Transfer} {Models} of {Mid}-{Infrared} {H2O} {Lines} in the {Planet}-{Forming} {Region} of {Circumstellar} {Disks}},
	volume = {704},
	issn = {0004-637X},
	url = {https://ui.adsabs.harvard.edu/abs/2009ApJ...704.1471M},
	doi = {10.1088/0004-637X/704/2/1471},
	urldate = {2024-04-28},
	journal = {The Astrophysical Journal},
	publisher = {IOP},
	author = {Meijerink, R. and Pontoppidan, K. M. and Blake, G. A. and Poelman, D. R. and Dullemond, C. P.},
	month = oct,
	year = {2009},
	note = {ADS Bibcode: 2009ApJ...704.1471M},
	pages = {1471--1481},
}

@article{schneider_how_2021,
	title = {How drifting and evaporating pebbles shape giant planets. {II}. {Volatiles} and refractories in atmospheres},
	volume = {654},
	issn = {0004-6361},
	url = {https://ui.adsabs.harvard.edu/abs/2021A&A...654A..72S},
	doi = {10.1051/0004-6361/202141096},
	urldate = {2024-05-01},
	journal = {Astronomy and Astrophysics},
	author = {Schneider, Aaron David and Bitsch, Bertram},
	month = oct,
	year = {2021},
	note = {ADS Bibcode: 2021A\&A...654A..72S},
	pages = {A72},
}

@article{banzatti_jwst_2023,
	title = {{JWST} {Reveals} {Excess} {Cool} {Water} near the {Snow} {Line} in {Compact} {Disks}, {Consistent} with {Pebble} {Drift}},
	volume = {957},
	issn = {0004-637X},
	url = {https://ui.adsabs.harvard.edu/abs/2023ApJ...957L..22B},
	doi = {10.3847/2041-8213/acf5ec},
	urldate = {2024-05-01},
	journal = {The Astrophysical Journal},
	publisher = {IOP},
	author = {Banzatti, Andrea and Pontoppidan, Klaus M. and Carr, John S. and Jellison, Evan and Pascucci, Ilaria and Najita, Joan R. and Muñoz-Romero, Carlos E. and Öberg, Karin I. and Kalyaan, Anusha and Pinilla, Paola and Krijt, Sebastiaan and Long, Feng and Lambrechts, Michiel and Rosotti, Giovanni and Herczeg, Gregory J. and Salyk, Colette and Zhang, Ke and Bergin, Edwin A. and Ballering, Nicholas P. and Meyer, Michael R. and Bruderer, Simon and {Jdiscs Collaboration}},
	month = nov,
	year = {2023},
	note = {ADS Bibcode: 2023ApJ...957L..22B},
	pages = {L22},
}

@article{oberg_excess_2016,
	title = {Excess {C}/{O} and {C}/{H} in {Outer} {Protoplanetary} {Disk} {Gas}},
	volume = {831},
	issn = {0004-637X},
	url = {https://ui.adsabs.harvard.edu/abs/2016ApJ...831L..19O},
	doi = {10.3847/2041-8205/831/2/L19},
	urldate = {2024-05-01},
	journal = {The Astrophysical Journal},
	publisher = {IOP},
	author = {Öberg, Karin I. and Bergin, Edwin A.},
	month = nov,
	year = {2016},
	note = {ADS Bibcode: 2016ApJ...831L..19O},
	pages = {L19},
}

@article{booth_chemical_2017,
	title = {Chemical enrichment of giant planets and discs due to pebble drift},
	volume = {469},
	issn = {0035-8711},
	url = {https://ui.adsabs.harvard.edu/abs/2017MNRAS.469.3994B},
	doi = {10.1093/mnras/stx1103},
	urldate = {2024-05-01},
	journal = {Monthly Notices of the Royal Astronomical Society},
	publisher = {OUP},
	author = {Booth, Richard A. and Clarke, Cathie J. and Madhusudhan, Nikku and Ilee, John D.},
	month = aug,
	year = {2017},
	note = {ADS Bibcode: 2017MNRAS.469.3994B},
	pages = {3994--4011},
}

@article{krijt_co_2020,
	title = {{CO} {Depletion} in {Protoplanetary} {Disks}: {A} {Unified} {Picture} {Combining} {Physical} {Sequestration} and {Chemical} {Processing}},
	volume = {899},
	issn = {0004-637X},
	shorttitle = {{CO} {Depletion} in {Protoplanetary} {Disks}},
	url = {https://ui.adsabs.harvard.edu/abs/2020ApJ...899..134K},
	doi = {10.3847/1538-4357/aba75d},
	urldate = {2024-05-01},
	journal = {The Astrophysical Journal},
	publisher = {IOP},
	author = {Krijt, Sebastiaan and Bosman, Arthur D. and Zhang, Ke and Schwarz, Kamber R. and Ciesla, Fred J. and Bergin, Edwin A.},
	month = aug,
	year = {2020},
	note = {ADS Bibcode: 2020ApJ...899..134K},
	pages = {134},
}

@article{notsu_composition_2020,
	title = {The composition of hot {Jupiter} atmospheres assembled within chemically evolved protoplanetary discs},
	volume = {499},
	issn = {0035-8711},
	url = {https://ui.adsabs.harvard.edu/abs/2020MNRAS.499.2229N},
	doi = {10.1093/mnras/staa2944},
	urldate = {2024-05-01},
	journal = {Monthly Notices of the Royal Astronomical Society},
	publisher = {OUP},
	author = {Notsu, Shota and Eistrup, Christian and Walsh, Catherine and Nomura, Hideko},
	month = dec,
	year = {2020},
	note = {ADS Bibcode: 2020MNRAS.499.2229N},
	pages = {2229--2244},
}

@article{jewitt_interstellar_2023,
	title = {The {Interstellar} {Interlopers}},
	volume = {61},
	issn = {0066-4146},
	url = {https://ui.adsabs.harvard.edu/abs/2023ARA&A..61..197J},
	doi = {10.1146/annurev-astro-071221-054221},
	urldate = {2024-05-01},
	journal = {Annual Review of Astronomy and Astrophysics},
	author = {Jewitt, David and Seligman, Darryl Z.},
	month = aug,
	year = {2023},
	note = {ADS Bibcode: 2023ARA\&A..61..197J},
	pages = {197--236},
}

@article{seligman_volatile_2022,
	title = {The {Volatile} {Carbon}-to-oxygen {Ratio} as a {Tracer} for the {Formation} {Locations} of {Interstellar} {Comets}},
	volume = {3},
	issn = {2632-3338},
	url = {https://ui.adsabs.harvard.edu/abs/2022PSJ.....3..150S},
	doi = {10.3847/PSJ/ac75b5},
	urldate = {2024-05-01},
	journal = {The Planetary Science Journal},
	publisher = {IOP},
	author = {Seligman, Darryl Z. and Rogers, Leslie A. and Cabot, Samuel H. C. and Noonan, John W. and Kareta, Theodore and Mandt, Kathleen E. and Ciesla, Fred and McKay, Adam and Feinstein, Adina D. and Levine, W. Garrett and Bean, Jacob L. and Nordlander, Thomas and Krumholz, Mark R. and Mansfield, Megan and Hoover, Devin J. and Van Clepper, Eric},
	month = jul,
	year = {2022},
	note = {ADS Bibcode: 2022PSJ.....3..150S},
	pages = {150},
}

@article{bodewits_carbon_2020,
	title = {The carbon monoxide-rich interstellar comet {2I}/{Borisov}},
	volume = {4},
	copyright = {2020 The Author(s), under exclusive licence to Springer Nature Limited},
	issn = {2397-3366},
	url = {https://www.nature.com/articles/s41550-020-1095-2},
	doi = {10.1038/s41550-020-1095-2},
	language = {en},
	number = {9},
	urldate = {2024-05-02},
	journal = {Nat Astron},
	publisher = {Nature Publishing Group},
	author = {Bodewits, D. and Noonan, J. W. and Feldman, P. D. and Bannister, M. T. and Farnocchia, D. and Harris, W. M. and Li, J.-Y. and Mandt, K. E. and Parker, J. Wm and Xing, Z.-X.},
	month = sep,
	year = {2020},
	pages = {867--871},
}

@article{cordiner_unusually_2020,
	title = {Unusually high {CO} abundance of the first active interstellar comet},
	volume = {4},
	copyright = {2020 The Author(s), under exclusive licence to Springer Nature Limited},
	issn = {2397-3366},
	url = {https://www.nature.com/articles/s41550-020-1087-2},
	doi = {10.1038/s41550-020-1087-2},
	language = {en},
	number = {9},
	urldate = {2024-05-02},
	journal = {Nat Astron},
	publisher = {Nature Publishing Group},
	author = {Cordiner, M. A. and Milam, S. N. and Biver, N. and Bockelée-Morvan, D. and Roth, N. X. and Bergin, E. A. and Jehin, E. and Remijan, A. J. and Charnley, S. B. and Mumma, M. J. and Boissier, J. and Crovisier, J. and Paganini, L. and Kuan, Y.-J. and Lis, D. C.},
	month = sep,
	year = {2020},
	pages = {861--866},
}

@article{biver_extraordinary_2018,
	title = {The extraordinary composition of the blue comet {C}/2016 {R2} ({PanSTARRS})},
	volume = {619},
	copyright = {© ESO 2018},
	issn = {0004-6361, 1432-0746},
	url = {https://www.aanda.org/articles/aa/abs/2018/11/aa33449-18/aa33449-18.html},
	doi = {10.1051/0004-6361/201833449},
	language = {en},
	urldate = {2024-05-02},
	journal = {A\&A},
	publisher = {EDP Sciences},
	author = {Biver, N. and Bockelée-Morvan, D. and Paubert, G. and Moreno, R. and Crovisier, J. and Boissier, J. and Bertrand, E. and Boussier, H. and Kugel, F. and McKay, A. and Russo, N. Dello and DiSanti, M. A.},
	month = nov,
	year = {2018},
	pages = {A127},
}

@article{mckay_peculiar_2019,
	title = {The {Peculiar} {Volatile} {Composition} of {CO}-dominated {Comet} {C}/2016 {R2} ({PanSTARRS})},
	volume = {158},
	issn = {1538-3881},
	url = {https://dx.doi.org/10.3847/1538-3881/ab32e4},
	doi = {10.3847/1538-3881/ab32e4},
	language = {en},
	number = {3},
	urldate = {2024-05-02},
	journal = {AJ},
	publisher = {The American Astronomical Society},
	author = {McKay, Adam J. and DiSanti, Michael A. and Kelley, Michael S. P. and Knight, Matthew M. and Womack, Maria and Wierzchos, Kacper and Pinto, Olga Harrington and Bonev, Boncho and Villanueva, Geronimo L. and Russo, Neil Dello and Cochran, Anita L. and Biver, Nicolas and Bauer, James and Ronald J. Vervack, Jr. and Gibb, Erika and Roth, Nathan and Kawakita, Hideyo},
	month = aug,
	year = {2019},
	pages = {128},
}

@article{pinto_survey_2022,
	title = {A {Survey} of {CO}, {CO2}, and {H2O} in {Comets} and {Centaurs}},
	volume = {3},
	issn = {2632-3338},
	url = {https://iopscience.iop.org/article/10.3847/PSJ/ac960d/meta},
	doi = {10.3847/PSJ/ac960d},
	language = {en},
	number = {11},
	urldate = {2024-05-02},
	journal = {Planet. Sci. J.},
	publisher = {IOP Publishing},
	author = {Pinto, Olga Harrington and Womack, Maria and Fernandez, Yanga and Bauer, James},
	month = nov,
	year = {2022},
	pages = {247},
}

@article{simon_entrapment_2023,
	title = {Entrapment of {Hypervolatiles} in {Interstellar} and {Cometary} {H2O} and {CO2} {Ice} {Analogs}},
	volume = {955},
	issn = {0004-637X},
	url = {https://ui.adsabs.harvard.edu/abs/2023ApJ...955....5S},
	doi = {10.3847/1538-4357/aceaf8},
	urldate = {2024-05-04},
	journal = {The Astrophysical Journal},
	publisher = {IOP},
	author = {Simon, Alexia and Rajappan, Mahesh and Öberg, Karin I.},
	month = sep,
	year = {2023},
	note = {ADS Bibcode: 2023ApJ...955....5S},
	pages = {5},
}

@article{ahearn_cometary_2012,
	title = {{COMETARY} {VOLATILES} {AND} {THE} {ORIGIN} {OF} {COMETS}},
	volume = {758},
	issn = {0004-637X},
	url = {https://dx.doi.org/10.1088/0004-637X/758/1/29},
	doi = {10.1088/0004-637X/758/1/29},
	language = {en},
	number = {1},
	urldate = {2024-05-06},
	journal = {ApJ},
	publisher = {The American Astronomical Society},
	author = {A'Hearn, Michael F. and Feaga, Lori M. and Keller, H. Uwe and Kawakita, Hideyo and Hampton, Donald L. and Kissel, Jochen and Klaasen, Kenneth P. and McFadden, Lucy A. and Meech, Karen J. and Schultz, Peter H. and Sunshine, Jessica M. and Thomas, Peter C. and Veverka, Joseph and Yeomans, Donald K. and Besse, Sebastien and Bodewits, Dennis and Farnham, Tony L. and Groussin, Olivier and Kelley, Michael S. and Lisse, Carey M. and Merlin, Frederic and Protopapa, Silvia and Wellnitz, Dennis D.},
	month = sep,
	year = {2012},
	pages = {29},
}

@article{hahn_orbital_1999,
	title = {Orbital {Evolution} of {Planets} {Embedded} in a {Planetesimal} {Disk}},
	volume = {117},
	issn = {1538-3881},
	url = {https://iopscience.iop.org/article/10.1086/300891/meta},
	doi = {10.1086/300891},
	language = {en},
	number = {6},
	urldate = {2024-05-06},
	journal = {AJ},
	publisher = {IOP Publishing},
	author = {Hahn, Joseph M. and Malhotra, Renu},
	month = jun,
	year = {1999},
	pages = {3041},
}

@article{eistrup_cometary_2019,
	title = {Cometary compositions compared with protoplanetary disk midplane chemical evolution. {An} emerging chemical evolution taxonomy for comets},
	volume = {629},
	issn = {0004-6361},
	url = {https://ui.adsabs.harvard.edu/abs/2019A&A...629A..84E},
	doi = {10.1051/0004-6361/201935812},
	urldate = {2024-05-06},
	journal = {Astronomy and Astrophysics},
	author = {Eistrup, Christian and Walsh, Catherine and van Dishoeck, Ewine F.},
	month = sep,
	year = {2019},
	note = {ADS Bibcode: 2019A\&A...629A..84E},
	pages = {A84},
}

@article{bockelee-morvan_composition_2017,
	title = {The composition of cometary ices},
	volume = {375},
	issn = {1364-503X0080-46140962-8436},
	url = {https://ui.adsabs.harvard.edu/abs/2017RSPTA.37560252B},
	doi = {10.1098/rsta.2016.0252},
	urldate = {2024-05-06},
	journal = {Philosophical Transactions of the Royal Society of London Series A},
	author = {Bockelée-Morvan, D. and Biver, N.},
	month = may,
	year = {2017},
	note = {ADS Bibcode: 2017RSPTA.37560252B},
	pages = {20160252},
}

@article{chiang_spectral_1997,
	title = {Spectral {Energy} {Distributions} of {T} {Tauri} {Stars} with {Passive} {Circumstellar} {Disks}},
	volume = {490},
	issn = {0004-637X},
	url = {https://ui.adsabs.harvard.edu/abs/1997ApJ...490..368C},
	doi = {10.1086/304869},
	urldate = {2024-06-13},
	journal = {The Astrophysical Journal},
	publisher = {IOP},
	author = {Chiang, E. I. and Goldreich, P.},
	month = nov,
	year = {1997},
	note = {ADS Bibcode: 1997ApJ...490..368C},
	pages = {368--376},
}

@article{cuzzi_material_2004,
	title = {Material {Enhancement} in {Protoplanetary} {Nebulae} by {Particle} {Drift} through {Evaporation} {Fronts}},
	volume = {614},
	issn = {0004-637X},
	url = {https://ui.adsabs.harvard.edu/abs/2004ApJ...614..490C},
	doi = {10.1086/423611},
	urldate = {2024-07-17},
	journal = {The Astrophysical Journal},
	publisher = {IOP},
	author = {Cuzzi, Jeffrey N. and Zahnle, Kevin J.},
	month = oct,
	year = {2004},
	note = {ADS Bibcode: 2004ApJ...614..490C},
	pages = {490--496},
}

@article{walsh_chemical_2012,
	title = {Chemical {Processes} in {Protoplanetary} {Disks}. {II}. {On} the {Importance} of {Photochemistry} and {X}-{Ray} {Ionization}},
	volume = {747},
	issn = {0004-637X},
	url = {https://ui.adsabs.harvard.edu/abs/2012ApJ...747..114W},
	doi = {10.1088/0004-637X/747/2/114},
	urldate = {2024-10-28},
	journal = {The Astrophysical Journal},
	publisher = {IOP},
	author = {Walsh, Catherine and Nomura, Hideko and Millar, T. J. and Aikawa, Yuri},
	month = mar,
	year = {2012},
	note = {ADS Bibcode: 2012ApJ...747..114W},
	pages = {114},
}

@article{bergner_ice_2021,
	title = {Ice {Inheritance} in {Dynamical} {Disk} {Models}},
	volume = {919},
	issn = {0004-637X},
	url = {https://ui.adsabs.harvard.edu/abs/2021ApJ...919...45B},
	doi = {10.3847/1538-4357/ac0fd7},
	urldate = {2024-11-01},
	journal = {The Astrophysical Journal},
	publisher = {IOP},
	author = {Bergner, Jennifer B. and Ciesla, Fred},
	month = sep,
	year = {2021},
	note = {ADS Bibcode: 2021ApJ...919...45B},
	pages = {45},
}

@article{sturm_jwst_2023,
	title = {A {JWST} inventory of protoplanetary disk ices. {The} edge-on protoplanetary disk {HH} 48 {NE}, seen with the {Ice} {Age} {ERS} program},
	volume = {679},
	issn = {0004-6361},
	url = {https://ui.adsabs.harvard.edu/abs/2023A&A...679A.138S},
	doi = {10.1051/0004-6361/202347512},
	urldate = {2024-11-03},
	journal = {Astronomy and Astrophysics},
	author = {Sturm, J. A. and McClure, M. K. and Beck, T. L. and Harsono, D. and Bergner, J. B. and Dartois, E. and Boogert, A. C. A. and Chiar, J. E. and Cordiner, M. A. and Drozdovskaya, M. N. and Ioppolo, S. and Law, C. J. and Linnartz, H. and Lis, D. C. and Melnick, G. J. and McGuire, B. A. and Noble, J. A. and Öberg, K. I. and Palumbo, M. E. and Pendleton, Y. J. and Perotti, G. and Pontoppidan, K. M. and Qasim, D. and Rocha, W. R. M. and Terada, H. and Urso, R. G. and van Dishoeck, E. F.},
	month = nov,
	year = {2023},
	note = {ADS Bibcode: 2023A\&A...679A.138S},
	pages = {A138},
}

@misc{bergner_jwst_2024,
	title = {{JWST} ice band profiles reveal mixed ice compositions in the {HH} 48 {NE} disk},
	url = {https://ui.adsabs.harvard.edu/abs/2024arXiv240908117B},
	doi = {10.48550/arXiv.2409.08117},
	urldate = {2024-11-03},
	author = {Bergner, Jennifer B. and Sturm, J. A. and Piacentino, Elettra L. and McClure, M. K. and Oberg, Karin I. and Boogert, A. C. A. and Dartois, E. and Drozdovskaya, M. N. and Fraser, H. J. and Harsono, Daniel and Ioppolo, Sergio and Law, Charles J. and Lis, Dariusz C. and McGuire, Brett A. and Melnick, Gary J. and Noble, Jennifer A. and Palumbo, M. E. and Pendleton, Yvonne J. and Perotti, Giulia and Qasim, Danna and Rocha, W. R. M. and van Dishoeck, E. F.},
	month = sep,
	year = {2024},
	note = {Publication Title: arXiv e-prints
ADS Bibcode: 2024arXiv240908117B},
}

@article{oberg_astrochemistry_2021,
	title = {Astrochemistry and compositions of planetary systems},
	volume = {893},
	issn = {0370-1573},
	url = {https://ui.adsabs.harvard.edu/abs/2021PhR...893....1O},
	doi = {10.1016/j.physrep.2020.09.004},
	urldate = {2024-11-21},
	journal = {Physics Reports},
	author = {Öberg, Karin I. and Bergin, Edwin A.},
	month = jan,
	year = {2021},
	note = {ADS Bibcode: 2021PhR...893....1O},
	pages = {1--48},
}

@article{zhang_chemistry_2024,
	title = {Chemistry in {Protoplanetary} {Disks}},
	volume = {90},
	url = {https://ui.adsabs.harvard.edu/abs/2024RvMG...90...27Z},
	doi = {10.2138/rmg.2024.90.02},
	urldate = {2024-11-24},
	journal = {Reviews in Mineralogy and Geochemistry},
	author = {Zhang, Ke},
	month = jul,
	year = {2024},
	note = {ADS Bibcode: 2024RvMG...90...27Z},
	pages = {27--53},
}

@article{topchieva_ices_2024,
	title = {Ices on pebbles in protoplanetary discs},
	volume = {530},
	issn = {0035-8711},
	url = {https://ui.adsabs.harvard.edu/abs/2024MNRAS.530.2731T},
	doi = {10.1093/mnras/stae597},
	urldate = {2025-01-26},
	journal = {Monthly Notices of the Royal Astronomical Society},
	publisher = {OUP},
	author = {Topchieva, A. and Molyarova, T. and Akimkin, V. and Maksimova, L. and Vorobyov, E.},
	month = may,
	year = {2024},
	note = {ADS Bibcode: 2024MNRAS.530.2731T},
	pages = {2731--2748},
}

@article{molyarova_co_2025,
	title = {C/{O} ratios in self-gravitating protoplanetary discs with dust evolution},
	volume = {42},
	issn = {1323-3580},
	url = {https://ui.adsabs.harvard.edu/abs/2025PASA...42....4M},
	doi = {10.1017/pasa.2024.126},
	urldate = {2025-01-26},
	journal = {Publications of the Astronomical Society of Australia},
	author = {Molyarova, Tamara and Vorobyov, Eduard and Akimkin, Vitaly},
	month = jan,
	year = {2025},
	note = {ADS Bibcode: 2025PASA...42....4M},
	pages = {e004},
}

@article{chambers_analytic_2019,
	title = {An {Analytic} {Model} for an {Evolving} {Protoplanetary} {Disk} with a {Disk} {Wind}},
	volume = {879},
	issn = {0004-637X},
	url = {https://ui.adsabs.harvard.edu/abs/2019ApJ...879...98C},
	doi = {10.3847/1538-4357/ab2537},
	urldate = {2025-01-28},
	journal = {The Astrophysical Journal},
	publisher = {IOP},
	author = {Chambers, John},
	month = jul,
	year = {2019},
	note = {ADS Bibcode: 2019ApJ...879...98C},
	pages = {98},
}

@article{eistrup_molecular_2018,
	title = {Molecular abundances and {C}/{O} ratios in chemically evolving planet-forming disk midplanes},
	volume = {613},
	issn = {0004-6361},
	url = {https://ui.adsabs.harvard.edu/abs/2018A&A...613A..14E},
	doi = {10.1051/0004-6361/201731302},
	urldate = {2025-01-29},
	journal = {Astronomy and Astrophysics},
	author = {Eistrup, Christian and Walsh, Catherine and van Dishoeck, Ewine F.},
	month = may,
	year = {2018},
	note = {ADS Bibcode: 2018A\&A...613A..14E},
	pages = {A14},
}

@article{eistrup_chemical_2022,
	title = {Chemical evolution in ices on drifting, planet-forming pebbles},
	volume = {667},
	issn = {0004-6361},
	url = {https://ui.adsabs.harvard.edu/abs/2022A&A...667A.160E/abstract},
	doi = {10.1051/0004-6361/202243982},
	language = {en},
	urldate = {2025-01-29},
	journal = {Astronomy and Astrophysics},
	author = {Eistrup, Christian and Henning, Thomas},
	month = nov,
	year = {2022},
	pages = {A160},
}

@article{eistrup_chemical_2022-1,
	title = {Chemical evolution in planet-forming regions with growing grains},
	volume = {667},
	issn = {0004-6361},
	url = {https://ui.adsabs.harvard.edu/abs/2022A&A...667A.121E},
	doi = {10.1051/0004-6361/202243981},
	urldate = {2025-01-29},
	journal = {Astronomy and Astrophysics},
	author = {Eistrup, Christian and Cleeves, L. Ilsedore and Krijt, Sebastiaan},
	month = nov,
	year = {2022},
	note = {ADS Bibcode: 2022A\&A...667A.121E},
	pages = {A121},
}

@article{molyarova_gravitoviscous_2021,
	title = {Gravitoviscous {Protoplanetary} {Disks} with a {Dust} {Component}. {V}. {The} {Dynamic} {Model} for {Freeze}-out and {Sublimation} of {Volatiles}},
	volume = {910},
	issn = {0004-637X},
	url = {https://ui.adsabs.harvard.edu/abs/2021ApJ...910..153M},
	doi = {10.3847/1538-4357/abe2b0},
	urldate = {2025-02-03},
	journal = {The Astrophysical Journal},
	publisher = {IOP},
	author = {Molyarova, Tamara and Vorobyov, Eduard I. and Akimkin, Vitaly and Skliarevskii, Aleksandr and Wiebe, Dmitri and Güdel, Manuel},
	month = apr,
	year = {2021},
	note = {ADS Bibcode: 2021ApJ...910..153M},
	pages = {153},
}

@article{winters_solar_2019,
	title = {The {Solar} {Neighborhood}. {XLV}. {The} {Stellar} {Multiplicity} {Rate} of {M} {Dwarfs} {Within} 25 pc},
	volume = {157},
	issn = {0004-6256},
	url = {https://ui.adsabs.harvard.edu/abs/2019AJ....157..216W},
	doi = {10.3847/1538-3881/ab05dc},
	urldate = {2025-03-19},
	journal = {The Astronomical Journal},
	publisher = {IOP},
	author = {Winters, Jennifer G. and Henry, Todd J. and Jao, Wei-Chun and Subasavage, John P. and Chatelain, Joseph P. and Slatten, Ken and Riedel, Adric R. and Silverstein, Michele L. and Payne, Matthew J.},
	month = jun,
	year = {2019},
	note = {ADS Bibcode: 2019AJ....157..216W},
	pages = {216},
}

@article{gaidos_minimum_2017,
	title = {A minimum mass nebula for {M} dwarfs},
	volume = {470},
	issn = {0035-8711},
	url = {https://ui.adsabs.harvard.edu/abs/2017MNRAS.470L...1G},
	doi = {10.1093/mnrasl/slx063},
	urldate = {2025-03-19},
	journal = {Monthly Notices of the Royal Astronomical Society},
	publisher = {OUP},
	author = {Gaidos, E.},
	month = sep,
	year = {2017},
	note = {ADS Bibcode: 2017MNRAS.470L...1G},
	pages = {L1--L5},
}

@article{schoonenberg_planetesimal_2017,
	title = {Planetesimal formation near the snowline: in or out?},
	volume = {602},
	issn = {0004-6361},
	shorttitle = {Planetesimal formation near the snowline},
	url = {https://ui.adsabs.harvard.edu/abs/2017A&A...602A..21S},
	doi = {10.1051/0004-6361/201630013},
	urldate = {2025-03-28},
	journal = {Astronomy and Astrophysics},
	author = {Schoonenberg, Djoeke and Ormel, Chris W.},
	month = jun,
	year = {2017},
	note = {ADS Bibcode: 2017A\&A...602A..21S},
	pages = {A21},
}

@article{hyodo_formation_2019,
	title = {Formation of rocky and icy planetesimals inside and outside the snow line: effects of diffusion, sublimation, and back-reaction},
	volume = {629},
	issn = {0004-6361},
	shorttitle = {Formation of rocky and icy planetesimals inside and outside the snow line},
	url = {https://ui.adsabs.harvard.edu/abs/2019A&A...629A..90H},
	doi = {10.1051/0004-6361/201935935},
	urldate = {2025-03-28},
	journal = {Astronomy and Astrophysics},
	author = {Hyodo, Ryuki and Ida, Shigeru and Charnoz, Sébastien},
	month = sep,
	year = {2019},
	note = {ADS Bibcode: 2019A\&A...629A..90H},
	pages = {A90},
}

@article{asplund_chemical_2021,
	title = {The chemical make-up of the {Sun}: {A} 2020 vision},
	volume = {653},
	issn = {0004-6361},
	shorttitle = {The chemical make-up of the {Sun}},
	url = {https://ui.adsabs.harvard.edu/abs/2021A&A...653A.141A},
	doi = {10.1051/0004-6361/202140445},
	urldate = {2025-03-31},
	journal = {Astronomy and Astrophysics},
	author = {Asplund, M. and Amarsi, A. M. and Grevesse, N.},
	month = sep,
	year = {2021},
	note = {ADS Bibcode: 2021A\&A...653A.141A},
	pages = {A141},
}

@article{estrada_global_2022,
	title = {Global {Modeling} of {Nebulae} with {Particle} {Growth}, {Drift}, and {Evaporation} {Fronts}. {III}. {Redistribution} of {Refractories} and {Volatiles}},
	volume = {936},
	issn = {0004-637X},
	url = {https://ui.adsabs.harvard.edu/abs/2022ApJ...936...40E},
	doi = {10.3847/1538-4357/ac81c6},
	urldate = {2025-03-31},
	journal = {The Astrophysical Journal},
	publisher = {IOP},
	author = {Estrada, Paul R. and Cuzzi, Jeffrey N.},
	month = sep,
	year = {2022},
	note = {ADS Bibcode: 2022ApJ...936...40E},
	pages = {40},
}

@article{stammler_redistribution_2017,
	title = {Redistribution of {CO} at the location of the {CO} ice line in evolving gas and dust disks},
	volume = {600},
	issn = {0004-6361},
	url = {https://ui.adsabs.harvard.edu/abs/2017A&A...600A.140S},
	doi = {10.1051/0004-6361/201629041},
	urldate = {2025-04-01},
	journal = {Astronomy and Astrophysics},
	author = {Stammler, Sebastian Markus and Birnstiel, Tilman and Panić, Olja and Dullemond, Cornelis Petrus and Dominik, Carsten},
	month = apr,
	year = {2017},
	note = {ADS Bibcode: 2017A\&A...600A.140S},
	pages = {A140},
}

@article{stevenson_rapid_1988,
	title = {Rapid formation of {Jupiter} by diffusive redistribution of water vapor in the solar nebula},
	volume = {75},
	issn = {0019-1035},
	url = {https://ui.adsabs.harvard.edu/abs/1988Icar...75..146S},
	doi = {10.1016/0019-1035(88)90133-9},
	urldate = {2025-04-01},
	journal = {Icarus},
	author = {Stevenson, David J. and Lunine, Jonathan I.},
	month = jul,
	year = {1988},
	note = {ADS Bibcode: 1988Icar...75..146S},
	pages = {146--155},
}

@article{ciesla_evolution_2006,
	title = {The evolution of the water distribution in a viscous protoplanetary disk},
	volume = {181},
	issn = {0019-1035},
	url = {https://ui.adsabs.harvard.edu/abs/2006Icar..181..178C},
	doi = {10.1016/j.icarus.2005.11.009},
	urldate = {2025-04-01},
	journal = {Icarus},
	author = {Ciesla, Fred J. and Cuzzi, Jeffrey N.},
	month = mar,
	year = {2006},
	note = {ADS Bibcode: 2006Icar..181..178C},
	pages = {178--204},
}

@article{wang_early_2025,
	title = {Early {Accretion} of {Large} {Amounts} of {Solids} for {Directly} {Imaged} {Exoplanets}},
	volume = {981},
	issn = {0004-637X},
	url = {https://ui.adsabs.harvard.edu/abs/2025ApJ...981..138W},
	doi = {10.3847/1538-4357/adb42c},
	urldate = {2025-04-04},
	journal = {The Astrophysical Journal},
	publisher = {IOP},
	author = {Wang, Ji},
	month = mar,
	year = {2025},
	note = {ADS Bibcode: 2025ApJ...981..138W},
	pages = {138},
}

@article{bitsch_how_2022,
	title = {How drifting and evaporating pebbles shape giant planets. {III}. {The} formation of {WASP}-{77A} b and τ {Boötis} b},
	volume = {665},
	issn = {0004-6361},
	url = {https://ui.adsabs.harvard.edu/abs/2022A&A...665A.138B},
	doi = {10.1051/0004-6361/202243345},
	urldate = {2025-04-07},
	journal = {Astronomy and Astrophysics},
	author = {Bitsch, Bertram and Schneider, Aaron David and Kreidberg, Laura},
	month = sep,
	year = {2022},
	note = {ADS Bibcode: 2022A\&A...665A.138B},
	pages = {A138},
}

@article{turrini_tracing_2021,
	title = {Tracing the {Formation} {History} of {Giant} {Planets} in {Protoplanetary} {Disks} with {Carbon}, {Oxygen}, {Nitrogen}, and {Sulfur}},
	volume = {909},
	issn = {0004-637X},
	url = {https://ui.adsabs.harvard.edu/abs/2021ApJ...909...40T},
	doi = {10.3847/1538-4357/abd6e5},
	urldate = {2025-04-07},
	journal = {The Astrophysical Journal},
	publisher = {IOP},
	author = {Turrini, D. and Schisano, E. and Fonte, S. and Molinari, S. and Politi, R. and Fedele, D. and Panić, O. and Kama, M. and Changeat, Q. and Tinetti, G.},
	month = mar,
	year = {2021},
	note = {ADS Bibcode: 2021ApJ...909...40T},
	pages = {40},
}

@article{venturini_planet_2016,
	title = {Planet formation with envelope enrichment: new insights on planetary diversity},
	volume = {596},
	issn = {0004-6361},
	shorttitle = {Planet formation with envelope enrichment},
	url = {https://ui.adsabs.harvard.edu/abs/2016A&A...596A..90V},
	doi = {10.1051/0004-6361/201628828},
	urldate = {2025-04-07},
	journal = {Astronomy and Astrophysics},
	author = {Venturini, Julia and Alibert, Yann and Benz, Willy},
	month = dec,
	year = {2016},
	note = {ADS Bibcode: 2016A\&A...596A..90V},
	pages = {A90},
}

@article{madhusudhan_exoplanetary_2019,
	title = {Exoplanetary {Atmospheres}: {Key} {Insights}, {Challenges}, and {Prospects}},
	volume = {57},
	issn = {0066-4146},
	shorttitle = {Exoplanetary {Atmospheres}},
	url = {https://ui.adsabs.harvard.edu/abs/2019ARA&A..57..617M},
	doi = {10.1146/annurev-astro-081817-051846},
	urldate = {2025-04-07},
	journal = {Annual Review of Astronomy and Astrophysics},
	author = {Madhusudhan, Nikku},
	month = aug,
	year = {2019},
	note = {ADS Bibcode: 2019ARA\&A..57..617M},
	pages = {617--663},
}

@article{jiang_chemical_2023,
	title = {Chemical footprints of giant planet formation. {Role} of planet accretion in shaping the {C}/{O} ratio of protoplanetary disks},
	volume = {678},
	issn = {0004-6361},
	url = {https://ui.adsabs.harvard.edu/abs/2023A&A...678A..33J},
	doi = {10.1051/0004-6361/202346637},
	urldate = {2025-04-07},
	journal = {Astronomy and Astrophysics},
	author = {Jiang, Haochang and Wang, Yu and Ormel, Chris W. and Krijt, Sebastiaan and Dong, Ruobing},
	month = oct,
	year = {2023},
	note = {ADS Bibcode: 2023A\&A...678A..33J},
	pages = {A33},
}

@article{demirci_destruction_2020,
	title = {Destruction of eccentric planetesimals by ram pressure and erosion},
	volume = {644},
	issn = {0004-6361},
	url = {https://ui.adsabs.harvard.edu/abs/2020A&A...644A..20D},
	doi = {10.1051/0004-6361/202039312},
	urldate = {2025-04-11},
	journal = {Astronomy and Astrophysics},
	author = {Demirci, Tunahan and Schneider, Niclas and Teiser, Jens and Wurm, Gerhard},
	month = dec,
	year = {2020},
	note = {ADS Bibcode: 2020A\&A...644A..20D},
	pages = {A20},
}

@article{mcclure_ice_2023,
	title = {An {Ice} {Age} {JWST} inventory of dense molecular cloud ices},
	volume = {7},
	issn = {2397-3366},
	url = {http://arxiv.org/abs/2301.09140},
	doi = {10.1038/s41550-022-01875-w},
	number = {4},
	urldate = {2025-04-12},
	journal = {Nat Astron},
	author = {McClure, M. K. and Rocha, W. R. M. and Pontoppidan, K. M. and Crouzet, N. and Chu, L. E. U. and Dartois, E. and Lamberts, T. and Noble, J. A. and Pendleton, Y. J. and Perotti, G. and Qasim, D. and Rachid, M. G. and Smith, Z. L. and Sun, Fengwu and Beck, Tracy L. and Boogert, A. C. A. and Brown, W. A. and Caselli, P. and Charnley, S. B. and Cuppen, Herma M. and Dickinson, H. and Drozdovskaya, M. N. and Egami, E. and Erkal, J. and Fraser, H. and Garrod, R. T. and Harsono, D. and Ioppolo, S. and Jimenez-Serra, I. and Jin, M. and Jørgensen, J. K. and Kristensen, L. E. and Lis, D. C. and McCoustra, M. R. S. and McGuire, Brett A. and Melnick, G. J. and Oberg, Karin I. and Palumbo, M. E. and Shimonishi, T. and Sturm, J. A. and Dishoeck, E. F. van and Linnartz, H.},
	month = jan,
	year = {2023},
	note = {arXiv:2301.09140 [astro-ph]},
	pages = {431--443},
}

@misc{houge_burned_2025,
	title = {Burned to ashes: {How} the thermal decomposition of refractory organics in the inner protoplanetary disc impacts the gas-phase {C}/{O} ratio},
	shorttitle = {Burned to ashes},
	url = {https://ui.adsabs.harvard.edu/abs/2025arXiv250520427H},
	doi = {10.48550/arXiv.2505.20427},
	urldate = {2025-05-31},
	publisher = {arXiv},
	author = {Houge, Adrien and Johansen, Anders and Bergin, Edwin and Ciesla, Fred J. and Bitsch, Bertram and Lambrechts, Michiel and Henning, Thomas and Perotti, Giulia},
	month = may,
	year = {2025},
	note = {ADS Bibcode: 2025arXiv250520427H},
}

@article{cochran_composition_2015,
	title = {The {Composition} of {Comets}},
	volume = {197},
	issn = {1572-9672},
	url = {https://doi.org/10.1007/s11214-015-0183-6},
	doi = {10.1007/s11214-015-0183-6},
	language = {en},
	number = {1},
	urldate = {2025-06-02},
	journal = {Space Sci Rev},
	author = {Cochran, Anita L. and Levasseur-Regourd, Anny-Chantal and Cordiner, Martin and Hadamcik, Edith and Lasue, Jérémie and Gicquel, Adeline and Schleicher, David G. and Charnley, Steven B. and Mumma, Michael J. and Paganini, Lucas and Bockelée-Morvan, Dominique and Biver, Nicolas and Kuan, Yi-Jehng},
	month = dec,
	year = {2015},
	pages = {9--46},
}

@article{blum_formation_2022,
	title = {Formation of {Comets}},
	volume = {8},
	url = {https://ui.adsabs.harvard.edu/abs/2022Univ....8..381B},
	doi = {10.3390/universe8070381},
	urldate = {2025-06-02},
	journal = {Universe},
	author = {Blum, Jürgen and Bischoff, Dorothea and Gundlach, Bastian},
	month = jul,
	year = {2022},
	note = {ADS Bibcode: 2022Univ....8..381B},
	pages = {381},
}

@article{weissman_origin_2020,
	title = {Origin and {Evolution} of {Cometary} {Nuclei}},
	volume = {216},
	issn = {0038-6308},
	url = {https://ui.adsabs.harvard.edu/abs/2020SSRv..216....6W},
	doi = {10.1007/s11214-019-0625-7},
	urldate = {2025-06-02},
	journal = {Space Science Reviews},
	author = {Weissman, Paul and Morbidelli, Alessandro and Davidsson, Björn and Blum, Jürgen},
	month = jan,
	year = {2020},
	note = {ADS Bibcode: 2020SSRv..216....6W},
	pages = {6},
}

@article{malamud_are_2022,
	title = {Are there any pristine comets? {Constraints} from pebble structure},
	volume = {514},
	issn = {0035-8711},
	shorttitle = {Are there any pristine comets?},
	url = {https://ui.adsabs.harvard.edu/abs/2022MNRAS.514.3366M},
	doi = {10.1093/mnras/stac1535},
	urldate = {2025-06-02},
	journal = {Monthly Notices of the Royal Astronomical Society},
	publisher = {OUP},
	author = {Malamud, Uri and Landeck, Wolf A. and Bischoff, Dorothea and Kreuzig, Christopher and Perets, Hagai B. and Gundlach, Bastian and Blum, Jürgen},
	month = aug,
	year = {2022},
	note = {ADS Bibcode: 2022MNRAS.514.3366M},
	pages = {3366--3394},
}

@article{rubin_origin_2020,
	title = {On the {Origin} and {Evolution} of the {Material} in {67P}/{Churyumov}-{Gerasimenko}},
	volume = {216},
	issn = {0038-6308},
	url = {https://ui.adsabs.harvard.edu/abs/2020SSRv..216..102R},
	doi = {10.1007/s11214-020-00718-2},
	urldate = {2025-06-03},
	journal = {Space Science Reviews},
	author = {Rubin, Martin and Engrand, Cécile and Snodgrass, Colin and Weissman, Paul and Altwegg, Kathrin and Busemann, Henner and Morbidelli, Alessandro and Mumma, Michael},
	month = jul,
	year = {2020},
	note = {ADS Bibcode: 2020SSRv..216..102R},
	pages = {102},
}

@article{lippi_investigation_2021,
	title = {Investigation of the {Origins} of {Comets} as {Revealed} through {Infrared} {High}-resolution {Spectroscopy} {I}. {Molecular} {Abundances}},
	volume = {162},
	issn = {0004-6256},
	url = {https://ui.adsabs.harvard.edu/abs/2021AJ....162...74L},
	doi = {10.3847/1538-3881/abfdb7},
	urldate = {2025-06-03},
	journal = {The Astronomical Journal},
	publisher = {IOP},
	author = {Lippi, M. and Villanueva, G. L. and Mumma, M. J. and Faggi, S.},
	month = aug,
	year = {2021},
	note = {ADS Bibcode: 2021AJ....162...74L},
	pages = {74},
}

@article{biver_chemical_2024,
	title = {Chemical composition of comets {C}/2021 {A1} ({Leonard}) and {C}/2022 {E3} ({ZTF}) from radio spectroscopy and the abundance of {HCOOH} and {HNCO} in comets},
	volume = {690},
	issn = {0004-6361},
	url = {https://ui.adsabs.harvard.edu/abs/2024A&A...690A.271B},
	doi = {10.1051/0004-6361/202450921},
	urldate = {2025-06-03},
	journal = {Astronomy and Astrophysics},
	publisher = {EDP},
	author = {Biver, N. and Bockelée-Morvan, D. and Handzlik, B. and Sandqvist, Aa. and Boissier, J. and Drozdovskaya, M. N. and Moreno, R. and Crovisier, J. and Lis, D. C. and Cordiner, M. and Milam, S. and Roth, N. X. and Bonev, B. P. and Dello Russo, N. and Vervack, R. and Opitom, C. and Kawakita, H.},
	month = oct,
	year = {2024},
	note = {ADS Bibcode: 2024A\&A...690A.271B},
	pages = {A271},
}

@article{lippi_chemical_2024,
	title = {The chemical composition of {CO}-rich comet {C}/2023 {H2} ({Lemmon})},
	volume = {689},
	issn = {0004-6361},
	url = {https://ui.adsabs.harvard.edu/abs/2024A&A...689A..77L},
	doi = {10.1051/0004-6361/202450634},
	urldate = {2025-06-03},
	journal = {Astronomy and Astrophysics},
	publisher = {EDP},
	author = {Lippi, M. and Ferellec, L. and Opitom, C. and Faggi, S. and Mumma, M. J. and Villanueva, G. L.},
	month = sep,
	year = {2024},
	note = {ADS Bibcode: 2024A\&A...689A..77L},
	pages = {A77},
}

@article{willacy_comets_2022,
	title = {Comets in {Context}: {Comparing} {Comet} {Compositions} with {Protosolar} {Nebula} {Models}},
	volume = {931},
	issn = {0004-637X},
	shorttitle = {Comets in {Context}},
	url = {https://ui.adsabs.harvard.edu/abs/2022ApJ...931..164W},
	doi = {10.3847/1538-4357/ac67e3},
	urldate = {2025-06-03},
	journal = {The Astrophysical Journal},
	publisher = {IOP},
	author = {Willacy, Karen and Turner, Neal and Bonev, Boncho and Gibb, Erika and Dello Russo, Neil and DiSanti, Michael and Vervack, Jr., Ronald J. and Roth, Nathan X.},
	month = jun,
	year = {2022},
	note = {ADS Bibcode: 2022ApJ...931..164W},
	pages = {164},
}

@article{marceta_synthetic_2023,
	title = {Synthetic {Detections} of {Interstellar} {Objects} with the {Rubin} {Observatory} {Legacy} {Survey} of {Space} and {Time}},
	volume = {4},
	issn = {2632-3338},
	url = {https://ui.adsabs.harvard.edu/abs/2023PSJ.....4..230M},
	doi = {10.3847/PSJ/ad08c1},
	urldate = {2025-06-06},
	journal = {The Planetary Science Journal},
	publisher = {IOP},
	author = {Marčeta, Dušan and Seligman, Darryl Z.},
	month = dec,
	year = {2023},
	note = {ADS Bibcode: 2023PSJ.....4..230M},
	pages = {230},
}

@article{ootsubo_akari_2012,
	title = {{AKARI} {Near}-infrared {Spectroscopic} {Survey} for {CO2} in 18 {Comets}},
	volume = {752},
	issn = {0004-637X},
	url = {https://ui.adsabs.harvard.edu/abs/2012ApJ...752...15O},
	doi = {10.1088/0004-637X/752/1/15},
	urldate = {2025-06-06},
	journal = {The Astrophysical Journal},
	publisher = {IOP},
	author = {Ootsubo, Takafumi and Kawakita, Hideyo and Hamada, Saki and Kobayashi, Hitomi and Yamaguchi, Mitsuru and Usui, Fumihiko and Nakagawa, Takao and Ueno, Munetaka and Ishiguro, Masateru and Sekiguchi, Tomohiko and Watanabe, Jun-ichi and Sakon, Itsuki and Shimonishi, Takashi and Onaka, Takashi},
	month = jun,
	year = {2012},
	note = {ADS Bibcode: 2012ApJ...752...15O},
	pages = {15},
}

@misc{youdin_formation_2025,
	title = {Formation of {Giant} {Planets}},
	url = {https://ui.adsabs.harvard.edu/abs/2025arXiv250113214Y},
	doi = {10.48550/arXiv.2501.13214},
	urldate = {2025-06-11},
	publisher = {arXiv},
	author = {Youdin, Andrew N. and Zhu, Zhaohuan},
	month = jan,
	year = {2025},
	note = {ADS Bibcode: 2025arXiv250113214Y},
}

@article{knierim_constraining_2022,
	title = {Constraining the origin of giant exoplanets via elemental abundance measurements},
	volume = {665},
	issn = {0004-6361},
	url = {https://ui.adsabs.harvard.edu/abs/2022A&A...665L...5K},
	doi = {10.1051/0004-6361/202244516},
	urldate = {2025-06-11},
	journal = {Astronomy and Astrophysics},
	author = {Knierim, H. and Shibata, S. and Helled, R.},
	month = sep,
	year = {2022},
	note = {ADS Bibcode: 2022A\&A...665L...5K},
	pages = {L5},
}

@article{cuppen_laboratory_2024,
	title = {Laboratory and {Computational} {Studies} of {Interstellar} {Ices}},
	volume = {62},
	issn = {0066-4146, 1545-4282},
	url = {https://www.annualreviews.org/content/journals/10.1146/annurev-astro-071221-052732},
	doi = {10.1146/annurev-astro-071221-052732},
	language = {en},
	number = {Volume 62, 2024},
	urldate = {2025-06-26},
	journal = {Annual Review of Astronomy and Astrophysics},
	publisher = {Annual Reviews},
	author = {Cuppen, Herma M. and Linnartz, H. and Ioppolo, S.},
	month = sep,
	year = {2024},
	pages = {243--286},
}

@article{ikoma_formation_2025,
	title = {Formation of {Giant} {Planets}},
	url = {https://www.annualreviews.org/content/journals/10.1146/annurev-astro-052722-094843},
	doi = {10.1146/annurev-astro-052722-094843},
	language = {en},
	urldate = {2025-06-26},
	publisher = {Annual Reviews},
	author = {Ikoma, Masahiro and Kobayashi, Hiroshi},
	month = may,
	year = {2025},
}

@article{birnstiel_dust_2024,
	title = {Dust {Growth} and {Evolution} in {Protoplanetary} {Disks}},
	volume = {62},
	issn = {0066-4146, 1545-4282},
	url = {https://www.annualreviews.org/content/journals/10.1146/annurev-astro-071221-052705},
	doi = {10.1146/annurev-astro-071221-052705},
	language = {en},
	number = {Volume 62, 2024},
	urldate = {2025-06-26},
	journal = {Annual Review of Astronomy and Astrophysics},
	publisher = {Annual Reviews},
	author = {Birnstiel, Tilman},
	month = sep,
	year = {2024},
	pages = {157--202},
}

@article{mckay_quantifying_2021,
	title = {Quantifying the {Hypervolatile} {Abundances} in {Jupiter}-family {Comet} {46P}/{Wirtanen}},
	volume = {2},
	issn = {2632-3338},
	url = {https://ui.adsabs.harvard.edu/abs/2021PSJ.....2...21M},
	doi = {10.3847/PSJ/abd71d},
	urldate = {2025-06-26},
	journal = {The Planetary Science Journal},
	publisher = {IOP},
	author = {McKay, Adam J. and DiSanti, Michael A. and Cochran, Anita L. and Bonev, Boncho P. and Dello Russo, Neil and Vervack, Jr., Ronald J. and Gibb, Erika and Roth, Nathan X. and Saki, Mohammad and Khan, Younas and Kawakita, Hideyo},
	month = feb,
	year = {2021},
	note = {ADS Bibcode: 2021PSJ.....2...21M},
	pages = {21},
}

@article{mckay_evolution_2015,
	title = {Evolution of {H2O}, {CO}, and {CO2} production in {Comet} {C}/2009 {P1} {Garradd} during the 2011-2012 apparition},
	volume = {250},
	issn = {0019-1035},
	url = {https://ui.adsabs.harvard.edu/abs/2015Icar..250..504M},
	doi = {10.1016/j.icarus.2014.12.023},
	urldate = {2025-06-26},
	journal = {Icarus},
	author = {McKay, Adam J. and Cochran, Anita L. and DiSanti, Michael A. and Villanueva, Geronimo and Russo, Neil Dello and Vervack, Ronald J. and Morgenthaler, Jeffrey P. and Harris, Walter M. and Chanover, Nancy J.},
	month = apr,
	year = {2015},
	note = {ADS Bibcode: 2015Icar..250..504M},
	pages = {504--515},
}

@article{mckay_co2_2016,
	title = {The {CO2} abundance in {Comets} {C}/2012 {K1} ({PanSTARRS}), {C}/2012 {K5} ({LINEAR}), and {290P}/{Jäger} as measured with {Spitzer}},
	volume = {266},
	issn = {0019-1035},
	url = {https://ui.adsabs.harvard.edu/abs/2016Icar..266..249M},
	doi = {10.1016/j.icarus.2015.11.004},
	urldate = {2025-06-26},
	journal = {Icarus},
	author = {McKay, Adam J. and Kelley, Michael S. P. and Cochran, Anita L. and Bodewits, Dennis and DiSanti, Michael A. and Russo, Neil Dello and Lisse, Carey M.},
	month = mar,
	year = {2016},
	note = {ADS Bibcode: 2016Icar..266..249M},
	pages = {249--260},
}

@article{mckay_evolution_2018,
	title = {Evolution of {H2O} production in comet {C}/2012 {S1} ({ISON}) as inferred from forbidden oxygen and {OH} emission},
	volume = {309},
	issn = {0019-1035},
	url = {https://ui.adsabs.harvard.edu/abs/2018Icar..309....1M},
	doi = {10.1016/j.icarus.2018.02.024},
	urldate = {2025-06-26},
	journal = {Icarus},
	author = {McKay, Adam J. and Cochran, Anita L. and DiSanti, Michael A. and Dello Russo, Neil and Weaver, Harold and Vervack, Ronald J. and Harris, Walter M. and Kawakita, Hideyo},
	month = jul,
	year = {2018},
	note = {ADS Bibcode: 2018Icar..309....1M},
	pages = {1--12},
}

@article{virtanen_scipy_2020,
	title = {{SciPy} 1.0: fundamental algorithms for scientific computing in {Python}},
	volume = {17},
	shorttitle = {{SciPy} 1.0},
	url = {https://ui.adsabs.harvard.edu/abs/2020NatMe..17..261V},
	doi = {10.1038/s41592-019-0686-2},
	urldate = {2025-07-03},
	journal = {Nature Methods},
	author = {Virtanen, Pauli and Gommers, Ralf and Oliphant, Travis E. and Haberland, Matt and Reddy, Tyler and Cournapeau, David and Burovski, Evgeni and Peterson, Pearu and Weckesser, Warren and Bright, Jonathan and van der Walt, Stéfan J. and Brett, Matthew and Wilson, Joshua and Millman, K. Jarrod and Mayorov, Nikolay and Nelson, Andrew R. J. and Jones, Eric and Kern, Robert and Larson, Eric and Carey, C. J. and Polat, İlhan and Feng, Yu and Moore, Eric W. and VanderPlas, Jake and Laxalde, Denis and Perktold, Josef and Cimrman, Robert and Henriksen, Ian and Quintero, E. A. and Harris, Charles R. and Archibald, Anne M. and Ribeiro, Antônio H. and Pedregosa, Fabian and van Mulbregt, Paul and {SciPy 1. 0 Contributors}},
	month = feb,
	year = {2020},
	note = {ADS Bibcode: 2020NatMe..17..261V},
	pages = {261--272},
}

@article{astropy_collaboration_astropy_2022,
	title = {The {Astropy} {Project}: {Sustaining} and {Growing} a {Community}-oriented {Open}-source {Project} and the {Latest} {Major} {Release} (v5.0) of the {Core} {Package}},
	volume = {935},
	issn = {0004-637X},
	shorttitle = {The {Astropy} {Project}},
	url = {https://ui.adsabs.harvard.edu/abs/2022ApJ...935..167A},
	doi = {10.3847/1538-4357/ac7c74},
	urldate = {2025-07-03},
	journal = {The Astrophysical Journal},
	publisher = {IOP},
	author = {{Astropy Collaboration} and Price-Whelan, Adrian M. and Lim, Pey Lian and Earl, Nicholas and Starkman, Nathaniel and Bradley, Larry and Shupe, David L. and Patil, Aarya A. and Corrales, Lia and Brasseur, C. E. and Nöthe, Maximilian and Donath, Axel and Tollerud, Erik and Morris, Brett M. and Ginsburg, Adam and Vaher, Eero and Weaver, Benjamin A. and Tocknell, James and Jamieson, William and van Kerkwijk, Marten H. and Robitaille, Thomas P. and Merry, Bruce and Bachetti, Matteo and Günther, H. Moritz and Aldcroft, Thomas L. and Alvarado-Montes, Jaime A. and Archibald, Anne M. and Bódi, Attila and Bapat, Shreyas and Barentsen, Geert and Bazán, Juanjo and Biswas, Manish and Boquien, Médéric and Burke, D. J. and Cara, Daria and Cara, Mihai and Conroy, Kyle E. and Conseil, Simon and Craig, Matthew W. and Cross, Robert M. and Cruz, Kelle L. and D'Eugenio, Francesco and Dencheva, Nadia and Devillepoix, Hadrien A. R. and Dietrich, Jörg P. and Eigenbrot, Arthur Davis and Erben, Thomas and Ferreira, Leonardo and Foreman-Mackey, Daniel and Fox, Ryan and Freij, Nabil and Garg, Suyog and Geda, Robel and Glattly, Lauren and Gondhalekar, Yash and Gordon, Karl D. and Grant, David and Greenfield, Perry and Groener, Austen M. and Guest, Steve and Gurovich, Sebastian and Handberg, Rasmus and Hart, Akeem and Hatfield-Dodds, Zac and Homeier, Derek and Hosseinzadeh, Griffin and Jenness, Tim and Jones, Craig K. and Joseph, Prajwel and Kalmbach, J. Bryce and Karamehmetoglu, Emir and Kałuszyński, Mikołaj and Kelley, Michael S. P. and Kern, Nicholas and Kerzendorf, Wolfgang E. and Koch, Eric W. and Kulumani, Shankar and Lee, Antony and Ly, Chun and Ma, Zhiyuan and MacBride, Conor and Maljaars, Jakob M. and Muna, Demitri and Murphy, N. A. and Norman, Henrik and O'Steen, Richard and Oman, Kyle A. and Pacifici, Camilla and Pascual, Sergio and Pascual-Granado, J. and Patil, Rohit R. and Perren, Gabriel I. and Pickering, Timothy E. and Rastogi, Tanuj and Roulston, Benjamin R. and Ryan, Daniel F. and Rykoff, Eli S. and Sabater, Jose and Sakurikar, Parikshit and Salgado, Jesús and Sanghi, Aniket and Saunders, Nicholas and Savchenko, Volodymyr and Schwardt, Ludwig and Seifert-Eckert, Michael and Shih, Albert Y. and Jain, Anany Shrey and Shukla, Gyanendra and Sick, Jonathan and Simpson, Chris and Singanamalla, Sudheesh and Singer, Leo P. and Singhal, Jaladh and Sinha, Manodeep and Sipőcz, Brigitta M. and Spitler, Lee R. and Stansby, David and Streicher, Ole and Šumak, Jani and Swinbank, John D. and Taranu, Dan S. and Tewary, Nikita and Tremblay, Grant R. and de Val-Borro, Miguel and Van Kooten, Samuel J. and Vasović, Zlatan and Verma, Shresth and de Miranda Cardoso, José Vinícius and Williams, Peter K. G. and Wilson, Tom J. and Winkel, Benjamin and Wood-Vasey, W. M. and Xue, Rui and Yoachim, Peter and Zhang, Chen and Zonca, Andrea and {Astropy Project Contributors}},
	month = aug,
	year = {2022},
	note = {ADS Bibcode: 2022ApJ...935..167A},
	pages = {167},
}

@article{hunter_matplotlib_2007,
	title = {Matplotlib: {A} {2D} {Graphics} {Environment}},
	volume = {9},
	shorttitle = {Matplotlib},
	url = {https://ui.adsabs.harvard.edu/abs/2007CSE.....9...90H},
	doi = {10.1109/MCSE.2007.55},
	urldate = {2025-07-03},
	journal = {Computing in Science and Engineering},
	author = {Hunter, John D.},
	month = may,
	year = {2007},
	note = {ADS Bibcode: 2007CSE.....9...90H},
	pages = {90--95},
}

@article{harris_array_2020,
	title = {Array programming with {NumPy}},
	volume = {585},
	issn = {0028-0836},
	url = {https://ui.adsabs.harvard.edu/abs/2020Natur.585..357H},
	doi = {10.1038/s41586-020-2649-2},
	urldate = {2025-07-03},
	journal = {Nature},
	author = {Harris, Charles R. and Millman, K. Jarrod and van der Walt, Stéfan J. and Gommers, Ralf and Virtanen, Pauli and Cournapeau, David and Wieser, Eric and Taylor, Julian and Berg, Sebastian and Smith, Nathaniel J. and Kern, Robert and Picus, Matti and Hoyer, Stephan and van Kerkwijk, Marten H. and Brett, Matthew and Haldane, Allan and del Río, Jaime Fernández and Wiebe, Mark and Peterson, Pearu and Gérard-Marchant, Pierre and Sheppard, Kevin and Reddy, Tyler and Weckesser, Warren and Abbasi, Hameer and Gohlke, Christoph and Oliphant, Travis E.},
	month = sep,
	year = {2020},
	note = {ADS Bibcode: 2020Natur.585..357H},
	pages = {357--362},
}

@article{mckinney_data_2010,
	title = {Data {Structures} for {Statistical} {Computing} in {Python}},
	url = {https://proceedings.scipy.org/articles/Majora-92bf1922-00a},
	doi = {10.25080/Majora-92bf1922-00a},
	language = {en},
	urldate = {2025-07-03},
	journal = {scipy},
	author = {McKinney, Wes},
	month = may,
	year = {2010},
}

@misc{cadiou_matplotlib_2022,
	title = {Matplotlib label lines},
	url = {https://zenodo.org/records/7428071},
	doi = {10.5281/zenodo.7428071},
	urldate = {2025-07-03},
	publisher = {Zenodo},
	author = {Cadiou, Corentin},
	month = dec,
	year = {2022},
	note = {Language: eng},
}

@article{collings_laboratory_2004,
	title = {A laboratory survey of the thermal desorption of astrophysically relevant molecules},
	volume = {354},
	issn = {0035-8711},
	url = {https://ui.adsabs.harvard.edu/abs/2004MNRAS.354.1133C},
	doi = {10.1111/j.1365-2966.2004.08272.x},
	urldate = {2025-08-27},
	journal = {Monthly Notices of the Royal Astronomical Society},
	author = {Collings, Mark P. and Anderson, Mark A. and Chen, Rui and Dever, John W. and Viti, Serena and Williams, David A. and McCoustra, Martin R. S.},
	month = nov,
	year = {2004},
	note = {ADS Bibcode: 2004MNRAS.354.1133C},
	pages = {1133--1140},
}

@article{visser_chemical_2009,
	title = {The chemical history of molecules in circumstellar disks. {I}. {Ices}},
	volume = {495},
	issn = {0004-6361},
	url = {https://ui.adsabs.harvard.edu/abs/2009A&A...495..881V},
	doi = {10.1051/0004-6361/200810846},
	urldate = {2025-09-04},
	journal = {Astronomy and Astrophysics},
	author = {Visser, R. and van Dishoeck, E. F. and Doty, S. D. and Dullemond, C. P.},
	month = mar,
	year = {2009},
	note = {ADS Bibcode: 2009A\&A...495..881V},
	pages = {881--897},
}

@article{visser_chemical_2011,
	title = {The chemical history of molecules in circumstellar disks. {II}. {Gas}-phase species},
	volume = {534},
	issn = {0004-6361},
	url = {https://ui.adsabs.harvard.edu/abs/2011A&A...534A.132V},
	doi = {10.1051/0004-6361/201117249},
	urldate = {2025-09-04},
	journal = {Astronomy and Astrophysics},
	author = {Visser, R. and Doty, S. D. and van Dishoeck, E. F.},
	month = oct,
	year = {2011},
	note = {ADS Bibcode: 2011A\&A...534A.132V},
	pages = {A132},
}

@article{hincelin_survival_2013,
	title = {Survival of {Interstellar} {Molecules} to {Prestellar} {Dense} {Core} {Collapse} and {Early} {Phases} of {Disk} {Formation}},
	volume = {775},
	issn = {0004-637X},
	url = {https://ui.adsabs.harvard.edu/abs/2013ApJ...775...44H},
	doi = {10.1088/0004-637X/775/1/44},
	urldate = {2025-09-04},
	journal = {The Astrophysical Journal},
	author = {Hincelin, U. and Wakelam, V. and Commerçon, B. and Hersant, F. and Guilloteau, S.},
	month = sep,
	year = {2013},
	note = {ADS Bibcode: 2013ApJ...775...44H},
	pages = {44},
}

@article{morbidelli_formation_2024,
	title = {Formation and evolution of a protoplanetary disk: {Combining} observations, simulations, and cosmochemical constraints},
	volume = {691},
	issn = {0004-6361},
	shorttitle = {Formation and evolution of a protoplanetary disk},
	url = {https://ui.adsabs.harvard.edu/abs/2024A&A...691A.147M},
	doi = {10.1051/0004-6361/202451388},
	urldate = {2025-09-04},
	journal = {Astronomy and Astrophysics},
	author = {Morbidelli, Alessandro and Marrocchi, Yves and Ahmad, Adnan Ali and Bhandare, Asmita and Charnoz, Sébastien and Commerçon, Benoît and Dullemond, Cornelis P. and Guillot, Tristan and Hennebelle, Patrick and Lee, Yueh-Ning and Lovascio, Francesco and Marschall, Raphael and Marty, Bernard and Maury, Anaëlle and Tamami, Okamoto},
	month = nov,
	year = {2024},
	note = {ADS Bibcode: 2024A\&A...691A.147M},
	pages = {A147},
}

@article{bolin_interstellar_2025,
	title = {Interstellar comet {3I}/{ATLAS}: discovery and physical description},
	volume = {542},
	issn = {0035-8711},
	shorttitle = {Interstellar comet {3I}/{ATLAS}},
	url = {https://ui.adsabs.harvard.edu/abs/2025MNRAS.542L.139B},
	doi = {10.1093/mnrasl/slaf078},
	urldate = {2025-09-26},
	journal = {Monthly Notices of the Royal Astronomical Society},
	publisher = {OUP},
	author = {Bolin, Bryce T. and Belyakov, Matthew and Fremling, Christoffer and Graham, Matthew J. and Abdelaziz, Ahmed M. and Elhosseiny, Eslam and Gray, Candace L. and Ingebretsen, Carl and Jewett, Gracyn and Lisse, Carey M. and Karpov, Sergey and Kilic, Mukremin and Mašek, Martin and Molham, Mona and Roderick, Diana and Takey, Ali and Abron, Laura-May and Coughlin, Michael W. and Hsieh, Cheng-Han and Noll, Keith S. and Wong, Ian},
	month = sep,
	year = {2025},
	note = {ADS Bibcode: 2025MNRAS.542L.139B},
	pages = {L139--L143},
}

@article{jewitt_hubble_2025,
	title = {Hubble {Space} {Telescope} {Observations} of the {Interstellar} {Interloper} {3I}/{ATLAS}},
	volume = {990},
	issn = {0004-637X},
	url = {https://ui.adsabs.harvard.edu/abs/2025ApJ...990L...2J},
	doi = {10.3847/2041-8213/adf8d8},
	urldate = {2025-09-26},
	journal = {The Astrophysical Journal},
	publisher = {IOP},
	author = {Jewitt, David and Hui, Man-To and Mutchler, Max and Kim, Yoonyoung and Agarwal, Jessica},
	month = sep,
	year = {2025},
	note = {ADS Bibcode: 2025ApJ...990L...2J},
	pages = {L2},
}

@article{hopkins_different_2025,
	title = {From a {Different} {Star}: {3I}/{ATLAS} in the {Context} of the Ōtautahi–{Oxford} {Interstellar} {Object} {Population} {Model}},
	volume = {990},
	issn = {0004-637X},
	shorttitle = {From a {Different} {Star}},
	url = {https://ui.adsabs.harvard.edu/abs/2025ApJ...990L..30H},
	doi = {10.3847/2041-8213/adfbf4},
	urldate = {2025-09-26},
	journal = {The Astrophysical Journal},
	publisher = {IOP},
	author = {Hopkins, Matthew J. and Dorsey, Rosemary C. and Forbes, John C. and Bannister, Michele T. and Lintott, Chris J. and Leicester, Brayden},
	month = sep,
	year = {2025},
	note = {ADS Bibcode: 2025ApJ...990L..30H},
	pages = {L30},
}

@article{taylor_kinematic_2025,
	title = {The {Kinematic} {Age} of {3I}/{ATLAS} and {Its} {Implications} for {Early} {Planet} {Formation}},
	volume = {990},
	issn = {0004-637X},
	url = {https://ui.adsabs.harvard.edu/abs/2025ApJ...990L..14T},
	doi = {10.3847/2041-8213/adfa28},
	urldate = {2025-09-26},
	journal = {The Astrophysical Journal},
	publisher = {IOP},
	author = {Taylor, Aster G. and Seligman, Darryl Z.},
	month = sep,
	year = {2025},
	note = {ADS Bibcode: 2025ApJ...990L..14T},
	pages = {L14},
}

@article{meech_brief_2017,
	title = {A brief visit from a red and extremely elongated interstellar asteroid},
	volume = {552},
	issn = {0028-0836},
	url = {https://ui.adsabs.harvard.edu/abs/2017Natur.552..378M},
	doi = {10.1038/nature25020},
	urldate = {2025-10-14},
	journal = {Nature},
	author = {Meech, Karen J. and Weryk, Robert and Micheli, Marco and Kleyna, Jan T. and Hainaut, Olivier R. and Jedicke, Robert and Wainscoat, Richard J. and Chambers, Kenneth C. and Keane, Jacqueline V. and Petric, Andreea and Denneau, Larry and Magnier, Eugene and Berger, Travis and Huber, Mark E. and Flewelling, Heather and Waters, Chris and Schunova-Lilly, Eva and Chastel, Serge},
	month = dec,
	year = {2017},
	note = {ADS Bibcode: 2017Natur.552..378M},
	pages = {378--381},
}

@article{jewitt_initial_2019,
	title = {Initial {Characterization} of {Interstellar} {Comet} {2I}/2019 {Q4} ({Borisov})},
	volume = {886},
	issn = {0004-637X},
	url = {https://ui.adsabs.harvard.edu/abs/2019ApJ...886L..29J},
	doi = {10.3847/2041-8213/ab530b},
	urldate = {2025-10-14},
	journal = {The Astrophysical Journal},
	publisher = {IOP},
	author = {Jewitt, David and Luu, Jane},
	month = dec,
	year = {2019},
	note = {ADS Bibcode: 2019ApJ...886L..29J},
	pages = {L29},
}

@misc{drazkowska_planet_2023,
	address = {eprint: arXiv:2203.09759},
	title = {Planet {Formation} {Theory} in the {Era} of {ALMA} and {Kepler}: from {Pebbles} to {Exoplanets}},
	shorttitle = {Planet {Formation} {Theory} in the {Era} of {ALMA} and {Kepler}},
	url = {https://ui.adsabs.harvard.edu/abs/2023ASPC..534..717D},
	doi = {10.48550/arXiv.2203.09759},
	urldate = {2025-10-14},
	publisher = {arXiv},
	author = {Drążkowska, J. and Bitsch, B. and Lambrechts, M. and Mulders, G. D. and Harsono, D. and Vazan, A. and Liu, B. and Ormel, C. W. and Kretke, K. and Morbidelli, A.},
	month = jul,
	year = {2023},
	note = {Conference Name: Protostars and Planets VII
Volume: 534
ADS Bibcode: 2023ASPC..534..717D},
}

@article{pacetti_planet_2025,
	title = {Planet formation in chemically diverse and evolving discs: {I}. {Composition} of planetary building blocks},
	volume = {701},
	issn = {0004-6361},
	shorttitle = {Planet formation in chemically diverse and evolving discs},
	url = {https://ui.adsabs.harvard.edu/abs/2025A&A...701A.194P},
	doi = {10.1051/0004-6361/202554012},
	urldate = {2025-10-14},
	journal = {Astronomy and Astrophysics},
	publisher = {EDP},
	author = {Pacetti, E. and Schisano, E. and Turrini, D. and Dullemond, C. P. and Molinari, S. and Walsh, C. and Fonte, S. and Lebreuilly, U. and Klessen, R. S. and Hennebelle, P. and Ivanovski, S. L. and Politi, R. and Polychroni, D. and Simonetti, P. and Testi, L.},
	month = sep,
	year = {2025},
	note = {ADS Bibcode: 2025A\&A...701A.194P},
	pages = {A194},
}

@article{xenos_how_2025,
	title = {How the chemical composition of solids influences the formation of planetesimals},
	volume = {701},
	issn = {0004-6361},
	url = {https://ui.adsabs.harvard.edu/abs/2025A&A...701A..47X},
	doi = {10.1051/0004-6361/202555591},
	urldate = {2025-10-14},
	journal = {Astronomy and Astrophysics},
	publisher = {EDP},
	author = {Xenos, Konstantinos Odysseas and Bitsch, Bertram and Andama, Geoffrey},
	month = sep,
	year = {2025},
	note = {ADS Bibcode: 2025A\&A...701A..47X},
	pages = {A47},
}

@article{miley_impact_2021,
	title = {The impact of pre-main sequence stellar evolution on mid-plane snowline locations and {C}/{O} in planet forming discs},
	volume = {500},
	issn = {0035-8711},
	url = {https://ui.adsabs.harvard.edu/abs/2021MNRAS.500.4658M},
	doi = {10.1093/mnras/staa3517},
	urldate = {2025-10-15},
	journal = {Monthly Notices of the Royal Astronomical Society},
	publisher = {OUP},
	author = {Miley, James M. and Panić, Olja and Booth, Richard A. and Ilee, John D. and Ida, Shigeru and Kunitomo, Masanobu},
	month = jan,
	year = {2021},
	note = {ADS Bibcode: 2021MNRAS.500.4658M},
	pages = {4658--4670},
}

@article{penzlin_bowie-align_2024,
	title = {{BOWIE}-{ALIGN}: how formation and migration histories of giant planets impact atmospheric compositions},
	volume = {535},
	issn = {0035-8711},
	shorttitle = {{BOWIE}-{ALIGN}},
	url = {https://ui.adsabs.harvard.edu/abs/2024MNRAS.535..171P},
	doi = {10.1093/mnras/stae2362},
	urldate = {2025-10-15},
	journal = {Monthly Notices of the Royal Astronomical Society},
	publisher = {OUP},
	author = {Penzlin, Anna B. T. and Booth, Richard A. and Kirk, James and Owen, James E. and Ahrer, E. and Christie, Duncan A. and Claringbold, Alastair B. and Esparza-Borges, Emma and López-Morales, M. and Mayne, N. J. and McCormack, Mason and Meech, Annabella and Panwar, Vatsal and Powell, Diana and Sergeev, Denis E. and Taylor, Jake and Wheatley, Peter J. and Zamyatina, Maria},
	month = nov,
	year = {2024},
	note = {ADS Bibcode: 2024MNRAS.535..171P},
	pages = {171--186},
}

@misc{miotello_setting_2023,
	address = {eprint: arXiv:2203.09818},
	title = {Setting the {Stage} for {Planet} {Formation}: {Measurements} and {Implications} of the {Fundamental} {Disk} {Properties}},
	shorttitle = {Setting the {Stage} for {Planet} {Formation}},
	url = {https://ui.adsabs.harvard.edu/abs/2023ASPC..534..501M},
	doi = {10.48550/arXiv.2203.09818},
	urldate = {2025-10-17},
	publisher = {arXiv},
	author = {Miotello, A. and Kamp, I. and Birnstiel, T. and Cleeves, L. C. and Kataoka, A.},
	month = jul,
	year = {2023},
	note = {Conference Name: Protostars and Planets VII
Volume: 534
ADS Bibcode: 2023ASPC..534..501M},
}

@article{lee_viscous_1991,
	title = {Viscous excretion discs around {Be} stars},
	volume = {250},
	issn = {0035-8711},
	url = {https://ui.adsabs.harvard.edu/abs/1991MNRAS.250..432L},
	doi = {10.1093/mnras/250.2.432},
	urldate = {2025-10-20},
	journal = {Monthly Notices of the Royal Astronomical Society},
	publisher = {OUP},
	author = {Lee, Umin and Saio, Hideyuki and Osaki, Yoji},
	month = may,
	year = {1991},
	note = {ADS Bibcode: 1991MNRAS.250..432L},
	pages = {432--437},
}

@article{tabone_alma_2025,
	title = {The {ALMA} {Survey} of {Gas} {Evolution} of {PROtoplanetary} {Disks} ({AGE}-{PRO}). {VII}. {Testing} {Accretion} {Mechanisms} from {Disk} {Population} {Synthesis}},
	volume = {989},
	issn = {0004-637X},
	url = {https://ui.adsabs.harvard.edu/abs/2025ApJ...989....7T},
	doi = {10.3847/1538-4357/adc7b1},
	urldate = {2025-11-18},
	journal = {The Astrophysical Journal},
	publisher = {IOP},
	author = {Tabone, Benoît and Rosotti, Giovanni P. and Trapman, Leon and Pinilla, Paola and Pascucci, Ilaria and Somigliana, Alice and Alexander, Richard and Vioque, Miguel and Anania, Rossella and Kuznetsova, Aleksandra and Zhang, Ke and Pérez, Laura M. and Cieza, Lucas A. and Carpenter, John and Deng, Dingshan and Agurto-Gangas, Carolina and Ruiz-Rodriguez, Dary A. and Sierra, Anibal and Kurtovic, Nicolás T. and Miley, James and González-Ruilova, Camilo and TorresVillanueva, Estephani and Hogerheijde, Michiel R. and Schwarz, Kamber and Toci, Claudia and Testi, Leonardo and Lodato, Giuseppe},
	month = aug,
	year = {2025},
	note = {ADS Bibcode: 2025ApJ...989....7T},
	pages = {7},
}

@article{tabone_secular_2022,
	title = {Secular evolution of {MHD} wind-driven discs: analytical solutions in the expanded α-framework},
	volume = {512},
	issn = {0035-8711},
	shorttitle = {Secular evolution of {MHD} wind-driven discs},
	url = {https://ui.adsabs.harvard.edu/abs/2022MNRAS.512.2290T},
	doi = {10.1093/mnras/stab3442},
	urldate = {2025-11-18},
	journal = {Monthly Notices of the Royal Astronomical Society},
	publisher = {OUP},
	author = {Tabone, Benoît and Rosotti, Giovanni P. and Cridland, Alexander J. and Armitage, Philip J. and Lodato, Giuseppe},
	month = may,
	year = {2022},
	note = {ADS Bibcode: 2022MNRAS.512.2290T},
	pages = {2290--2309},
}

@article{williams_locked_2025,
	title = {Locked {In} {Ice}: how {Pebble} {Drift} and {Volatile} {Entrapment} can {Significantly} {Impact} {Carbon} and {Oxygen} {Ratios} in {Evolving} {Protoplanetary} {Discs}},
	issn = {0035-8711},
	shorttitle = {Locked {In} {Ice}},
	url = {https://ui.adsabs.harvard.edu/abs/2025MNRAS.tmp.1736W},
	doi = {10.1093/mnras/staf1839},
	urldate = {2025-11-23},
	journal = {Monthly Notices of the Royal Astronomical Society},
	publisher = {OUP},
	author = {Williams, Joe and Krijt, Sebastiaan and Bitsch, Bertram and Houge, Adrien and Bergner, Jennifer},
	month = oct,
	year = {2025},
	note = {ADS Bibcode: 2025MNRAS.tmp.1736W},
}

@article{drazkowska_planetesimal_2017,
	title = {Planetesimal formation starts at the snow line},
	volume = {608},
	issn = {0004-6361},
	url = {https://ui.adsabs.harvard.edu/abs/2017A&A...608A..92D},
	doi = {10.1051/0004-6361/201731491},
	urldate = {2025-12-12},
	journal = {Astronomy and Astrophysics},
	publisher = {EDP},
	author = {Drążkowska, J. and Alibert, Y.},
	month = dec,
	year = {2017},
	note = {ADS Bibcode: 2017A\&A...608A..92D},
	pages = {A92},
}

@article{ormel_catching_2018,
	title = {Catching drifting pebbles. {II}. {A} stochastic equation of motion for pebbles},
	volume = {615},
	issn = {0004-6361},
	url = {https://ui.adsabs.harvard.edu/abs/2018A&A...615A.178O},
	doi = {10.1051/0004-6361/201732562},
	urldate = {2025-12-15},
	journal = {Astronomy and Astrophysics},
	publisher = {EDP},
	author = {Ormel, Chris W. and Liu, Beibei},
	month = aug,
	year = {2018},
	note = {ADS Bibcode: 2018A\&A...615A.178O},
	pages = {A178},
}

@article{semenov_rosseland_2003,
	title = {Rosseland and {Planck} mean opacities for protoplanetary discs},
	volume = {410},
	issn = {0004-6361},
	url = {https://ui.adsabs.harvard.edu/abs/2003A&A...410..611S},
	doi = {10.1051/0004-6361:20031279},
	urldate = {2025-12-19},
	journal = {Astronomy and Astrophysics},
	publisher = {EDP},
	author = {Semenov, D. and Henning, Th. and Helling, Ch. and Ilgner, M. and Sedlmayr, E.},
	month = nov,
	year = {2003},
	note = {ADS Bibcode: 2003A\&A...410..611S},
	pages = {611--621},
}

@article{woitke_consistent_2016,
	title = {Consistent dust and gas models for protoplanetary disks. {I}. {Disk} shape, dust settling, opacities, and {PAHs}},
	volume = {586},
	issn = {0004-6361},
	url = {https://ui.adsabs.harvard.edu/abs/2016A&A...586A.103W},
	doi = {10.1051/0004-6361/201526538},
	urldate = {2025-12-19},
	journal = {Astronomy and Astrophysics},
	publisher = {EDP},
	author = {Woitke, P. and Min, M. and Pinte, C. and Thi, W.-F. and Kamp, I. and Rab, C. and Anthonioz, F. and Antonellini, S. and Baldovin-Saavedra, C. and Carmona, A. and Dominik, C. and Dionatos, O. and Greaves, J. and Güdel, M. and Ilee, J. D. and Liebhart, A. and Ménard, F. and Rigon, L. and Waters, L. B. F. M. and Aresu, G. and Meijerink, R. and Spaans, M.},
	month = feb,
	year = {2016},
	note = {ADS Bibcode: 2016A\&A...586A.103W},
	pages = {A103},
}

@article{cuzzi_utilitarian_2014,
	title = {Utilitarian {Opacity} {Model} for {Aggregate} {Particles} in {Protoplanetary} {Nebulae} and {Exoplanet} {Atmospheres}},
	volume = {210},
	issn = {0067-0049},
	url = {https://ui.adsabs.harvard.edu/abs/2014ApJS..210...21C},
	doi = {10.1088/0067-0049/210/2/21},
	urldate = {2025-12-19},
	journal = {The Astrophysical Journal Supplement Series},
	publisher = {IOP},
	author = {Cuzzi, Jeffrey N. and Estrada, Paul R. and Davis, Sanford S.},
	month = feb,
	year = {2014},
	note = {ADS Bibcode: 2014ApJS..210...21C},
	pages = {21},
}

@article{molliere_interpreting_2022,
	title = {Interpreting the {Atmospheric} {Composition} of {Exoplanets}: {Sensitivity} to {Planet} {Formation} {Assumptions}},
	volume = {934},
	issn = {0004-637X},
	shorttitle = {Interpreting the {Atmospheric} {Composition} of {Exoplanets}},
	url = {https://ui.adsabs.harvard.edu/abs/2022ApJ...934...74M},
	doi = {10.3847/1538-4357/ac6a56},
	urldate = {2026-03-04},
	journal = {The Astrophysical Journal},
	publisher = {IOP},
	author = {Mollière, Paul and Molyarova, Tamara and Bitsch, Bertram and Henning, Thomas and Schneider, Aaron and Kreidberg, Laura and Eistrup, Christian and Burn, Remo and Nasedkin, Evert and Semenov, Dmitry and Mordasini, Christoph and Schlecker, Martin and Schwarz, Kamber R. and Lacour, Sylvestre and Nowak, Mathias and Schulik, Matthäus},
	month = jul,
	year = {2022},
	note = {ADS Bibcode: 2022ApJ...934...74M},
	pages = {74},
}

@article{mah_close-ice_2023,
	title = {Close-in ice lines and the super-stellar {C}/{O} ratio in discs around very low-mass stars},
	volume = {677},
	issn = {0004-6361},
	url = {https://ui.adsabs.harvard.edu/abs/2023A&A...677L...7M},
	doi = {10.1051/0004-6361/202347169},
	urldate = {2026-03-05},
	journal = {Astronomy and Astrophysics},
	publisher = {EDP},
	author = {Mah, Jingyi and Bitsch, Bertram and Pascucci, Ilaria and Henning, Thomas},
	month = sep,
	year = {2023},
	note = {ADS Bibcode: 2023A\&A...677L...7M},
	pages = {L7},
}

@article{knierim_convective_2024,
	title = {Convective {Mixing} in {Gas} {Giant} {Planets} with {Primordial} {Composition} {Gradients}},
	volume = {977},
	issn = {0004-637X},
	url = {https://ui.adsabs.harvard.edu/abs/2024ApJ...977..227K},
	doi = {10.3847/1538-4357/ad8dd0},
	urldate = {2026-03-05},
	journal = {The Astrophysical Journal},
	publisher = {IOP},
	author = {Knierim, Henrik and Helled, Ravit},
	month = dec,
	year = {2024},
	note = {ADS Bibcode: 2024ApJ...977..227K},
	pages = {227},
}

@article{lienert_changing_2025,
	title = {Changing disc compositions via internal photoevaporation: {II}. {M} dwarf systems},
	volume = {700},
	issn = {0004-6361},
	shorttitle = {Changing disc compositions via internal photoevaporation},
	url = {https://ui.adsabs.harvard.edu/abs/2025A&A...700A..67L},
	doi = {10.1051/0004-6361/202553904},
	urldate = {2026-03-05},
	journal = {Astronomy and Astrophysics},
	publisher = {EDP},
	author = {Lienert, J. L. and Bitsch, B. and Henning, Th.},
	month = aug,
	year = {2025},
	note = {ADS Bibcode: 2025A\&A...700A..67L},
	pages = {A67},
}

@article{knierim_unraveling_2025,
	title = {Unraveling the origin of giant exoplanets: {Observational} implications of convective mixing},
	volume = {698},
	issn = {0004-6361},
	shorttitle = {Unraveling the origin of giant exoplanets},
	url = {https://ui.adsabs.harvard.edu/abs/2025A&A...698L...1K},
	doi = {10.1051/0004-6361/202554506},
	urldate = {2026-03-05},
	journal = {Astronomy and Astrophysics},
	publisher = {EDP},
	author = {Knierim, H. and Helled, R.},
	month = jun,
	year = {2025},
	note = {ADS Bibcode: 2025A\&A...698L...1K},
	pages = {L1},
}

@article{booth_planet-forming_2019,
	title = {Planet-forming material in a protoplanetary disc: the interplay between chemical evolution and pebble drift},
	volume = {487},
	issn = {0035-8711},
	shorttitle = {Planet-forming material in a protoplanetary disc},
	url = {https://doi.org/10.1093/mnras/stz1488},
	doi = {10.1093/mnras/stz1488},
	number = {3},
	urldate = {2026-03-05},
	journal = {Mon Not R Astron Soc},
	author = {Booth, R A and Ilee, J D},
	month = aug,
	year = {2019},
	pages = {3998--4011},
}

@article{danti_composition_2023,
	title = {Composition of giant planets: {The} roles of pebbles and planetesimals},
	volume = {679},
	issn = {0004-6361},
	shorttitle = {Composition of giant planets},
	url = {https://ui.adsabs.harvard.edu/abs/2023A&A...679L...7D},
	doi = {10.1051/0004-6361/202347501},
	urldate = {2026-03-27},
	journal = {Astronomy and Astrophysics},
	publisher = {EDP},
	author = {Danti, C. and Bitsch, B. and Mah, J.},
	month = nov,
	year = {2023},
	note = {ADS Bibcode: 2023A\&A...679L...7D},
	pages = {L7},
}

@article{guzman_franco_how_2026,
	title = {How initial disc conditions sculpt the atmospheric composition of giant planets},
	volume = {707},
	issn = {0004-6361},
	url = {https://ui.adsabs.harvard.edu/abs/2026A&A...707A.276G},
	doi = {10.1051/0004-6361/202556632},
	urldate = {2026-03-27},
	journal = {Astronomy and Astrophysics},
	publisher = {EDP},
	author = {Guzmán Franco, Angie Daniela and Savvidou, Sofia and Bitsch, Bertram},
	month = mar,
	year = {2026},
	note = {ADS Bibcode: 2026A\&A...707A.276G},
	pages = {A276},
}
\bibliographystyle{aasjournalv7}

\end{document}